\documentclass{article}

\usepackage{arxiv}
\usepackage[utf8]{inputenc}
\usepackage[T1]{fontenc}
\usepackage{hyperref}
\usepackage{url}
\usepackage{booktabs}
\usepackage{amsfonts}
\usepackage{nicefrac}
\usepackage{microtype}
\usepackage{graphicx}
\usepackage{natbib}
\usepackage{doi}
\usepackage{caption}
\usepackage{subcaption}
\usepackage{algorithm}
\usepackage{algorithmic}
\usepackage{amsmath}

\hypersetup{colorlinks=true,linkcolor=blue,citecolor=blue,urlcolor=blue}

\title{Hölder Dither: Free-Form Artistic Dithering Using Generalized Means}

\author{
\href{https://orcid.org/0000-0003-3618-4166}{\includegraphics[scale=0.06]{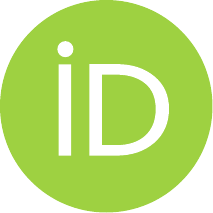}\hspace{1mm}Ergun Akleman} \\
Visual Computing \& Computational Media, SPVFA \\
Computer Science and Engineering, COE \\
Texas A\&M University, USA \\
\texttt{ergun@tamu.edu}
}

\date{November 2026}
\begin{document}

\maketitle

\begin{abstract}
We introduce \emph{Hölder Dither}, a functional framework for real-time artistic dithering inspired by engraving, hatching, crosshatching, and pointillism. Our approach replaces predefined dithering matrices with procedurally constructed dithering functions based on normalized distances to pencils of equidistant parallel lines. By combining these periodic functions using generalized Hölder averages, we obtain continuous families of patterns ranging from intersecting lines to isolated point-like shapes. The Hölder exponent controls both pattern geometry and tonal appearance, while spatially varying pencil parameters enable non-stationary patterns that adapt to local image features. The framework emphasizes artistic flexibility rather than faithful tonal reproduction. Its inherently parallel structure enables efficient GPU-based implementation, as demonstrated through a variety of monochromatic and color images generated in real time. A publicly available suite of interactive browser-based applications accompanies this work, enabling readers to explore the proposed framework and related statistical and geometric dithering methods using their own images.
\end{abstract}

\begin{figure}[htbp!]
\centering
\begin{subfigure}[b]{0.32\textwidth}
  \includegraphics[width=0.99\textwidth]{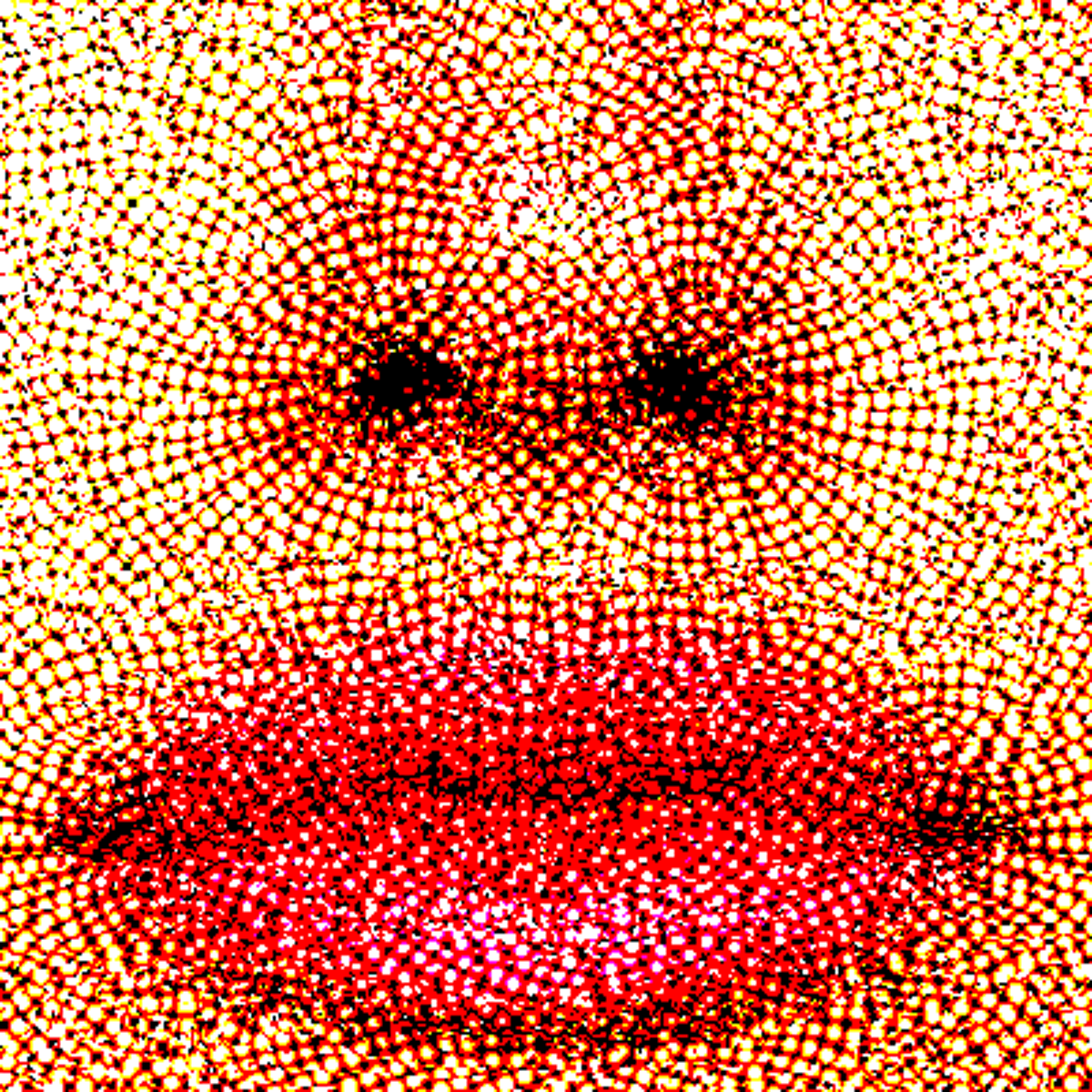}
        \caption{Crosshatching.}
        \label{fig_teaser1}
    \end{subfigure}   
          \hfill
    \begin{subfigure}[b]{0.32\textwidth}
  \includegraphics[width=0.99\textwidth]{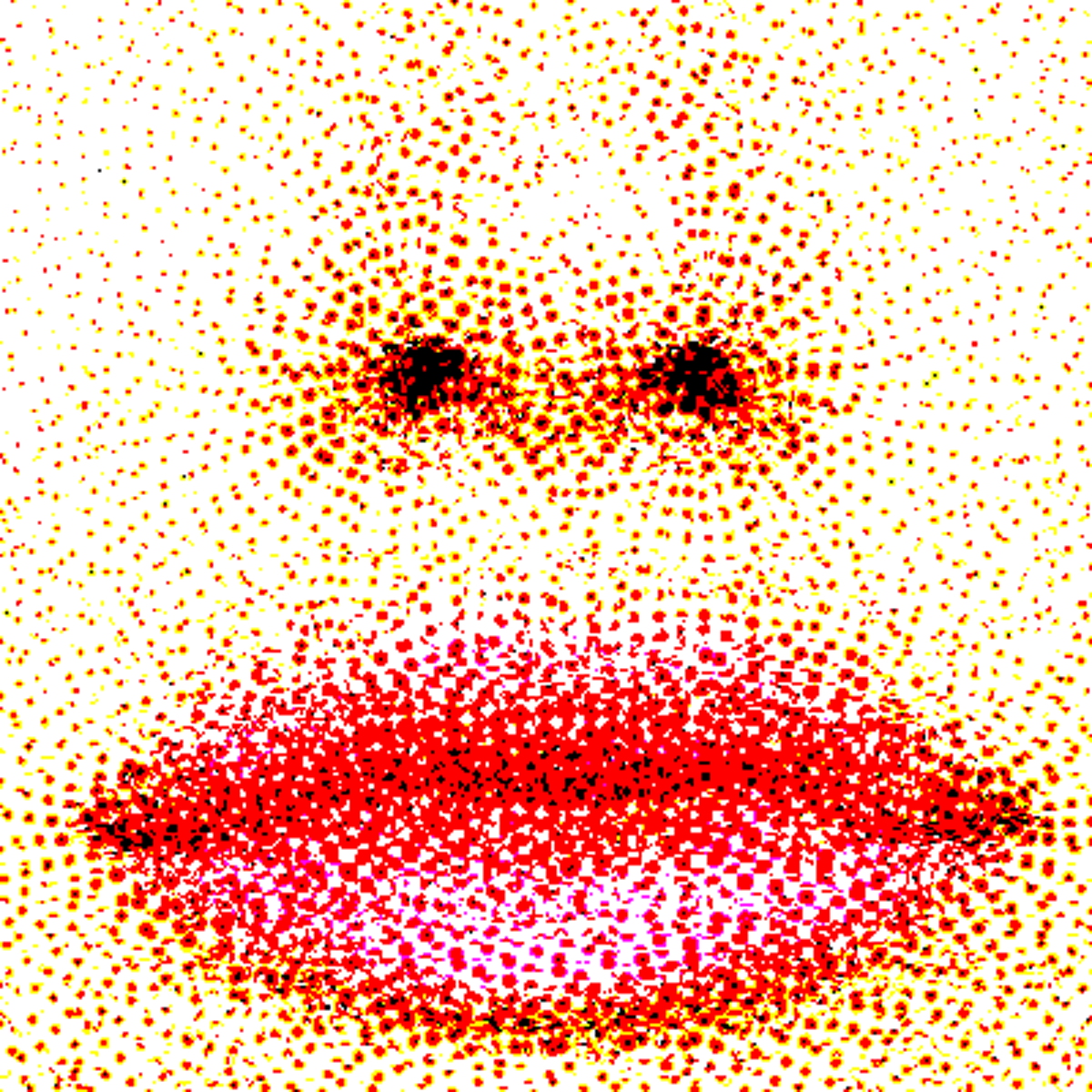}
        \caption{Pointillist with large dot sizes.}
        \label{fig_teaser2}
    \end{subfigure}   
          \hfill
\begin{subfigure}[b]{0.32\textwidth}
  \includegraphics[width=0.99\textwidth]{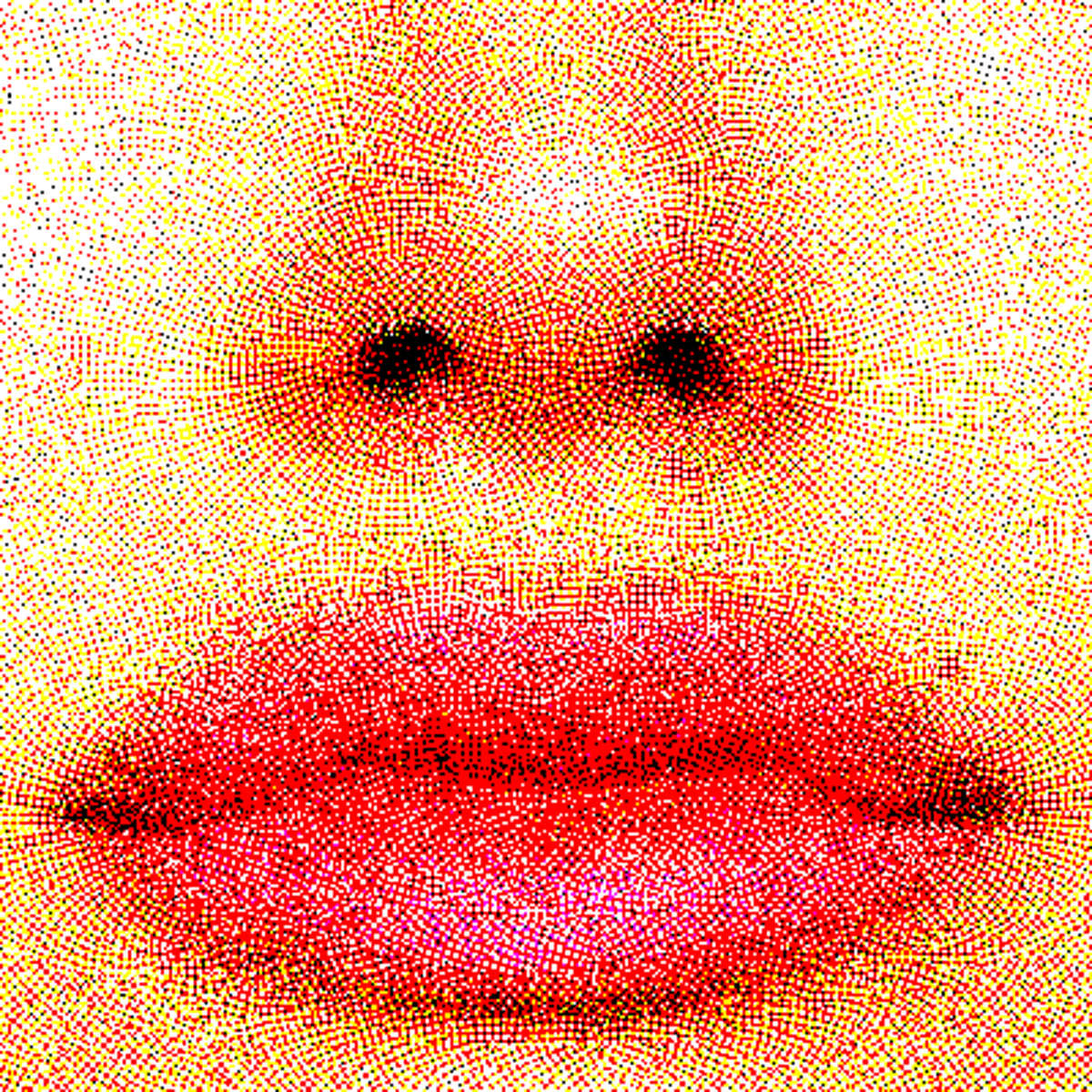}
        \caption{Pointillist with very small dot sizes.}
        \label{fig_teaser0}
    \end{subfigure}   
          \hfill
  \caption{By applying the Hölder average to the distance from a parallel set of lines, various artistic effects can be achieved. Notice how the crosshatching and dot patterns dynamically adapt to the local structure of the image while also introducing subtle human-like imperfections. These images were generated in real-time using Shader Playground, our software inspired by ShaderToy, with the final outputs captured directly from the screen. }
\label{fig_teaser}
\end{figure}

\section{Introduction and Motivation}

Our motivation in this paper is to develop a mathematical and computational framework for artistic image rendering inspired by traditional engraving, etching, hatching, crosshatching, and pointillism. In traditional artistic practices, the arrangement, orientation, spacing, and thickness of strokes and dots are not determined solely by the need to reproduce tonal values. Artists also manipulate these geometric properties to emphasize local image structures, introduce visual rhythms, and create distinctive artistic styles. Subtle irregularities in these patterns can further contribute to the handcrafted appearance of the resulting images.

\begin{figure}[htbp!]
\centering
\includegraphics[width=0.32\textwidth]{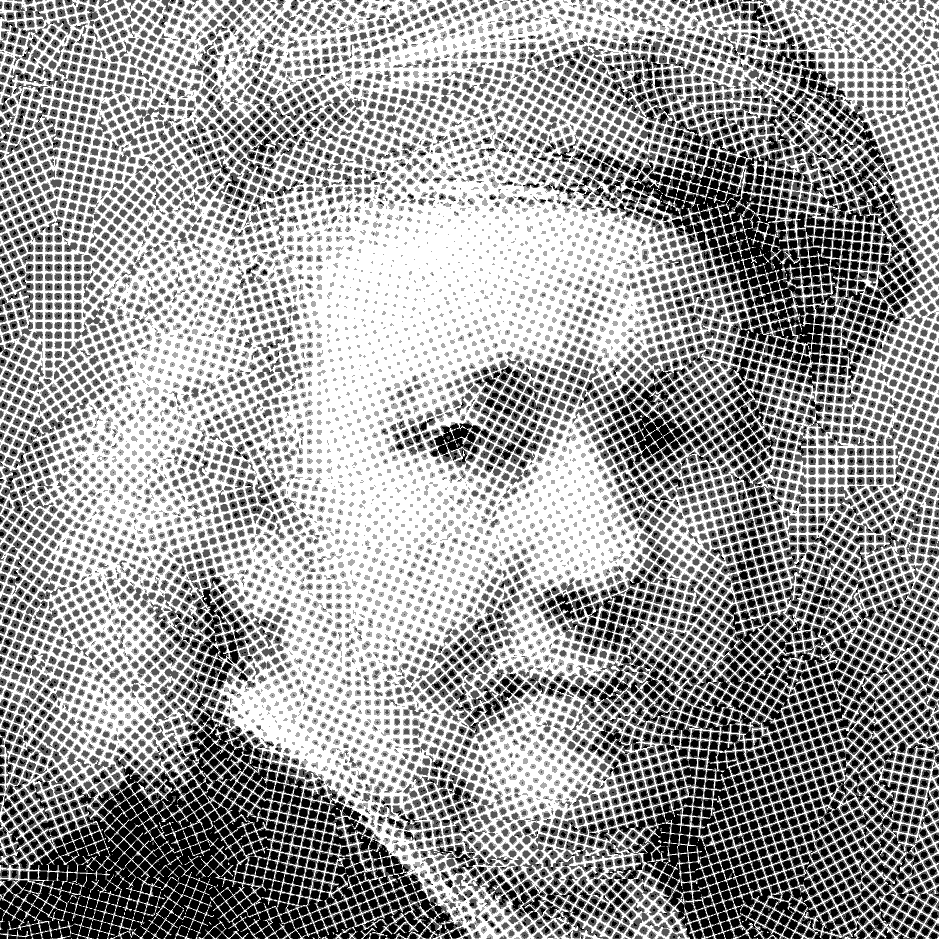} \hfill
\includegraphics[width=0.32\textwidth]{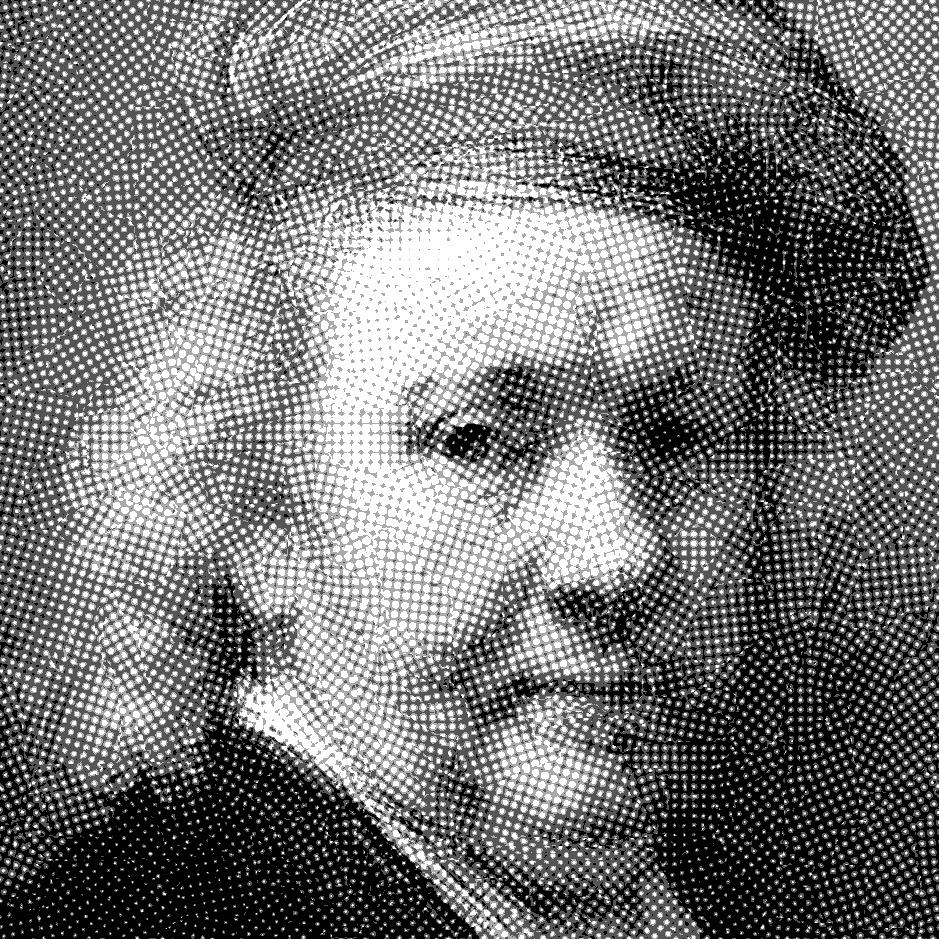} \hfill
\includegraphics[width=0.32\textwidth]{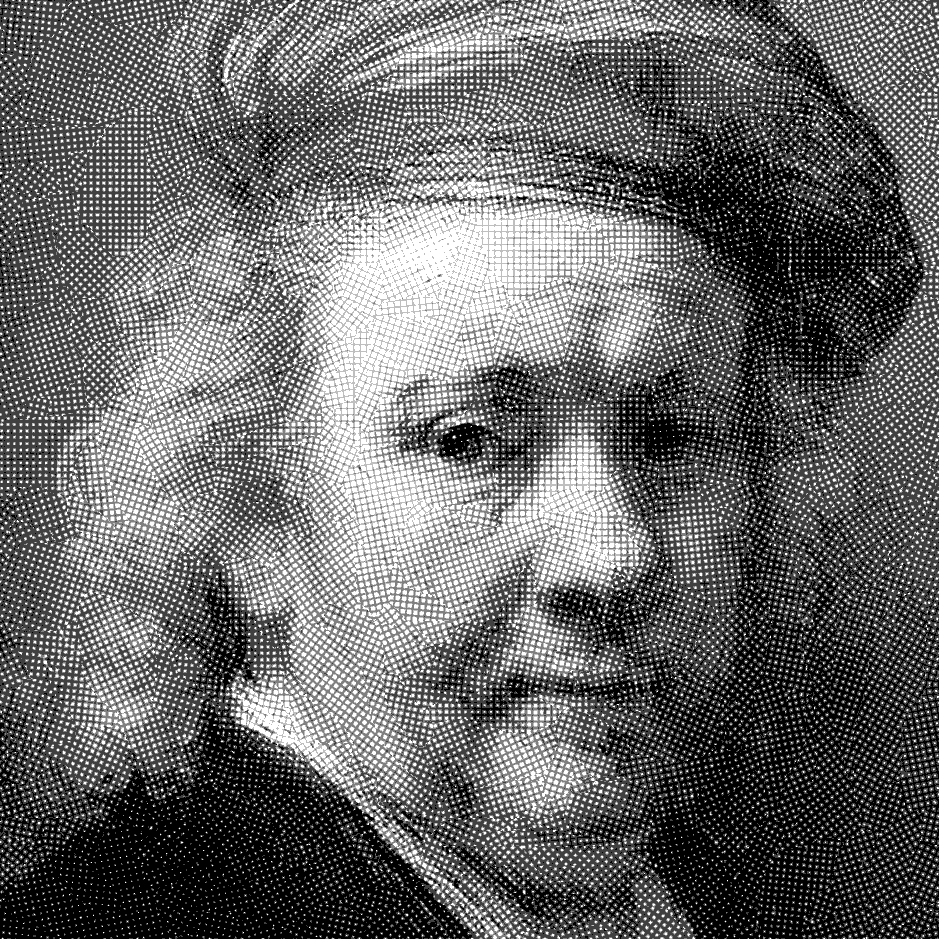}   
  \caption{Three examples of black-and-white crosshatching and pointillist images obtained by converting a Rembrandt painting using our dithering method. }
\label{fig_teaser4}
\end{figure}

A central objective of our framework is to combine \textbf{geometric precision} with \textbf{artistic variability}. We seek to construct mathematically controlled patterns whose geometric properties can be dynamically modified in response to local image features and artistic preferences. Figures~\ref{fig_teaser}, \ref{fig_teaser4}, and \ref{fig_teaser5} demonstrate the variety of artistic effects that can be obtained within this framework, ranging from directional hatching and crosshatching to pointillist and colored pencil effects.

\begin{figure}[htbp!]
\centering
\begin{subfigure}[b]{0.99\textwidth}
  \includegraphics[width=0.32\textwidth]{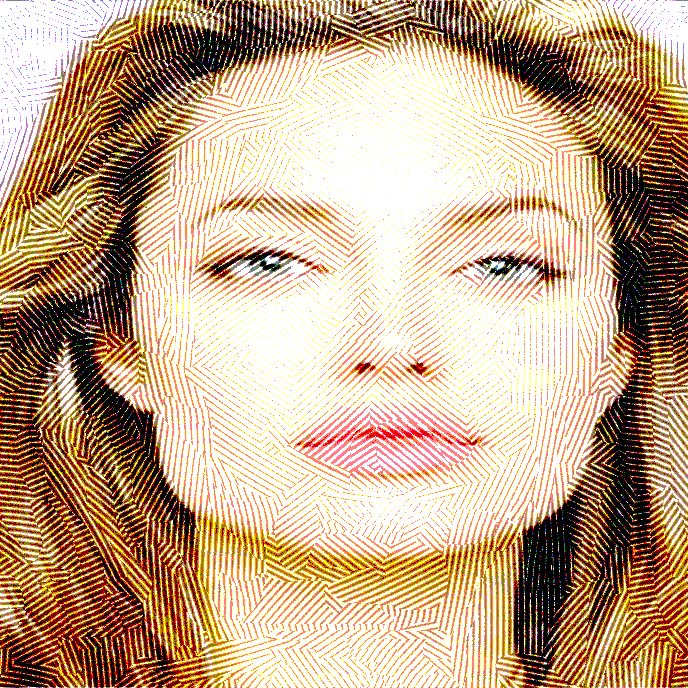} \hfill
\includegraphics[width=0.32\textwidth]{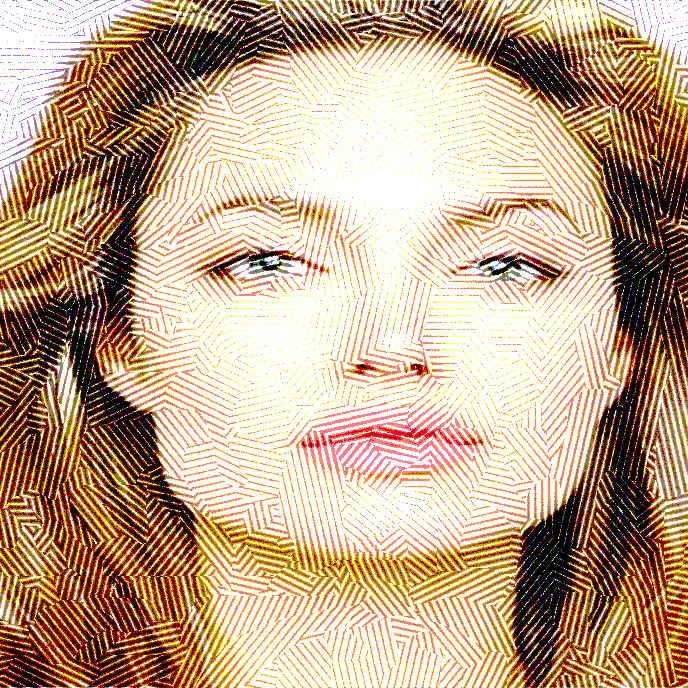} \hfill
\includegraphics[width=0.32\textwidth]{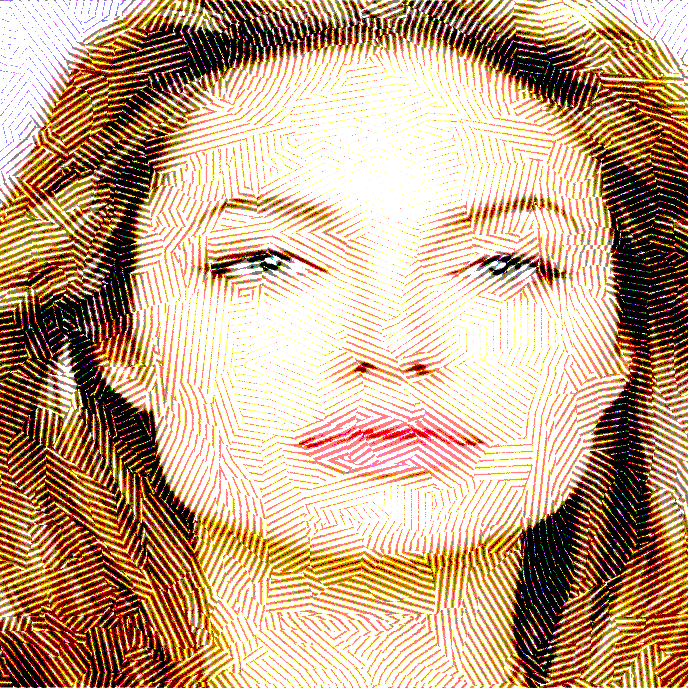} 
        \caption{Examples of hatching created by using equations of pencils of lines that are oriented based on image gradient and tangent.}
        \label{fig_jolie0506}
    \end{subfigure}   
          \hfill
\begin{subfigure}[b]{0.99\textwidth}
  \includegraphics[width=0.32\textwidth]{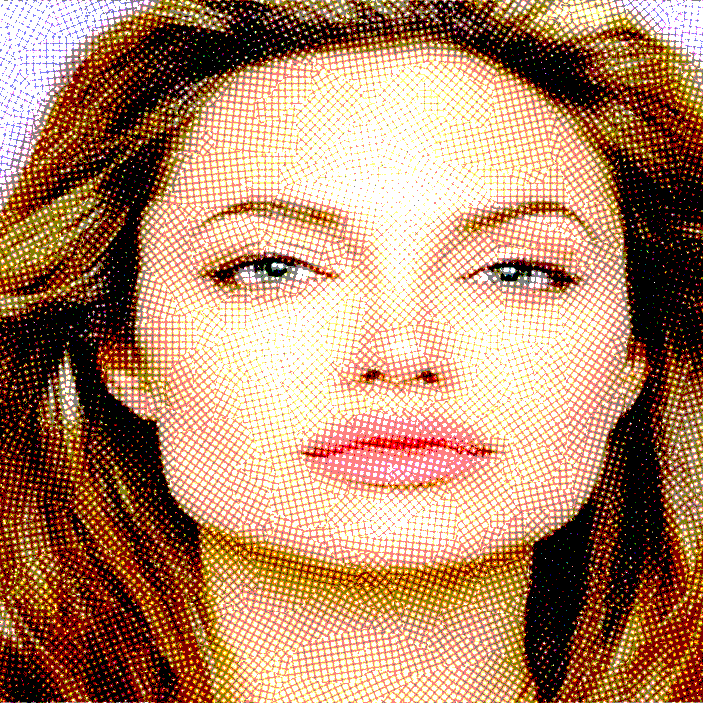} \hfill
\includegraphics[width=0.32\textwidth]{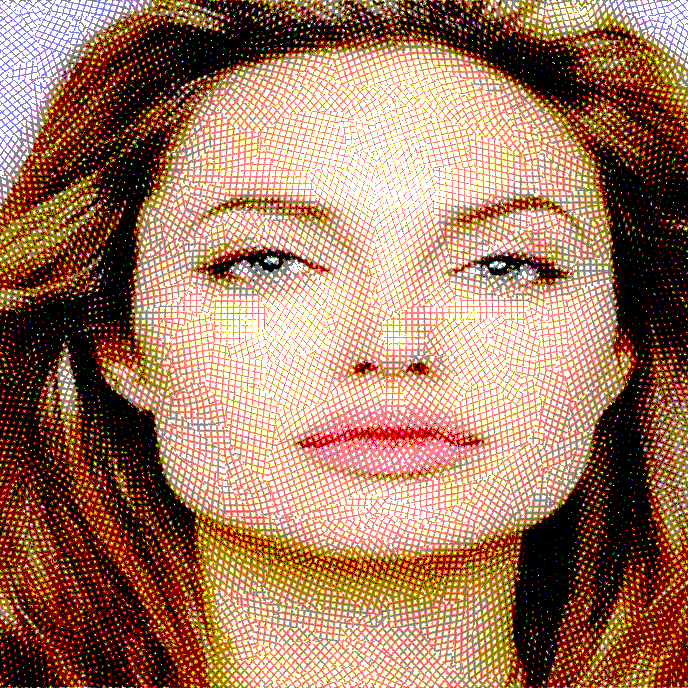} \hfill
 \includegraphics[width=0.32\textwidth]{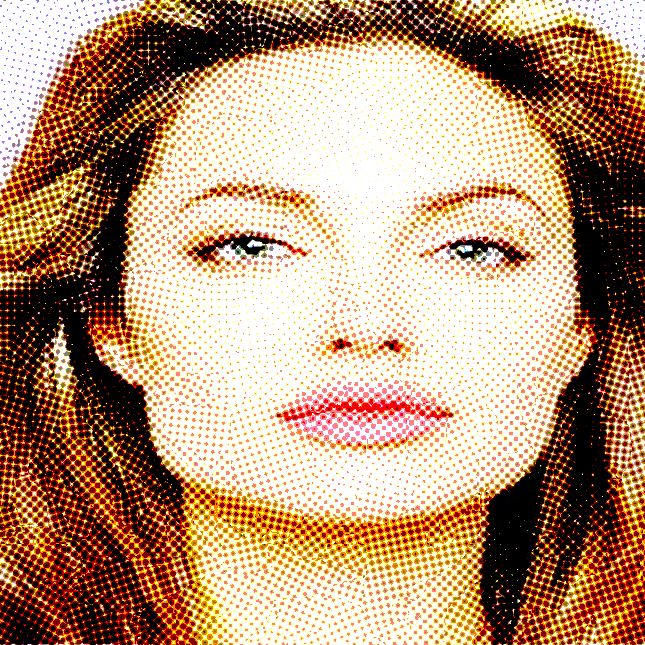} 
        \caption{Examples of Crosshatching and Pointillist images created as Hölder averages of two sets of pencil lines that are perpendicular to each other and oriented based on the image gradient and tangent.}
        \label{fig_jolie0304}
    \end{subfigure}   
          \hfill
  \caption{Examples of hatching and crosshatching. These images were also generated in real-time with the final outputs captured directly from the screen. }
\label{fig_teaser_jolie}
\end{figure}

\begin{figure}[htbp!]
    \centering
\begin{subfigure}[b]{0.31\textwidth}     \fbox{\includegraphics[width=\textwidth]{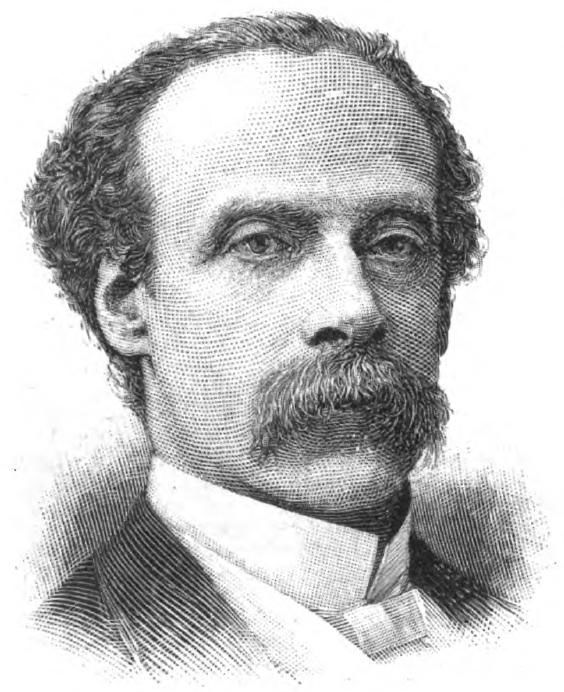}}
        \caption{Presidente de Chile (1886-1891) José Manuel Balmaceda.}
        \label{images/e0}
    \end{subfigure}
    \hfill
    \begin{subfigure}[b]{0.31\textwidth}
        \fbox{\includegraphics[width=\textwidth]{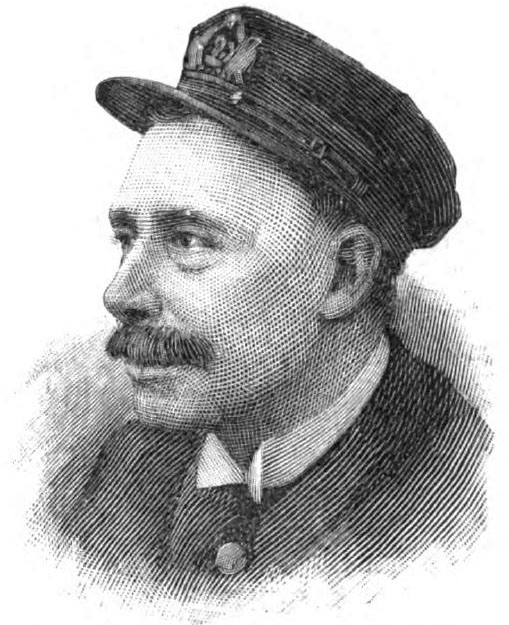}}
        \caption{Captain Carlos Moraga, of the "Almitante Condell"}
        \label{images/e1}
    \end{subfigure}
    \hfill
    \begin{subfigure}[b]{0.31\textwidth}       \fbox{\includegraphics[width=\textwidth]{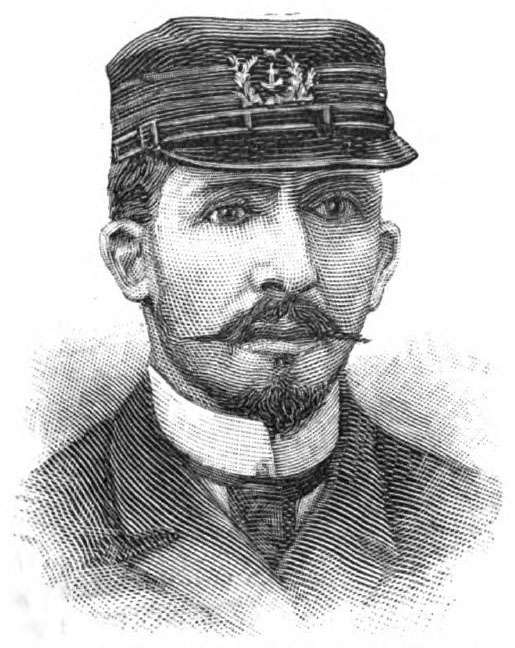}}
        \caption{Captain Alberto Fuentes, of the "Almirante Lynch".}
        \label{images/e2}
    \end{subfigure}
    \caption{Three hand-drawn engraved illustrations from Harper's Weekly published in 1891 (Illustrator is unknown).}
    \label{fig_historical_Engrave}
    \hfill
\end{figure}

Unlike conventional dithering applications, our primary objective is not faithful tonal reproduction. Instead, we view variations in tonal appearance as an additional degree of artistic freedom. For instance, changes in the geometry of dithering patterns can simultaneously produce lighter or darker images, as illustrated by the historical monochromatic examples in Figure~\ref{fig_historical_Engrave}. Rather than compensating for the tonal changes, we consider them useful properties that can be exploited during artistic exploration (see Figures~\ref{fig_teaser4} and~\ref{fig_teaser5}).

Another important objective is to provide these capabilities within a \textbf{real-time interactive system}. Interactive control allows artists and designers to explore different pattern geometries, tonal appearances, and stylistic variations without requiring computationally expensive preprocessing. All artistic rendering examples presented in this paper were generated interactively using GPU-based shader implementations, with the final images captured directly from the screen. Readers can interactively explore the proposed method using their own images through our publicly available standalone browser-based application \cite{akleman2026holderdithersoftware}: \url{https://people.tamu.edu/~ergun/software/dithering/Holder_Dither.html}.

\subsection{From Dithering Matrices to Dithering Functions}

Image dithering methods include sequential approaches, such as conventional error diffusion, and pointwise approaches, such as ordered dithering. The latter are particularly suitable for real-time GPU-based applications because dithering decisions can be made independently at each image location. In conventional ordered dithering, these decisions are typically made by comparing image intensities with threshold values stored in a predefined dithering matrix.

The geometric organization of such matrices plays an important role in determining the visual characteristics of the resulting images. Specialized halftoning and artistic screening approaches, including Donald Knuth's dot diffusion method \citep{knuth1987digital} and Victor Ostromoukhov's artistic screening techniques \citep{ostromoukhov1995artistic, ostromoukhov1996anticounterfeiting}, demonstrate the importance of spatial pattern design in digital halftoning. Our objective is complementary to these approaches: we seek to represent and manipulate dithering patterns directly as mathematical functions rather than primarily as discrete arrays of threshold values.

We refer to this approach as \emph{free-form dithering}. Replacing predefined dithering matrices with procedurally constructed dithering functions allows threshold values to be evaluated at arbitrary pixel or subpixel locations. Moreover, the parameters of these functions can vary spatially, enabling the construction of non-stationary dithering patterns whose local geometric properties respond to image content. Figures~\ref{fig_teaser} and~\ref{fig_teaser_jolie} illustrate this capability through hatching and crosshatching patterns whose orientations are determined by local image gradient and tangent directions.

This functional representation also separates the geometric definition of a dithering pattern from the sampling resolution of the final image. Consequently, changes in pattern frequency, orientation, position, and geometry can be performed directly by modifying functional parameters without constructing or storing additional dithering matrices.

\subsection{Implicit Shape Modeling for Artistic Dithering}

The fundamental observation underlying our framework is that dithering patterns can be interpreted as implicitly represented geometric shapes. From this perspective, the construction of artistic dithering patterns becomes a two-dimensional implicit shape modeling problem. This interpretation allows us to employ functional operations originally developed for geometric modeling to construct and manipulate artistic patterns.

As elementary building blocks, we use periodic functions representing normalized distances to pencils of equidistant parallel lines. Each pencil is characterized by its orientation, spacing, and position. These parameters provide direct geometric control over the underlying hatching patterns, while the distance-based representation facilitates control over the thickness of the resulting strokes.

More complex dithering functions are constructed by combining these elementary functions using generalized Hölder averages. A particularly useful property of these averages is that a single exponent provides a continuous transition between minimum and maximum operations. When applied to periodic distance functions, these operations produce families of geometric patterns ranging from unions of line structures to isolated point-like shapes. Intermediate exponent values generate a variety of transitional patterns.

\begin{figure}[htbp!]
\centering
\includegraphics[width=0.32\textwidth]{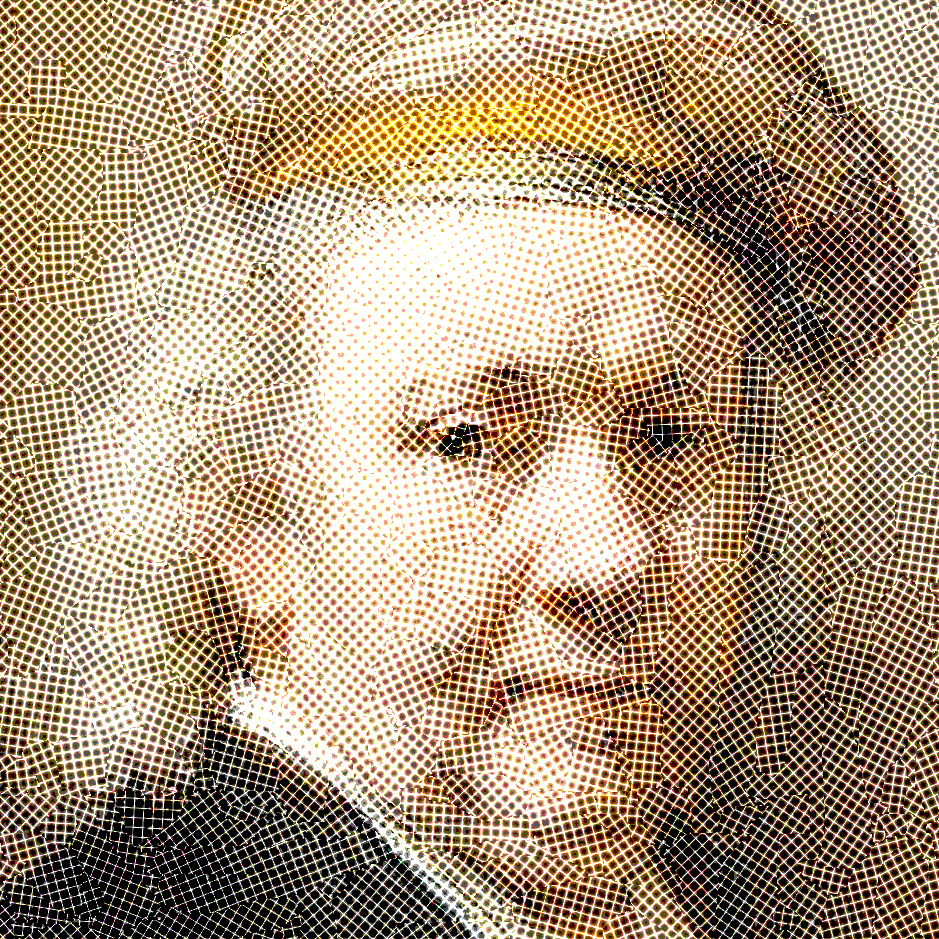} \hfill
\includegraphics[width=0.32\textwidth]{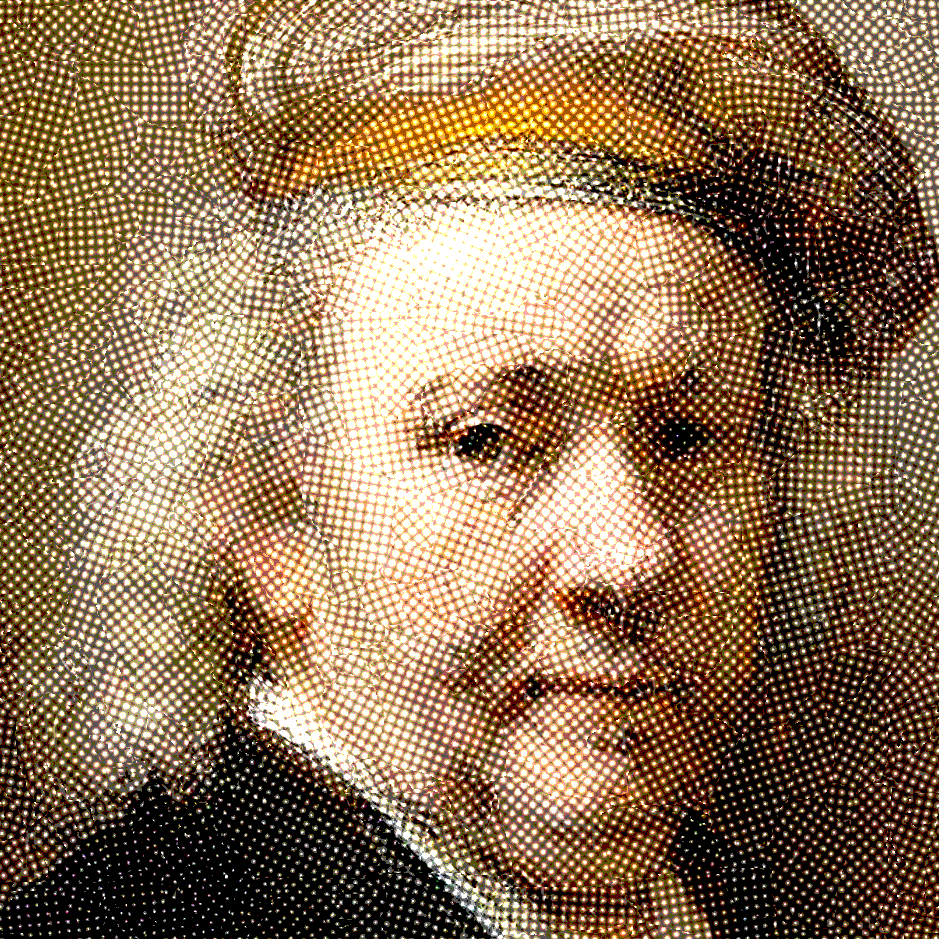} \hfill
\includegraphics[width=0.32\textwidth]{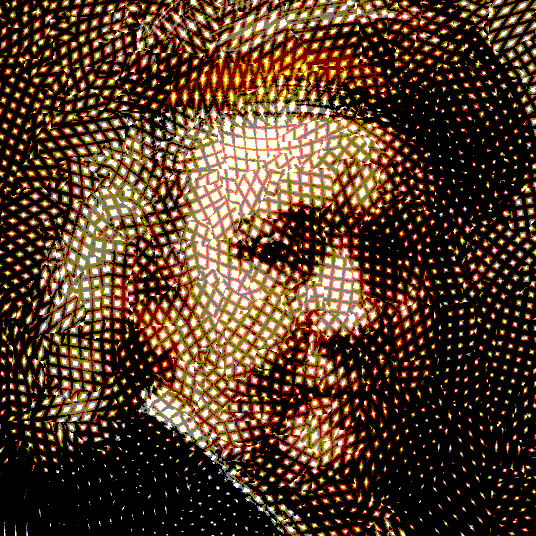}   
  \caption{Three examples of colored pencil crosshatching and pointillist images obtained by converting a Rembrandt painting using our dithering method. }
\label{fig_teaser5}
\end{figure}

An important consequence of this formulation is that the Hölder exponent simultaneously controls \textbf{pattern geometry and tonal appearance}. For a fixed input image, increasing or decreasing the exponent modifies both the shapes and the relative coverage of the resulting black and white regions. In the context of artistic dithering, this coupling is desirable because it provides an additional mechanism for controlling the overall visual character of an image.

Furthermore, spatial variations in the parameters of the underlying pencils and their Hölder averages enable the construction of locally adaptive patterns. Controlled variations in these parameters can also introduce geometric irregularities without requiring explicitly designed irregular tiles. The monochromatic examples in Figure~\ref{fig_teaser4} and the colored pencil examples in Figure~\ref{fig_teaser5} illustrate the artistic versatility of this functional approach.

\subsection{Contributions}

The main contributions of this paper can be summarized as follows:

\begin{itemize}

\item \textbf{A functional framework for artistic dithering:} We formulate artistic dithering as an implicit geometric modeling problem in which patterns are represented by procedurally evaluated functions rather than predefined matrices.

\item \textbf{Distance-based construction of elementary patterns:} We employ periodic functions representing normalized distances to pencils of equidistant parallel lines, providing direct control over pattern orientation, spacing, position, and stroke thickness.

\item \textbf{Hölder-based manipulation of pattern geometry:} We employ generalized Hölder averages to construct continuous families of artistic patterns, providing simultaneous control over their geometry and tonal appearance through a single exponent.

\item \textbf{Non-stationary and real-time artistic rendering:} We show how spatially varying functional parameters generate locally adaptive artistic patterns, while the pointwise formulation enables efficient parallel implementation in GPU-based shader languages.

\end{itemize}

Together, these contributions establish a connection between implicit geometric modeling and artistic dithering, providing a mathematical foundation for the interactive construction and manipulation of engraving-inspired image representations.

\begin{figure}[htbp!]
\centering
\begin{subfigure}[t]{0.52\textwidth}
    \centering
    \includegraphics[width=\linewidth]{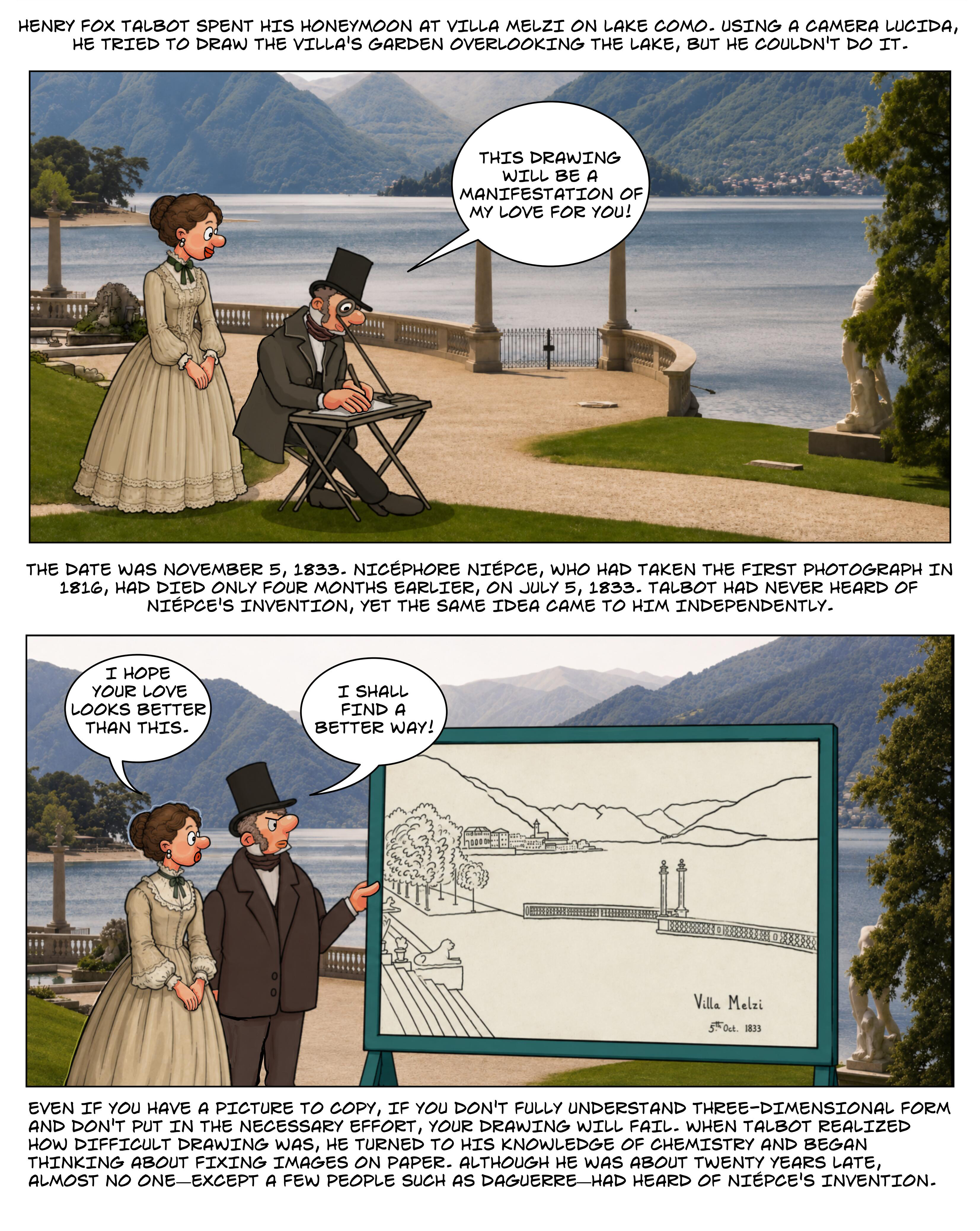}
    \caption{Talbot's artistic motivation for developing photography arose from his unsuccessful attempts to sketch the landscape around Lake Como using a camera lucida, inspiring him to seek a method by which nature could record its own images.}
    \label{fig_cartoon_honeymoon}
\end{subfigure}
\hfill
\begin{subfigure}[t]{0.44\textwidth}
    \centering
    \includegraphics[width=\linewidth]{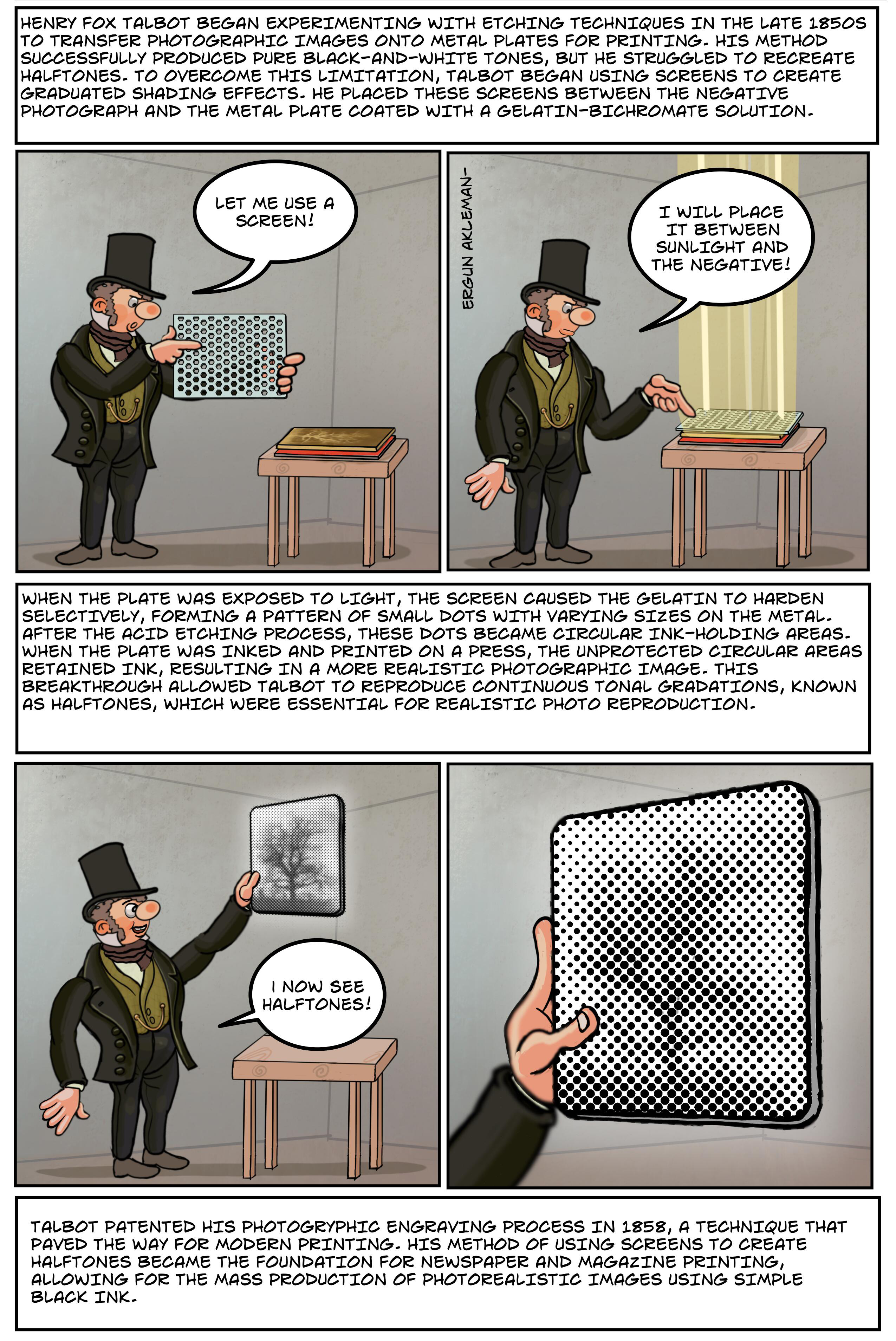}
    \caption{After developing the negative-positive photographic process, Talbot sought more economical ways to reproduce photographs, leading to his pioneering experiments with photographic engraving and screening.}
    \label{fig_cartoon_talbot}
\end{subfigure}
\caption{Two cartoons by the author illustrating how an artistic objective led to a series of technological and engineering innovations. Talbot's unsuccessful attempts to sketch the landscape around Lake Como using a camera lucida motivated his investigations into photography. After developing a negative-positive photographic process that enabled the production of multiple copies from a single negative, Talbot sought more economical ways to reproduce photographs using conventional printing technologies. His experiments with photographic engraving and screening established early principles of what would later become photographic halftoning and artistic dithering. The dialogues and scenes are humorous interpretations of historical developments.}
\label{fig_historical_cartoons0}
\end{figure}

\section{Historical Background and Related Work}

The development of artistic dithering can be viewed within the broader historical context of representing continuous visual information through spatially distributed discrete elements. Throughout this history, two complementary objectives have motivated the development of new image representation techniques: faithful reproduction of visual information and artistic interpretation through the geometric organization of visual elements. Our work is primarily motivated by the latter objective, while benefiting from the computational principles developed for the former. Rather than viewing artistic and engineering innovations as independent developments, we organize the historical discussion around three interconnected phases: artistic motivations leading to photography and photographic screening, the application of screening to mass communication, and the possible influence of photographic technologies on subsequent artistic experimentation. These phases illustrate how engineering and artistic techniques can develop in tandem, without implying that every historical connection is a documented causal relationship.

\subsection{Traditional Engraving and Artistic Interpretation}

Before the invention of photography, engraving and etching were among the principal techniques for reproducing images in printed publications. These techniques were mastered by renowned artists such as Albrecht Dürer, Francisco de Goya, William Hogarth, and Rembrandt van Rijn \citep{hind2011history,ostromoukhov1999digital}. By carefully manipulating the orientation, spacing, thickness, and arrangement of lines and dots, artists could create visual impressions of depth, shading, texture, and material properties using a limited range of printable intensities.

In traditional engraving, the geometric organization of individual marks serves purposes beyond tonal reproduction. For instance, the orientation of hatching lines can emphasize the geometry of a depicted object, while variations in spacing and thickness can contribute to its perceived material properties and visual character. Crosshatching introduces additional geometric relationships between strokes, providing artists with further control over both local appearance and overall composition.

These artistic practices provide the primary motivation for our functional dithering framework. Rather than treating individual marks merely as mechanisms for reproducing continuous tones, we consider their geometric properties as independently controllable elements of artistic expression.

\begin{figure}[htbp!]
\centering
\begin{subfigure}[b]{0.49\textwidth}
    \centering    \includegraphics[width=\linewidth]{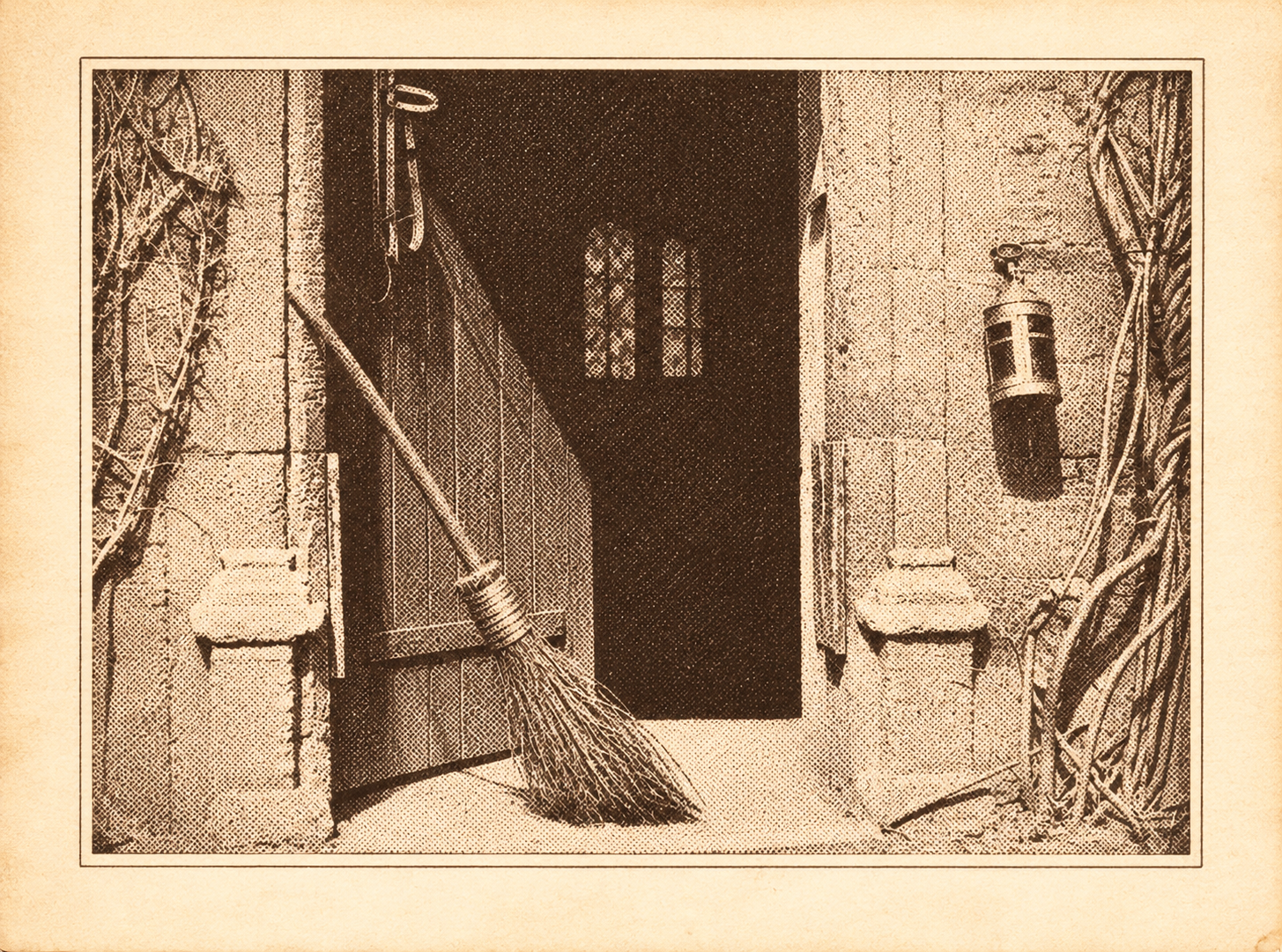}
    \caption{\emph{The Open Door}, Plate VI.}
    \label{fig_talbot_open_door}
\end{subfigure}
\hfill
\begin{subfigure}[b]{0.49\textwidth}
    \centering
    \includegraphics[width=\linewidth]{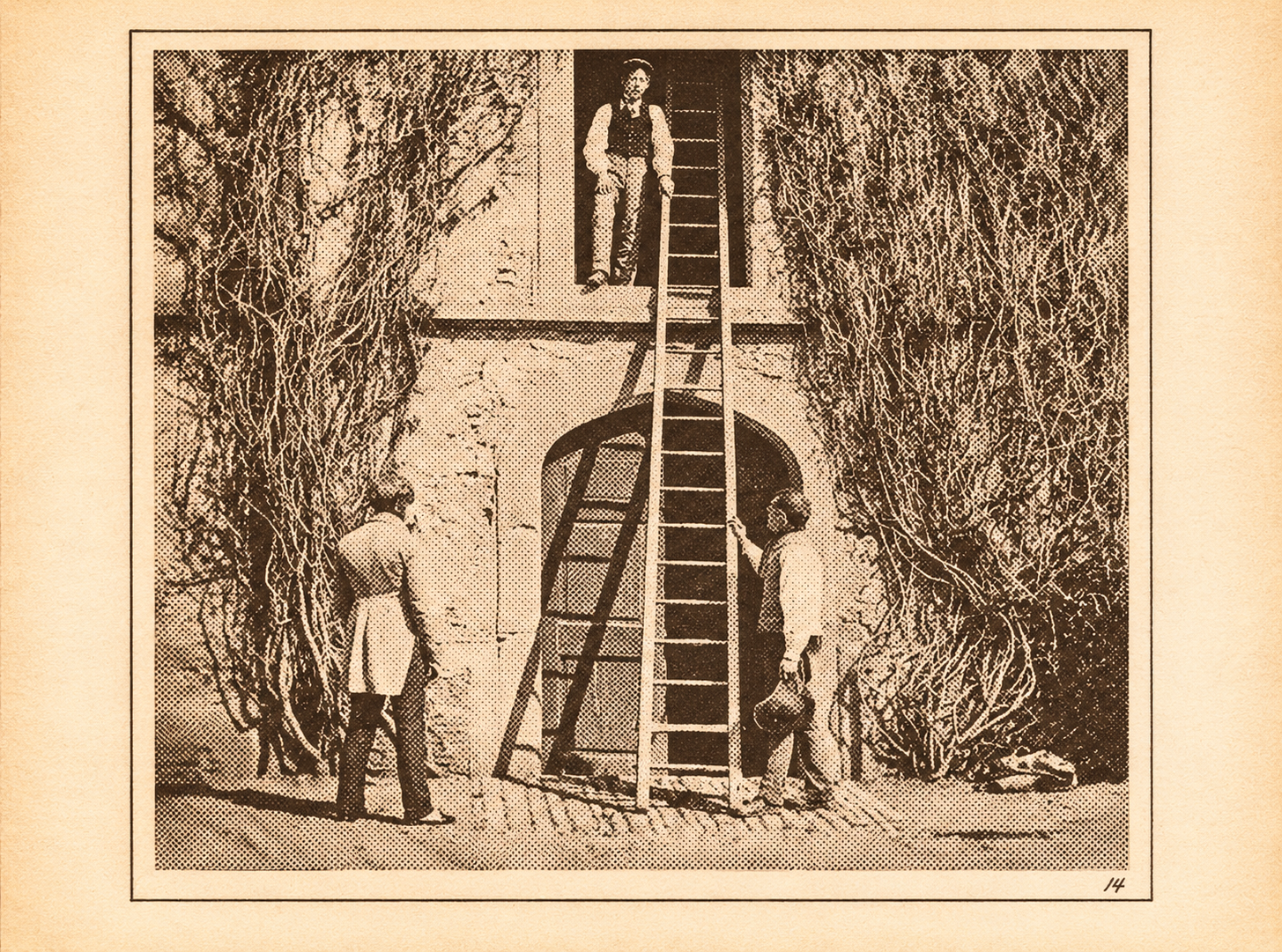}
    \caption{\emph{The Ladder}, Plate XIV.}
    \label{fig_talbot_ladder}
\end{subfigure}
\caption{Two examples from William Henry Fox Talbot's pioneering photographic book, \emph{The Pencil of Nature} (1844--1846) \citep{talbot1844pencil}. The original photographic prints were individually mounted onto the pages rather than reproduced directly using a printing press. These AI-assisted visual reconstructions incorporate illustrative screening patterns to demonstrate how these photographs might have appeared if reproduced using later photographic halftoning techniques. The screening patterns are illustrative and were not present in the original mounted photographic prints.}
\label{fig_talbot_pencil_of_nature}
\end{figure}

\subsection{From Artistic Motivation to Photographic Screening}

An especially interesting example of the interplay between art and engineering is William Henry Fox Talbot's account of his attempts to sketch Lake Como in Italy in October 1833. Frustrated by the difficulty of drawing even with the aid of a camera lucida, Talbot began to consider whether the images formed by optical instruments could be permanently recorded on paper. He recounts this artistic motivation in the introductory remarks of \emph{The Pencil of Nature} \citep{talbot1844pencil}. The cartoon in Figure~\ref{fig_cartoon_honeymoon} provides a humorous interpretation of this episode. His subsequent photographic experiments led to a negative-positive process in which multiple positive photographic prints could be obtained from a single negative. Thus, a problem initially motivated by artistic representation led to a technology with important implications for image reproduction.

The invention of photography introduced a different challenge: how to reproduce continuous-tone photographic images using printing technologies that were primarily designed for transferring ink from discrete regions of a printing surface. Henry Fox Talbot, who published actual photographic prints in his pioneering book \emph{The Pencil of Nature} \citep{talbot1844pencil}, subsequently investigated photographic engraving techniques that would allow photographic images to be reproduced through conventional printing processes. Figure~\ref{fig_talbot_pencil_of_nature} presents two examples from his book, \emph{The Open Door} and \emph{The Ladder}. Although the negative-positive process enabled multiple photographic copies, the individual prints still had to be produced and mounted onto the book pages. The later challenge was to reproduce photographs directly in ink, alongside printed text.

Talbot's experiments with screens during the 1850s represent an important step toward photographic halftoning. The fundamental idea was to transform continuous tonal variations into spatial variations in the geometry and coverage of printable elements. Figure~\ref{fig_cartoon_talbot} presents an illustrated account of this principle, emphasizing how apparent intermediate tones can emerge from spatial arrangements of black and white regions.

\subsection{Photographic Screening and Mass Communication}

\begin{figure}[htbp!]
\centering
\begin{subfigure}[t]{0.475\textwidth}
    \centering
    \includegraphics[width=\linewidth]{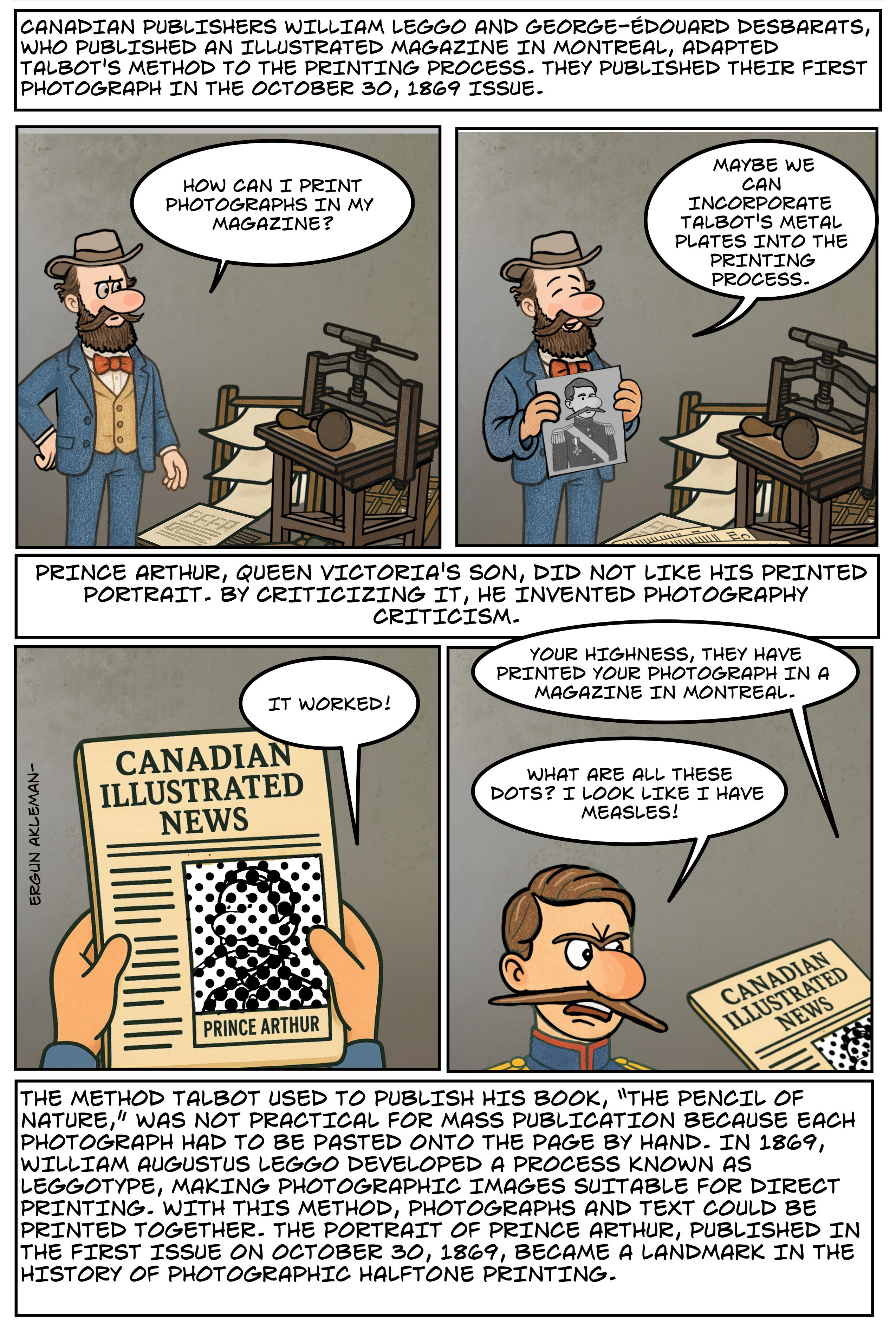}
    \caption{Leggo and Desbarats and the reproduction of photographs in illustrated publications.}
    \label{fig_cartoon_leggo}
\end{subfigure}
\hfill
\begin{subfigure}[t]{0.48\textwidth}
    \centering
    \includegraphics[width=\linewidth]{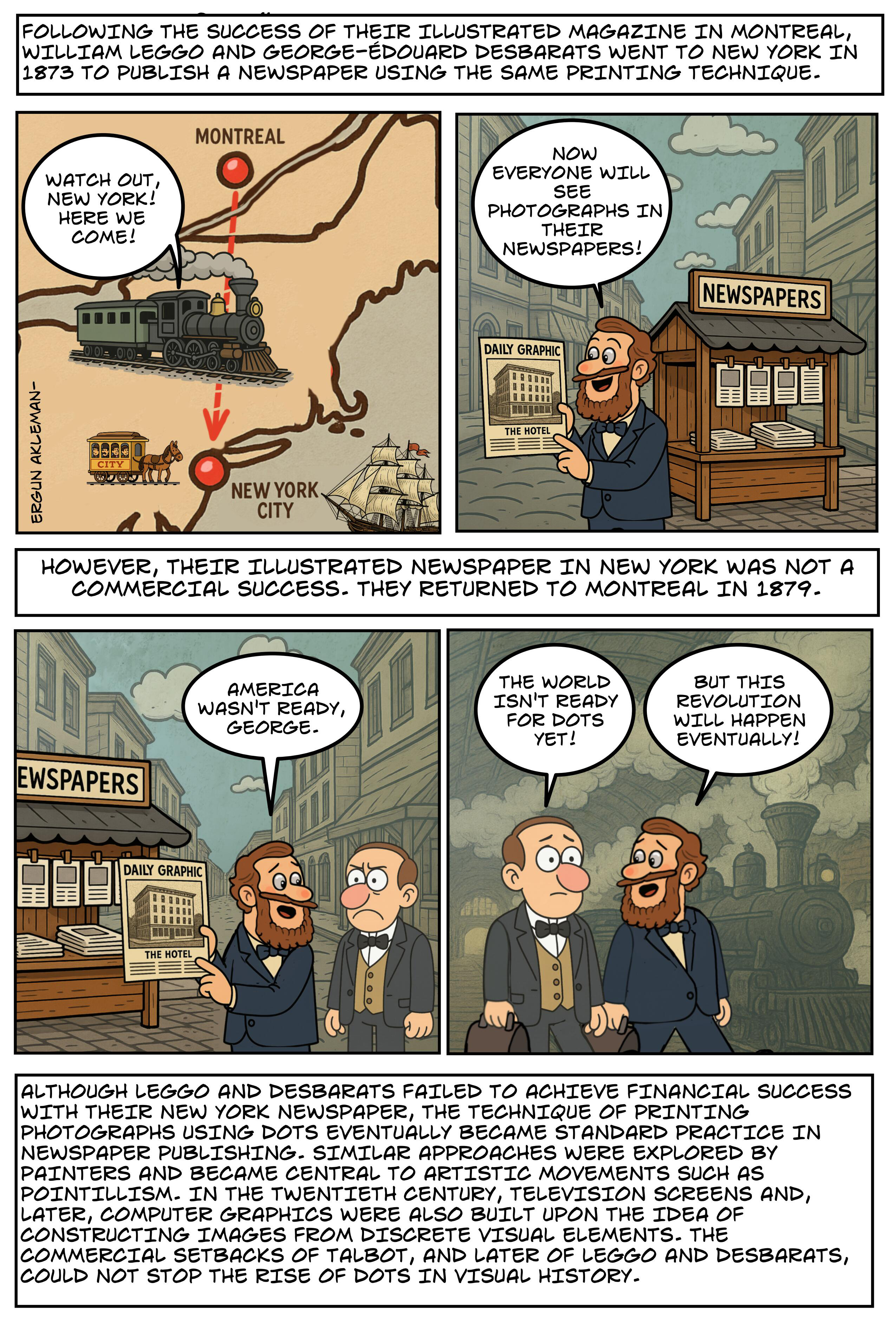}
    \caption{The introduction of photographic reproduction into illustrated daily newspapers.}
    \label{fig_cartoon_newspaper}
\end{subfigure}
\caption{Two cartoons by the author illustrating the transition from photographic screening to mass communication through illustrated magazines and newspapers. The dialogues are fictional.}
\label{fig_historical_cartoons1}
\end{figure}

The next phase concerned the integration of photographic reproduction into the ordinary printing and publishing workflow. Figures~\ref{fig_cartoon_leggo} and \ref{fig_cartoon_newspaper} illustrate this transition through the work of Leggo and Desbarats.

\begin{figure}[htp!]
    \centering
    \begin{subfigure}[b]{0.44\textwidth}        \fbox{\includegraphics[width=\textwidth]{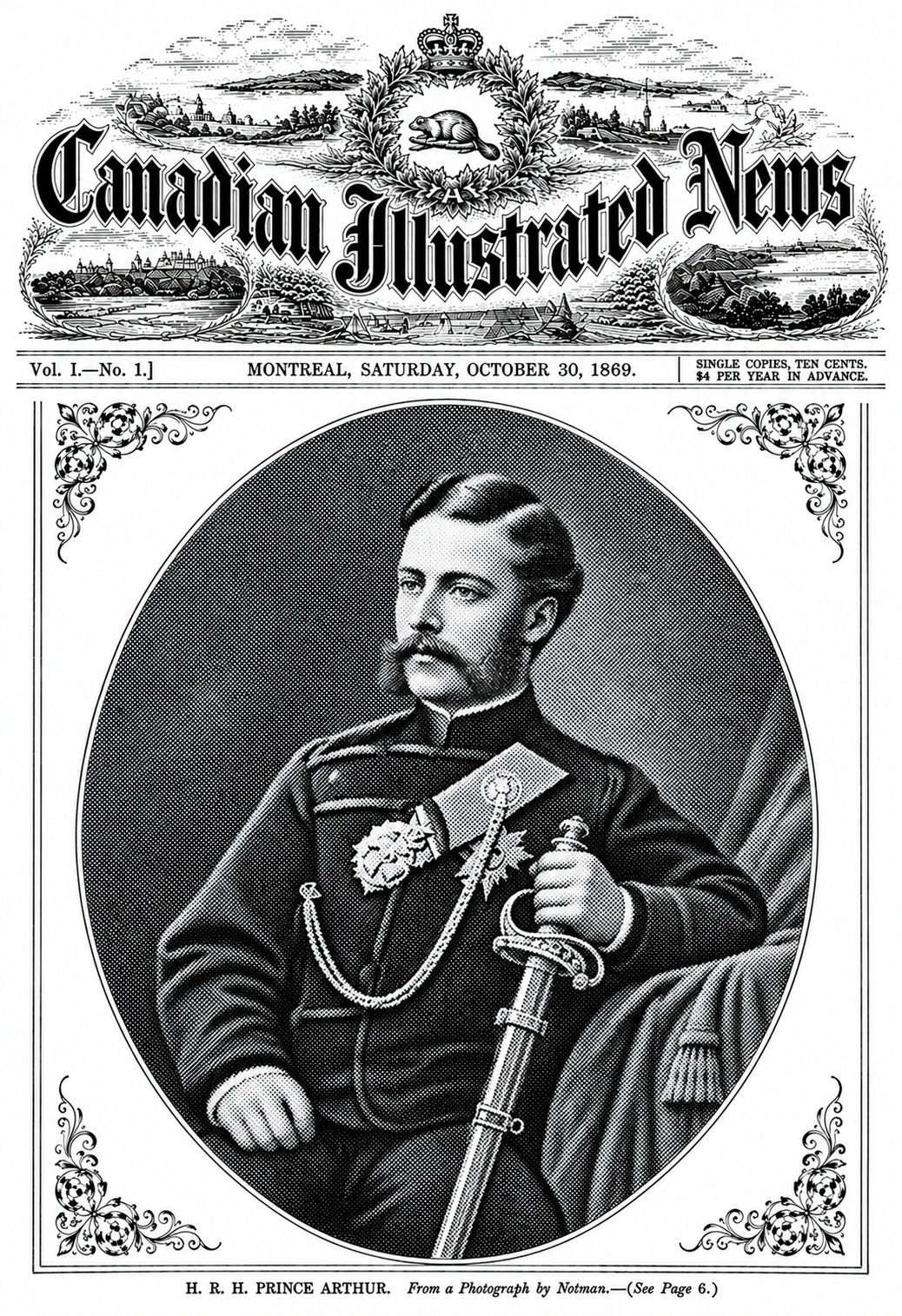}}
        \caption{Prince Arthur}
        \label{images/h0}
    \end{subfigure}
    \hfill
    \begin{subfigure}[b]{0.52\textwidth}      \fbox{\includegraphics[width=\textwidth]{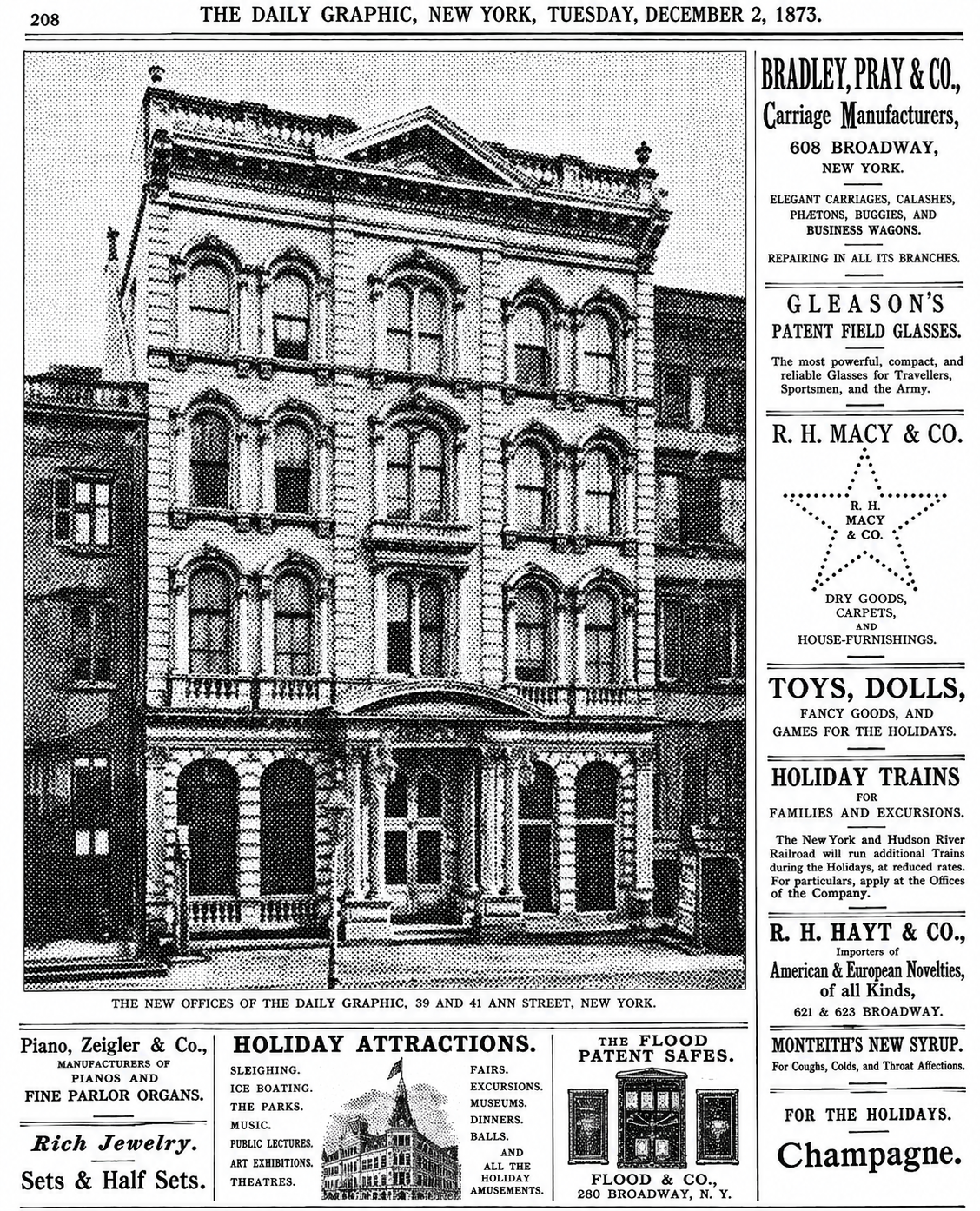}}
        \caption{Steinway Hall}
        \label{images/h1}
    \end{subfigure}
    \caption{Two AI-assisted visual reconstructions of early photomechanical reproductions associated with George-Édouard Desbarats and William Leggo \citep{lamonde2005history, galarneau1984presse}. The portrait of Prince Arthur in (a) appeared in the inaugural issue of the \emph{Canadian Illustrated News} on October 30, 1869. The image of Steinway Hall on East 14th Street, between University Place and Fifth Avenue in Manhattan, in (b) appeared in the New York \emph{Daily Graphic} on December 2, 1873, and has been identified as an early example of halftone photographic reproduction in a United States periodical. The reconstructed details and screening patterns are illustrative rather than archival evidence.}
    \label{fig_historical_Halftone0}
\end{figure}

During the late 1860s, William Augustus Leggo and George-Édouard Desbarats further developed photographic reproduction techniques for illustrated publications, including the \emph{Canadian Illustrated News} and \emph{L'Opinion publique} \citep{black2008canadian,lamonde2005history,galarneau1984presse}. A notable example was the photographic portrait of Prince Arthur published in the October 30, 1869 issue of the \emph{Canadian Illustrated News}. Figure~\ref{fig_cartoon_leggo} illustrates this historical development, and Figure~\ref{images/h0} presents an illustrative reconstruction of the published portrait.

\begin{figure}[htp!]
    \centering  
 \fbox{\includegraphics[width=\textwidth]{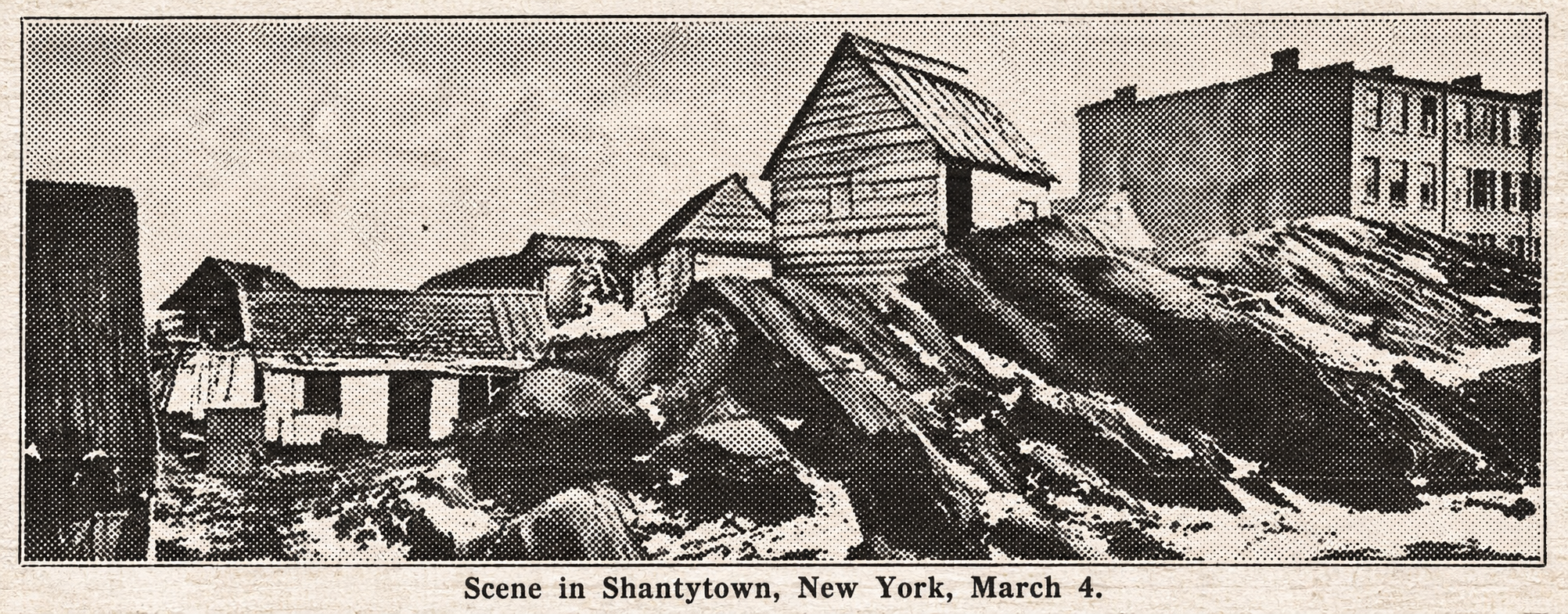}}
        \caption{\emph{A Scene in Shantytown}, New York, March 4, 1880. An AI-assisted visual reconstruction of an early newspaper halftone photograph published in the \emph{Daily Graphic}, the newspaper established in 1873 by Leggo and Desbarats. The screening pattern is illustrative rather than a reproduction of the original printing structure.}
        \label{fig_historical_Halftone1}
\end{figure}

Leggo and Desbarats subsequently extended their illustrated publishing activities to New York, where they participated in the establishment of the \emph{Daily Graphic} in 1873. Their efforts illustrate the technological and commercial challenges associated with introducing photographic reproduction into mass-market publications, as depicted in Figure~\ref{fig_cartoon_newspaper}. Despite these challenges, halftoning eventually became a fundamental technology for the widespread reproduction of photographic imagery. The December 2, 1873 reproduction of Steinway Hall in the \emph{Daily Graphic} (Figure~\ref{images/h1}) and the March 4, 1880 image \emph{A Scene in Shantytown} (Figure~\ref{fig_historical_Halftone1}) illustrate successive milestones in early photographic printing. The latter is commonly associated with Stephen Henry Horgan and is often cited as a pioneering screened newspaper photograph. Together with the earlier Canadian example, these images illustrate the gradual development of photomechanical reproduction rather than a single unambiguous priority claim.

From the perspective of our work, the important conceptual development was the recognition that continuous visual information could be represented through the spatial organization of discrete geometric elements. Unlike traditional engraving, however, photographic halftoning was primarily motivated by the faithful reproduction of photographic tones rather than the independent artistic manipulation of the geometry of individual marks.

\subsection{The Interplay Between Photographic Technology and Artistic Innovation}

\begin{figure}[htbp!]
\centering
\begin{subfigure}[t]{0.48\textwidth}
    \centering
    \includegraphics[width=\linewidth]{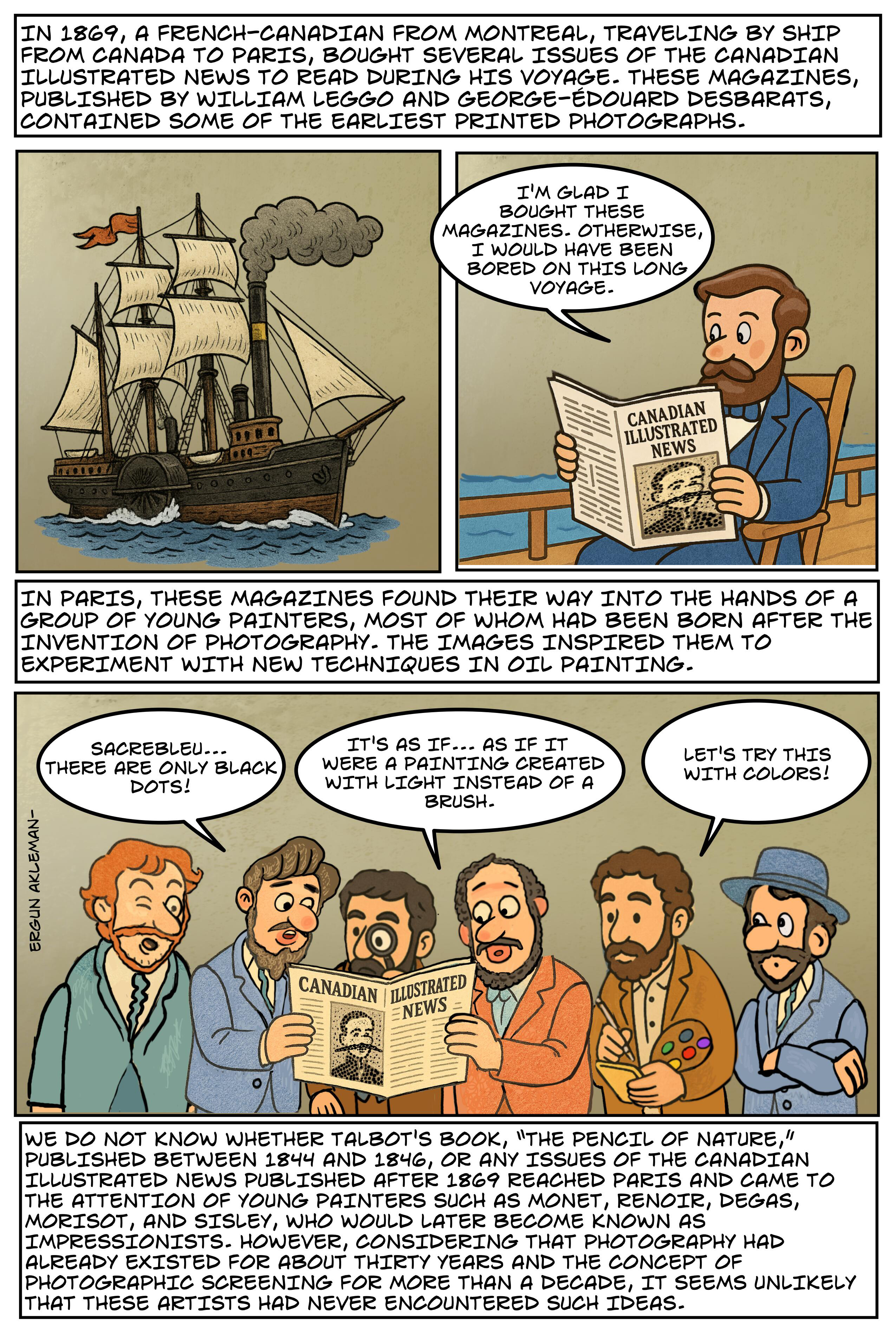}
    \caption{A speculative account of the possible influence of photographic screening on Impressionist painting.}
    \label{fig_cartoon_impressionism}
\end{subfigure}
\hfill
\begin{subfigure}[t]{0.48\textwidth}
    \centering
    \includegraphics[width=\linewidth]{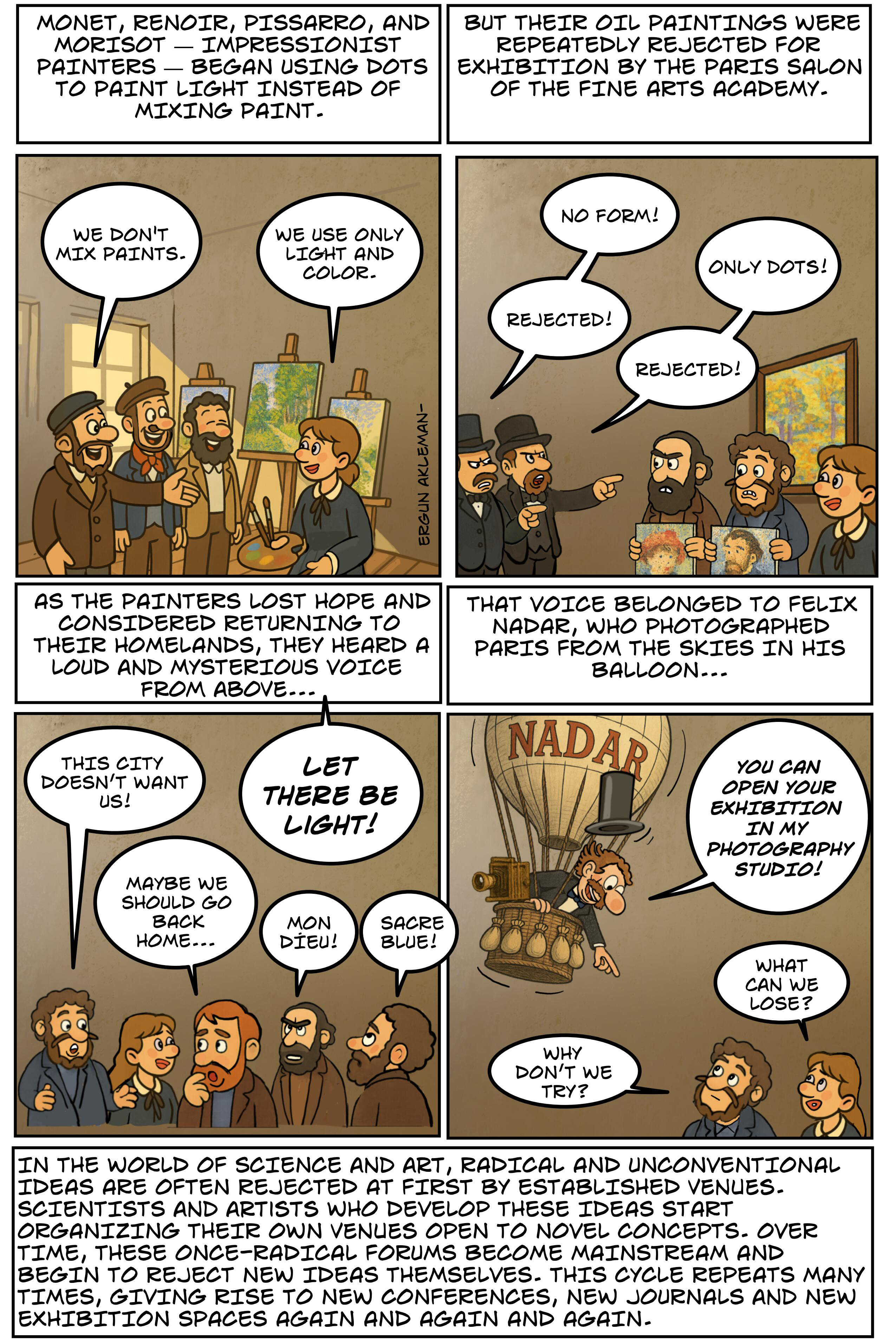}
    \caption{The connection between photography and the first Impressionist exhibition through Nadar.}
    \label{fig_cartoon_nadar}
\end{subfigure}
\caption{Two cartoons by the author illustrating possible interactions between photographic technology and artistic innovation. The imagined encounter in (a) is a historical hypothesis rather than a documented event; (b) humorously depicts the documented connection between the first Impressionist exhibition and the former studio of the photographer Nadar.}
\label{fig_historical_cartoons2}
\end{figure}

An intriguing historical question is whether the emergence of photographic screening influenced the development of painting techniques based on discrete visual elements. Photographic halftoning and Impressionist painting emerged within relatively close historical periods. While photographic screening demonstrated that continuous tonal impressions could be constructed from discrete black and white marks, Impressionist painters explored the construction of visual impressions through individually visible brushstrokes and spatially distributed colors.

This conceptual relationship becomes even clearer in the later development of Neo-Impressionism and Pointillism, where discrete colored marks became an explicit part of the painting technique. Although the artistic motivations and physical mechanisms of these approaches differ from those of photographic halftoning, both exploit the human visual system's ability to integrate spatially distributed visual elements into coherent perceptual impressions.

We speculate that the dissemination of photographic screening through illustrated publications may have contributed to the broader artistic awareness of the possibilities of constructing images from discrete visual elements. Figure~\ref{fig_cartoon_impressionism} presents a humorous illustration of this hypothesis, imagining how early photographic halftones might have been received by painters in Paris. We are not aware of direct historical evidence establishing such a causal relationship, and the illustrated encounter should therefore be interpreted as a fictional scenario rather than a historical reconstruction.

A documented connection between photography and Impressionism provides a complementary perspective. The first independent Impressionist exhibition opened on April 15, 1874, at 35 boulevard des Capucines in Paris, in the former studio of the photographer Gaspard-Félix Tournachon, better known as Nadar. The cartoon in Figure~\ref{fig_cartoon_nadar} humorously interprets this connection. The exhibition venue establishes a concrete association between the photographic and artistic communities, although it does not establish that photographic screening inspired Impressionist brushwork. This distinction is important: the historical narrative illustrates the possibility of mutual inspiration between engineering and art without treating their chronological proximity as evidence of causation.

Regardless of whether such a direct historical influence existed, the conceptual similarities between photographic halftoning and painting techniques based on discrete marks are particularly relevant to our work. Both demonstrate that the visual appearance of an image depends not only on the individual colors or intensities of its constituent elements but also on their geometric organization. Our framework seeks to exploit this relationship computationally by making the geometry of these elements directly and interactively controllable.

\subsection{Digital Dithering and Artistic Screening}

The emergence of digital imaging introduced new computational approaches to representing continuous visual information using discrete intensity and color values. Early investigations into the use of spatially distributed quantization thresholds and noise in image representation include the work of Goodall \citep{Goodall1951television}, Roberts \citep{roberts1962picture}, and Schuchman \citep{schuchman1964dither}. These developments established important foundations for the subsequent use of dithering in digital imaging and computer graphics.

Error diffusion, introduced by Floyd and Steinberg \citep{floyd1976an}, distributes quantization errors to neighboring pixels to improve the reproduction of continuous tonal variations. Color image quantization techniques, such as those developed by Heckbert \citep{heckbert1982color}, address related problems associated with representing images using limited color palettes. Although these approaches are primarily concerned with image reproduction, they also demonstrate how the spatial distribution of discrete visual elements influences perceived image quality.

A particularly relevant direction for our work is the deliberate geometric organization of halftoning patterns. Knuth's dot diffusion method \citep{knuth1987digital} demonstrates how spatial organization can be incorporated into digital halftoning while providing opportunities for parallel computation. Space-filling curve ordered dithering \citep{zhang1998space} and image-based dither screens \citep{verevka1999halftoning} further demonstrate the flexibility of computational approaches to constructing and organizing dithering patterns.

Among the most directly relevant contributions to artistic dithering are the artistic screening techniques developed by Ostromoukhov and Hersch \citep{ostromoukhov1995artistic,ostromoukhov1996anticounterfeiting}. Their work demonstrates that halftoning patterns can themselves become vehicles for artistic and geometric expression, rather than serving exclusively as mechanisms for tonal reproduction. Ostromoukhov's subsequent work on digital facial engraving \citep{ostromoukhov1999digital} is particularly relevant to our objective of generating engraving-inspired images whose patterns respond to the geometric characteristics of depicted objects.

The distinction between these approaches and our work lies primarily in the representation and manipulation of the underlying dithering patterns. Our objective is to construct dithering functions directly through implicit geometric representations and functional shape operations. In particular, generalized Hölder averages provide a continuous parameterization of pattern geometry, enabling transitions between hatching, crosshatching, pointillist, and intermediate geometric structures without explicitly constructing separate dithering matrices for each pattern.

The resulting functional formulation also enables spatial variations in pattern orientation, spacing, position, and geometry. Consequently, the geometric construction of patterns and their adaptation to local image properties can be performed within the same computational framework. This representation is particularly suitable for interactive GPU-based implementations, where dithering functions can be evaluated independently at each image location.

In this sense, Hölder Dither brings together two historically complementary approaches to image representation: the artistic manipulation of geometric marks characteristic of traditional engraving and painting, and the computational construction of discrete visual patterns characteristic of digital halftoning and dithering. In this respect, the framework continues the historical interplay between artistic objectives and engineering innovations: mathematical tools developed for implicit geometric modeling are repurposed to expand the possibilities of artistic image representation.

\section{Theoretical Framework}

Binary discretization provides a useful starting point for mathematically describing engraving-inspired hatching, crosshatching, and halftoning, although the physical marks used in these practices need not be binary in appearance. Discretization generally partitions a range of values into intervals \citep{kotsiantis2006discretization}; in its simplest binary form, an image is represented using two output values. We first develop this binary formulation to isolate the geometry of the resulting patterns and subsequently extend it to gray-level and color rendering. Our principal objective is to construct and manipulate the underlying geometric regions through functions rather than to manipulate already discretized images.

Let $\Omega=[0,1]^2$ denote a canonical image domain, which can be mapped to a rectangular image domain by an appropriate coordinate transformation. We define two families of functions:
\[
\mathcal F=\{f:\Omega\to[0,1]\},\qquad
\mathcal D=\{d:\Omega\to\{0,1\}\}.
\]
Thus, $\mathcal D\subseteq\mathcal F$. The former describes continuous-valued intensity images (not necessarily spatially continuous), whereas the latter describes binary images. Dithering can be viewed as a mapping from $\mathcal F$ to $\mathcal D$. Direct manipulation of binary images is less convenient for functional shape operations because nonconstant binary images on a connected domain are discontinuous. Instead, we manipulate real-valued implicit functions prior to discretization.

\subsection{Implicit Representation of Dithered Images}

For any real-valued function $h$ on $\Omega$, we define three disjoint sets:
\begin{align*}
S_0(h)&=\{\mathbf{x}\in\Omega:h(\mathbf{x})<0\},\\
\delta S(h)&=\{\mathbf{x}\in\Omega:h(\mathbf{x})=0\},\\
S_1(h)&=\{\mathbf{x}\in\Omega:h(\mathbf{x})>0\}.
\end{align*}
Their union is $\Omega$. For a continuous implicit function whose zero set separates positive and negative regions, $\delta S(h)$ describes their interface. More generally, zero sets may contain other points or entire regions. In binary rendering we assign the negative region to black and the positive region to white; equality must also be assigned a value. Here we assign it to black, consistent with thresholding by the ceiling function in the ideal binary case. We therefore distinguish the strict sign regions used in geometric arguments from the final binary classification.

Implicit geometric modeling provides a useful language for manipulating these regions. In solid modeling, regularized set operations remove lower-dimensional remnants by taking the closure of the interior of a set \citep{requicha1983solid}. Related functional constructions use minimum, maximum, and sign reversal \citep{akleman1996interactive,akleman1999constant}. For arbitrary real-valued functions $u$ and $v$, the strict sign regions satisfy
\begin{align*}
S_0(u)&=S_1(-u),\\
S_0(u)\cap S_0(v)&=S_0(\max(u,v)),\\
S_0(u)\cup S_0(v)&=S_0(\min(u,v)),\\
S_0(u)\cap S_1(v)&=S_0(\max(u,-v)).
\end{align*}
The last expression represents a strict-sign difference, excluding the zero set of $v$. It agrees with ordinary set difference away from that zero set. These identities motivate functional shape manipulation without requiring the explicit construction of boundaries; regularization, when needed, is a separate set-theoretic operation.

Let $f\in\mathcal F$ denote an input intensity image and $g\in\mathcal F$ a dithering function. We construct the implicit function $h(\mathbf{x})=f(\mathbf{x})-g(\mathbf{x})$ and define the binary rendering by
\begin{equation}
d(\mathbf{x})=\begin{cases}0,&h(\mathbf{x})\leq0,\\1,&h(\mathbf{x})>0.\end{cases}
\label{eq_binary_dither}
\end{equation}
The geometric regions are therefore obtained by comparing two functions at every location. If $g$ is periodic, applying it to a constant input $f(\mathbf{x})=a$, $0\leq a\leq1$, reveals its repeating patterns, defined by $S_0(a-g)$ and $S_1(a-g)$ together with the chosen boundary convention. Applying the same function to a spatially varying image generally produces nonperiodic regions whose coverage conveys the input image.

\subsection{Dithering Functions}

We use the term \emph{free-form dithering} for a functional approach in which the threshold function $g$ is constructed procedurally rather than stored in a predefined matrix. Stationary patterns use fixed geometric parameters, whereas non-stationary patterns allow these parameters to vary across the image. Although stationary pencils are periodic, combinations of pencils need not share a common translational period, and spatially varying patterns are generally not periodic. Given the input image and the parameters at a location, the dithering decision can be made independently of decisions at other locations. This property is particularly useful for GPU implementations, even when determining those parameters requires sampling a neighborhood of the input image.

Earlier work on implicit geometric modeling has primarily addressed the construction of three-dimensional shapes \citep{akleman1996interactive,akleman1996ray,akleman1999generalized}. Here we apply related functional ideas to the construction of two-dimensional visual elements and their spatial arrangements. Generalized Hölder averages are particularly useful because a single exponent connects exact minimum and maximum operations through a continuous family of intermediate operations.

\subsection{Generalized Hölder Averages as Shape Operations}

Given $N\geq1$ dithering functions $g_i\in\mathcal F$, the generalized Hölder average is evaluated pointwise. For strictly positive inputs and finite $p\neq0$, it is defined by
\begin{equation}
H_p(g_0,\ldots,g_{N-1})=\left(\frac{1}{N}\sum_{i=0}^{N-1}g_i^p\right)^{1/p}.
\label{eq_Hölder}
\end{equation}
Here $p$ is the exponent, not the spatial coordinate. The geometric mean $H_0=(\prod_i g_i)^{1/N}$ is obtained in the limit as $p\to0$. The arithmetic, harmonic, and root-mean-square means correspond to $p=1$, $p=-1$, and $p=2$, respectively \citep{akleman2024projective,yildiz2026holder}. The limits $p\to+\infty$ and $p\to-\infty$ are the maximum and minimum. For nonnegative inputs, the positive-exponent formula is valid directly; if any input is zero, the mean for $p\leq0$ is defined as zero by continuous extension in the input values. This convention is important because our distance functions vanish on their defining lines. For all these cases the output lies between the minimum and maximum inputs and hence belongs to $[0,1]$.

The connection to shape operations can be seen by considering the white regions of the binary renderings. For a fixed input image $f$, let $W_i=\{\mathbf{x}\in\Omega:g_i(\mathbf{x})<f(\mathbf{x})\}$. Then
\begin{align*}
\{\mathbf{x}:H_{-\infty}(\mathbf{x})<f(\mathbf{x})\}&=\bigcup_i W_i,\\
\{\mathbf{x}:H_{+\infty}(\mathbf{x})<f(\mathbf{x})\}&=\bigcap_i W_i.
\end{align*}
Here $H_p(\mathbf{x})$ abbreviates the mean of all $g_i(\mathbf{x})$. With our binary convention, the corresponding black regions are the complementary intersection and union, respectively. The exponent thus provides a continuous functional transition between these two operations, rather than requiring a separate pattern design for each intermediate configuration. Connectivity can change during this transition even though the underlying mean varies continuously with the exponent.

\subsection{Normalized Distance Functions for Pencils of Lines}

A simple elementary pattern is obtained from a pencil of equidistant parallel lines. Let $\mathbf{n}_i=(\cos\theta_i,\sin\theta_i)$ be the unit normal, $\mathbf{p}_i$ a reference point on one line, and $T_i>0$ the distance between adjacent lines. The signed distance to the reference line is $F_i(\mathbf{x})=\mathbf{n}_i\cdot(\mathbf{x}-\mathbf{p}_i)$. The pencil consists of the lines $F_i(\mathbf{x})=mT_i$, $m\in\mathbb Z$. Since $F_i$ is real-valued and is not generally restricted to $[0,1]$, we transform the dimensionless coordinate $t_i=F_i/T_i$ using a periodic function. For example, $(\sin(2\pi t_i)+1)/2$ is bounded in $[0,1]$, but we prefer a triangular wave that is directly related to the distance to the closest line:
\begin{equation}
M(t)=2\left|t-\left\lfloor t+\tfrac12\right\rfloor\right|,\qquad
 g_i(\mathbf{x})=M\left(\frac{\mathbf{n}_i\cdot(\mathbf{x}-\mathbf{p}_i)}{T_i}\right).
\label{eq_pencil}
\end{equation}
The floor function returns the largest integer not exceeding its argument, and the absolute value ensures $M(t)\in[0,1]$. The triangular wave has period one. Its zeros correspond to the lines of the pencil and its maxima occur midway between adjacent lines. In Euclidean coordinates, the distance to the nearest line is exactly $T_i g_i(\mathbf{x})/2$. This interpretation holds for fixed pencil parameters and in the metric in which $T_i$ is measured; implementations may use pixel coordinates to preserve the intended aspect ratio.

\begin{figure}[htbp!]
\centering
\begin{tabular}{ccc}
\fbox{\includegraphics[width=0.30\textwidth]{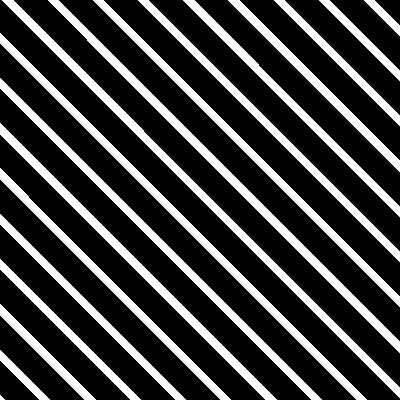}}&
\fbox{\includegraphics[width=0.30\textwidth]{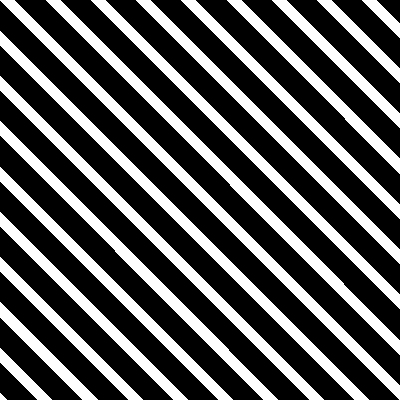}}&
\fbox{\includegraphics[width=0.30\textwidth]{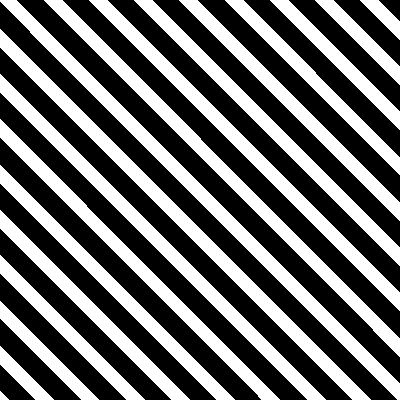}}\\
 $a=0.3$ & $a=0.4$ & $a=0.5$\\
\fbox{\includegraphics[width=0.30\textwidth]{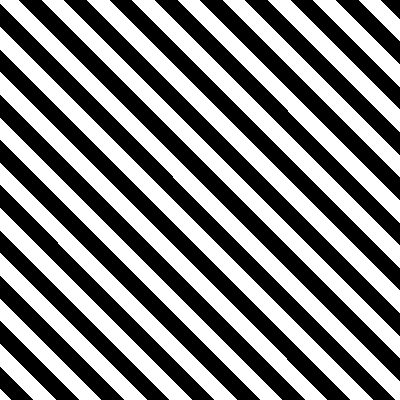}}&
\fbox{\includegraphics[width=0.30\textwidth]{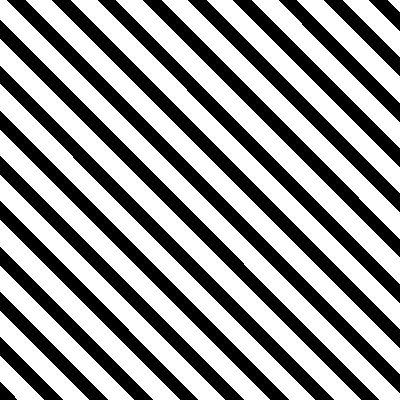}}&
\fbox{\includegraphics[width=0.30\textwidth]{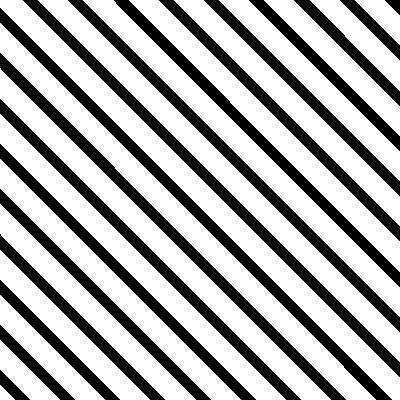}}\\  $a=0.6$ & $a=0.7$ &$a=0.8$ 
\end{tabular}
  \caption{Binary dithering of constant-valued images with a single pencil of equidistant parallel lines. The input intensity $a$ increases from $0.3$ to $0.8$ in increments of $0.1$, with $\mathbf{p}_0=(0,0)$ and $\theta=45^\circ$. The width of the white bands increases and that of the black bands decreases as $a$ increases.}
\label{fig_lines}
  \end{figure}

Each pencil is specified by three geometric parameters: its orientation $\theta_i$, period $T_i$, and reference point $\mathbf{p}_i$ (only the normal component of the latter affects phase). When a single pencil is applied to a constant intensity $a$ with $0<a<1$, its white bands satisfy $g_i<a$ and have width $aT_i$; the complementary black bands have width $(1-a)T_i$. Thus the input intensity controls stroke thickness, as illustrated in Figure~\ref{fig_lines}. The level sets $g_i=c$ for $0<c<1$ consist of two offset lines per period.

\subsection{Geometric Properties of Hölder Dithering Patterns}

Combining multiple pencils using Equation~\ref{eq_Hölder} creates families of hatching, crosshatching, dot-like structures, and intermediate forms. When two perpendicular pencils of equal period are used, their common line intersections form a square lattice. With three equal-period pencils at orientations separated by $\pi/3$ and a common reference point, the common intersections form a triangular lattice, often described visually as a hexagonal arrangement. For $p=2$, sufficiently small sublevel sets near these intersections are circular in the ideal equal-period configurations: locally, the squared mean is a sum of squared projections onto the pencil normals. Larger thresholds can produce different shapes and connectivity as the nearest-line branches change.

The effects of varying the exponent for two and three pencils are illustrated in Figures~\ref{fig_2lines} and~\ref{fig_3lines}. The minimum produces a union of white bands, whereas the maximum produces their intersection (and hence a union of black bands). Intermediate exponents produce a range of geometric configurations whose appearance also depends on the threshold, relative orientations, periods, and phases. Accordingly, $p=2$ does not guarantee isolated circular dots at every threshold or for every pencil arrangement.

The six-pencil example in Figure~\ref{fig_6lines} further illustrates the distinction between rotational symmetry and translational periodicity. For $N$ pencils with equal periods, a common reference point, and orientations separated by $\pi/N$, rotation by $\pi/N$ about the reference point permutes the pencil functions, since reversing a line normal does not change its distance function. The symmetric Hölder average therefore retains this rotational symmetry. A common reference point alone is insufficient to guarantee it. Moreover, a common translational period would have to satisfy $\mathbf{n}_i\cdot\mathbf{v}/T_i\in\mathbb Z$ for every pencil, where $\mathbf{v}$ is a translation vector. Such compatible periods need not exist for arbitrary orientations and spacings. Thus the same functional construction can support regular lattice patterns as well as arrangements without a common two-dimensional translational lattice, including the possibility of quasi-periodic patterns. The number of pencils and exponent offer complementary controls over shape, symmetry, connectivity, and arrangement.

\subsection{Relationship Between Pattern Geometry and Tonal Appearance}

A central property of the generalized mean is its monotonicity with respect to the exponent. For fixed nonnegative input functions and $p<q$, the generalized mean inequality gives $H_p(\mathbf{x})\leq H_q(\mathbf{x})$ at every point. Let $B_p$ denote the black region obtained by using $H_p$ as the dithering function in Equation~\ref{eq_binary_dither}. Then
\begin{equation}
 B_p=\{\mathbf{x}\in\Omega:f(\mathbf{x})\leq H_p(\mathbf{x})\},\qquad
 p<q\ \Longrightarrow\ B_p\subseteq B_q.
\label{eq_black_nesting}
\end{equation}
For measurable images and patterns, the area of $B_p$ is consequently nondecreasing with the exponent; the same ordering holds for the count of black pixels on a fixed sampling grid. This observation also holds for a fixed nonnegative threshold scaling and a fixed additive perturbation independent of the exponent. It does not require the dithering functions to be periodic. If other parameters change simultaneously, the ordering need not apply.

This property explains why changing the exponent affects both the geometry and the overall tonal appearance of a rendering. In conventional tone-preserving dithering, one might seek to compensate for changes in coverage. In our artistic setting, we intentionally retain the coupling as an additional form of artistic control. Figures~\ref{fig_2lines}, \ref{fig_3lines}, and~\ref{fig_6lines} illustrate this effect for constant inputs, and Figure~\ref{fig_asli} demonstrates it in a portrait rendering. The formal result concerns geometric coverage; perceived brightness also depends on color, spatial organization, and the viewing conditions.

\ifdefined\FloatBarrier\FloatBarrier\fi

\section{Computational Methodology}

The preceding formulation describes how to construct families of geometric patterns. We now describe how these functions can be adapted to image features and artistic preferences, including discrete directional sampling, local pencil construction, controlled irregularity, gray-level and color rendering, and function interpolation. These operations can be combined in a pointwise GPU implementation for interactive artistic exploration.

\subsection{Estimation of Local Gradient Directions}

Rather than first estimating the two Cartesian components of a gradient and subsequently quantizing its angle, our implementation directly searches a finite set of candidate directions. This is particularly useful for hatching because the objective is not necessarily a highly accurate numerical gradient, but a locally meaningful orientation selected from a controllable set of possibilities. 
We use a smoothed guide image to reduce sensitivity to fine-scale variations. In the standalone implementation, the input image itself serves as the default guide, with adjustable mipmap smoothing, although a separate guide image can optionally be supplied.

In the supplied shader, directional sampling is performed in pixel coordinates. Let $s(\mathbf{x})$ denote the mean of the three RGB components of the guide image sampled at pixel coordinate $\mathbf{x}$. For $K$ candidate directions, we use
\[
\phi_k=\frac{2\pi k}{K},\qquad \mathbf{v}_k=(\cos\phi_k,\sin\phi_k),\qquad k=0,\ldots,K-1.
\]
For a sampling radius $R>0$, the shader evaluates an asymmetric directional contrast:
\begin{equation}
 C_k(\mathbf{x})=s(\mathbf{x}-0.1R\mathbf{v}_k)-s(\mathbf{x}+R\mathbf{v}_k),\qquad
 k^*=\operatorname*{arg\,max}_{0\leq k<K}|C_k(\mathbf{x})|.
\label{eq_directional_contrast}
\end{equation}
The two sampling points are separated by $1.1R$. Unlike a centered finite difference, this stencil samples a longer distance on one side of the evaluation point than on the other. Since the separation is constant across directions, dividing the contrast by $1.1R$ would not change the maximizing direction. For a smooth guide and small $R$, the contrast approximates $-1.1R\nabla s\cdot\mathbf{v}_k$ to first order. At larger radii it instead measures directional contrast over an extended neighborhood. We therefore interpret it as a finite-scale gradient-orientation estimate rather than as an exact differential gradient.

The supplied implementation returns $\operatorname{sign}(C_{k^*})\mathbf{v}_{k^*}$ when the maximum absolute contrast is nonzero. Under the stated subtraction convention the sign is opposite to the increasing-intensity gradient in the infinitesimal limit, but this does not affect the orientation of a straight pencil because its distance function is invariant to normal reversal. Ties are resolved by retaining the first maximizing sample encountered, and if all contrasts are zero the first candidate direction is used. The returned vector has unit length; its magnitude does not represent gradient strength. The maximum contrast, if retained separately, could serve as a finite-scale measure of directional response.

The number of directions $K$ and radius $R$ determine different aspects of the estimate. The former controls the angular choices available to the local pencils, whereas the latter controls the spatial scale over which intensity variation is examined. For straight pencils, opposite directions represent the same orientation, so for an exactly evenly spaced full-circle sampling set with even $K$ there are $K/2$ distinct unoriented axes, and with odd $K$ there are $K$ axes. Nevertheless, opposite candidates may yield different contrast magnitudes because the sampling stencil is asymmetric. Neighboring points selecting the same orientation can produce groups of consistently oriented marks. A finer sampling set allows greater directional variation, while a coarser one can favor larger groups with a common orientation. Neither choice guarantees spatial coherence, especially near abrupt transitions or in flat regions, as discussed later.

\subsection{Locally Adaptive Dithering Functions}

Let $\psi(\mathbf{x})$ denote the orientation of the selected gradient axis. A pencil normal can be chosen as $\theta_0(\mathbf{x})=\psi(\mathbf{x})$ to generate lines perpendicular to the gradient, or $\theta_0(\mathbf{x})=\psi(\mathbf{x})+\pi/2$ to generate lines perpendicular to the tangent. Additional pencils can be obtained by rotating the initial normal through prescribed angles. For example, two mutually perpendicular pencils use a relative angle of $\pi/2$, whereas three equally distributed unoriented pencils use increments of $\pi/3$.

More generally, the orientation, period, and reference point of each pencil can vary with location, giving
\[
g_i(\mathbf{x})=M\left(\frac{\mathbf{n}_i(\mathbf{x})\cdot(\mathbf{x}-\mathbf{p}_i(\mathbf{x}))}{T_i(\mathbf{x})}\right),\qquad T_i(\mathbf{x})>0.
\]
These functions are constructed by evaluating locally specified pencils rather than by measuring exact distances to globally defined curved lines. Changes in orientation, spacing, and phase can therefore generate visually flowing patterns, but do not automatically ensure continuous strokes. In the standalone implementation, $N$ straight pencil functions are constructed using angular increments of $\pi/N$, a shared period, and a fixed origin in pixel coordinates. The shader also provides optional radial variants in which the linear projection is replaced by distance to a center, allowing concentric-circle patterns; spatially varying centers again need not define globally coherent curves.

The visual effects of these orientation choices are examined in Figure~\ref{fig_teaser8}.

\subsection{Controlled Geometric Irregularity}

The functional representation permits controlled variations in the periods $T_i$, reference points $\mathbf{p}_i$, or orientations $\theta_i$ of the pencils. Varying these parameters can change spacing, phase, or direction, respectively, and can produce asymmetric elements reminiscent of imperfections in handcrafted illustrations. Another way to introduce irregularity, used explicitly in the supplied shader, is to perturb the threshold comparison with a coordinate-dependent pseudorandom value. If $u(\mathbf{x})\in[0,1)$ is a pseudorandom scalar and $\rho$ controls its amplitude, the perturbation is $\eta(\mathbf{x})=\rho(u(\mathbf{x})-1/2)$. Adding this value to the comparison changes the local classification and can produce irregular edges and grain without explicitly moving the underlying lines.

These two mechanisms should be distinguished: geometric parameter variations alter the underlying pencil functions, whereas threshold perturbations alter the rendered coverage. Figures~\ref{fig_teaser0} and~\ref{fig_teaser4} illustrate the artistic possibilities of irregular patterns. The supplied shader uses deterministic coordinate-based noise, so a fixed sampling grid and fixed parameters yield a consistent perturbation rather than newly generated noise every frame.

\subsection{Gray-Level and Color Rendering}

Although the binary formulation provides a convenient model for geometric pattern construction, it does not restrict the final rendering to two intensity values. One approach is to define additional regions using intervals of an implicit function:
\begin{equation}
 S_{(a,b)}(f)=\{\mathbf{x}\in\Omega:b<f(\mathbf{x})<a\},\qquad 0\leq b<a\leq1.
\label{eq_ab}
\end{equation}
For a function taking values in $[0,1]$, successive intervals can be assigned distinct gray levels, with a consistent convention for shared endpoints. More generally, intervals of real-valued implicit functions may be used. Color rendering can be obtained by applying the comparisons separately to the red, green, and blue components of the input image. If a common geometric function is used across the channels, their resulting regions are correlated rather than independently positioned, which can contribute to a visual appearance reminiscent of colored pencil or mixed pigments. This is an artistic effect, not a physical model of pigment mixing.

The supplied shader implements multilevel rendering by averaging several threshold layers with different scales of the same geometric function. To describe this precisely, let $G=H_p(g_0,\ldots,g_{N-1})$ use the normalized pencil functions of Equation~\ref{eq_pencil}. The shader actually evaluates the half-scaled distance $\widetilde g_i=g_i/2$ and hence $\widetilde G=G/2$ by homogeneity of the mean. Let $A>0$ denote the adjustable threshold scale, $L\geq 1$ a real-valued layer-count parameter, and $m=\lfloor L\rfloor$ the number of layers evaluated. For a channel intensity $c$, the implemented layer average can be written as
\begin{equation}
 C_c(\mathbf{x})=\frac{1}{m}\sum_{j=0}^{m-1}\left\lceil c(\mathbf{x})-A w_j\widetilde G(\mathbf{x})+\eta_{c,j}(\mathbf{x})\right\rceil,
 \quad w_0=1,\quad w_j=j/L\ (j\geq1).
\label{eq_multilevel_dither}
\end{equation}
The shader computes this average incrementally. For arguments in $(-1,1]$, each ceiling comparison is binary (zero or one), and averaging the $m$ layers yields up to $m+1$ nominal levels. With sufficiently large perturbations the raw ceiling values may fall outside this binary range; the stored shader does not explicitly clamp each layer. Display into a normalized framebuffer limits the displayed channel values. A strictly bounded binary-layer variant could instead use explicit zero-or-one threshold tests; we distinguish this mathematical option from the existing implementation.

For monochromatic rendering the shader uses a weighted RGB intensity $0.33c_R+0.33c_G+0.34c_B$ for all channels. In color mode each input component is processed separately using the common geometric function. The first layer uses shared pseudorandom noise; later layers use the same coordinate-based noise for red and offset coordinates for green and blue, allowing subtle differences in the spatial color patterns. The colored pencil examples and gray-level results in Figures~\ref{fig_teaser}, \ref{fig_teaser4}, and~\ref{fig_teaser5} illustrate the broader possibilities of this kind of representation. The layer construction offers control beyond binary rendering but does not yet separately model the geometry, opacity, pressure, or internal texture of individual physical marks.

\subsection{Composite Dithering Functions}

Patterns obtained using positive and negative exponents can appear visually complementary, although they are not generally exact black-and-white inverses. For two strictly positive values $u,v$ and finite $p\neq0$, the power means satisfy $H_p(u,v)H_{-p}(u,v)=uv$, since $u^{-p}+v^{-p}=(u^p+v^p)/(uv)^p$. This identity also holds for the geometric mean and limiting minimum and maximum, and extends to nonnegative inputs by the stated conventions. It describes a multiplicative relationship between the functions, not an exact complement relationship between thresholded regions.

Motivated by the differing visual appearance of these patterns, we can interpolate two dithering functions $g_0$ and $g_1$ to construct a composite function. Let $t(\mathbf{x})\in[0,1]$ be a weight representing the average intensity in a local region, or an independently controlled artistic parameter. Then
\begin{equation}
 g(\mathbf{x})=(1-t(\mathbf{x}))g_0(\mathbf{x})+t(\mathbf{x})g_1(\mathbf{x}).
\label{eq_interpolate}
\end{equation}
The output remains in $[0,1]$ when both constituent functions do. Figure~\ref{fig_lincoln5} illustrates an earlier implementation using a weighted combination of the minimum and maximum of three equally rotated pencil functions, together with local direction estimation using seven sampling steps. No gray-level layers or deliberate randomness were used for that example. This construction enables transitions between patterns appropriate for different tonal regions without explicitly designing a separate screen for each region.

\subsection{Interactive GPU Implementation and Software Availability}

The experimental implementations used in this study were developed as GLSL fragment shaders in our Shader Playground environment. To make the proposed approach accessible to readers, artists, and designers, we also provide a freely accessible standalone browser-based implementation, Hölder Dither \cite{akleman2026holderdithersoftware}, at \url{https://people.tamu.edu/~ergun/software/dithering/Holder_Dither.html}. The application is implemented using WebGL~2 in a self-contained HTML file and requires no additional software installation in a compatible browser.

Each fragment samples an orientation guide in a finite set of candidate directions, selects a local orientation using the maximum absolute directional contrast, constructs the corresponding pencil functions, combines them using a generalized Hölder average, and assigns monochromatic or RGB values using the threshold layers described above. By default, the input image itself serves as the guide, with adjustable mipmap smoothing; a separate guide image can optionally be supplied. A fixed-orientation mode is also available for exploring stationary patterns. Each fragment reads the source textures independently of the dithering decisions at neighboring fragments. For $K$ sampled directions, $N$ pencils, and $m$ threshold layers, the corresponding per-fragment work is proportional to $K+N+m$, excluding texture and mipmap preparation.

The standalone implementation provides interactive controls for pattern spacing, number of pencils, Hölder exponent, orientation offset, feature scale (directional sampling radius), number of sampled directions, guide smoothing, darkness (threshold scale), tonal layers, randomness, and monochromatic or RGB rendering. The pencil orientations are separated by $\pi/N$ and use a common period and a fixed reference point in pixel coordinates. The exponent control includes the minimum and maximum limits at its endpoints and the geometric mean near zero. The implementation evaluates half-scaled pencil distances, as described in the preceding methodology, and handles zero distances for nonpositive exponents explicitly. Scaling the inputs by an appropriate extremum before evaluating finite nonzero power means also improves numerical robustness.

Users can load their own images, explore the effects of parameter changes interactively, and inspect the original image, smoothed guide, selected orientation field, or underlying dithering function. The preview adapts to the available screen size while the rendering and PNG export preserve the original image dimensions, subject to the graphics capabilities of the device. Image processing takes place locally in the browser. The publicly available application is intended to support independent exploration of the principal method rather than to reproduce every experimental shader variant presented in this paper. Our results demonstrate interactive rendering qualitatively; we do not report benchmark frame rates or claim a measured performance advantage over other methods.

To facilitate artistic exploration and further research, we have made publicly available a collection of interactive browser-based applications under the name \emph{Dithering Suite} \cite{akleman2026ditheringsuite}. The suite brings together several complementary approaches to image dithering, ranging from conventional threshold matrices and locally constructed statistical threshold distributions to procedurally generated geometric dithering functions. In addition to Hölder Dither, the suite includes Matrix-Based Dithering, Local Equal-Count Dithering, Local Line-Fitted Hatching, Bravais Lines and Hölder--Minkowski patterns, and Hölder--Minkowski Distance-Function Dithering.

These applications provide an interactive environment for exploring different mathematical and computational approaches to the construction of dithering patterns. Although the present paper primarily focuses on Hölder Dither, the accompanying applications illustrate the broader possibilities of statistical and geometric approaches to artistic image representation. The complete suite is available at \url{https://people.tamu.edu/~ergun/software/dithering/}.

\ifdefined\FloatBarrier\FloatBarrier\fi

\section{Geometric Coherence and Preservation of Stroke Thickness}
\label{sec_geometric_coherence}

An important consideration in artistic dithering is the preservation of the geometric characteristics of individual marks. In many forms of traditional hatching and etching, relatively consistent stroke thickness within groups of lines is desirable, even when the artist intentionally varies their orientation, spacing, length, density, or thickness to convey form and tone. Our experiments with continuously varying image-gradient directions revealed a difficulty that is not necessarily attributable to the accuracy of gradient estimation. Even a smooth and accurately estimated orientation field can introduce substantial unintended variations in stroke thickness and spacing when it is used directly to construct pencil functions. In this section, we explain the geometric origin of this phenomenon and motivate our alternative approach of selecting orientations from a finite set of candidates.

\subsection{Geometric Properties of Stationary Pencils}

Consider a single pencil of equidistant parallel lines with a fixed unit normal $\mathbf{n}$, reference point $\mathbf{p}$, and period $T>0$. As in Equation~\eqref{eq_pencil}, we construct the unwrapped pencil coordinate
\begin{equation}
F(\mathbf{x})=\mathbf{n}\cdot(\mathbf{x}-\mathbf{p}).
\label{eq_coherence_stationary_phase}
\end{equation}
The lines of the pencil are the level sets $F(\mathbf{x})=mT$, where $m\in\mathbb{Z}$. Since the normal and reference point are constant, we have
\begin{equation}
\nabla F(\mathbf{x})=\mathbf{n},\qquad \|\nabla F(\mathbf{x})\|=1.
\label{eq_coherence_stationary_gradient}
\end{equation}
Consequently, $F$ is the signed Euclidean distance to its reference line. The periodic function $g(\mathbf{x})=M(F(\mathbf{x})/T)$ is proportional to the unsigned distance to the closest line of the pencil. A band defined by $|F(\mathbf{x})-mT|<\epsilon$, with $0<\epsilon<T/2$, therefore has thickness $2\epsilon$, and adjacent centerlines are separated by exactly $T$. The complementary bands also have constant thickness $T-2\epsilon$. Thus, at a constant threshold, both the white and black portions of a stationary pencil have well-defined geometric widths. Variations in the image intensity can intentionally change these widths through thresholding, but the underlying pencil itself introduces no additional geometric distortion.

\subsection{Geometric Distortions Caused by Continuously Varying Orientations}

A seemingly natural way to adapt a pencil to an image is to compute the local image gradient, normalize it, and use the resulting unit vector as the normal of the pencil at each point. With a fixed reference point and period, this construction yields
\begin{equation}
F(\mathbf{x})=\mathbf{n}(\mathbf{x})\cdot(\mathbf{x}-\mathbf{p}),\qquad \|\mathbf{n}(\mathbf{x})\|=1.
\label{eq_coherence_variable_phase}
\end{equation}
Although the normal is normalized at every location, the resulting scalar function is not necessarily a signed distance function. At points where the orientation field is differentiable, the product rule gives
\begin{equation}
\boxed{\nabla F(\mathbf{x})=\mathbf{n}(\mathbf{x})+J_{\mathbf{n}}(\mathbf{x})^{\mathsf T}(\mathbf{x}-\mathbf{p})},
\label{eq_coherence_variable_gradient}
\end{equation}
where $J_{\mathbf{n}}$ is the Jacobian of the unit normal field. The additional term arises from spatial variations in the orientation and can change both the magnitude and direction of the gradient of the pencil coordinate. In particular, normalization of the image-gradient vector does not imply normalization of the gradient of the pencil coordinate. This distinction is essential because the latter determines the local geometry of the generated marks.

To make the dependence on orientation explicit, let $\mathbf{n}(\mathbf{x})=(\cos\theta(\mathbf{x}),\sin\theta(\mathbf{x}))$ and let $\mathbf{t}(\mathbf{x})=(-\sin\theta(\mathbf{x}),\cos\theta(\mathbf{x}))$ denote its perpendicular unit vector. Wherever a differentiable local angular representation exists, Equation~\eqref{eq_coherence_variable_gradient} can be written as
\begin{equation}
\nabla F(\mathbf{x})=\mathbf{n}(\mathbf{x})+\bigl[\mathbf{t}(\mathbf{x})\cdot(\mathbf{x}-\mathbf{p})\bigr]\nabla\theta(\mathbf{x}).
\label{eq_coherence_angular_gradient}
\end{equation}
Even relatively small changes in orientation can therefore produce substantial distortions at locations far from the reference point. For the dimensionless pencil coordinate $q=F/T$, the sensitivity to a change in angle at a fixed location satisfies
\begin{equation}
\left|\frac{\partial q}{\partial\theta}\right|=\frac{|\mathbf{t}\cdot(\mathbf{x}-\mathbf{p})|}{T}\leq\frac{\|\mathbf{x}-\mathbf{p}\|}{T}.
\label{eq_coherence_phase_sensitivity}
\end{equation}
This sensitivity is especially relevant for fine hatching, where a small period can cause a small angular change to traverse a substantial fraction of a pattern cycle. Smoothing the image may reduce some angular variations, but it does not enforce the distance-function condition $\|\nabla F\|=1$.

The consequences for stroke thickness can be understood by considering the same band $|F(\mathbf{x})-mT|<\epsilon$ around a regular point $\mathbf{x}_0$ on a level set. A first-order expansion along the unit normal to that level set gives the approximate geometric thickness
\begin{equation}
\boxed{w(\mathbf{x}_0)\approx\frac{2\epsilon}{\|\nabla F(\mathbf{x}_0)\|}}.
\label{eq_coherence_stroke_width}
\end{equation}
Likewise, when the field changes little across adjacent levels, their local separation is approximately $T/\|\nabla F\|$. Thus, a gradient magnitude of two approximately halves the intended thickness, whereas a magnitude of one-half approximately doubles it. Near critical points, where the gradient of the pencil coordinate vanishes, the first-order approximation breaks down and the level-set geometry can become degenerate. Uncontrolled variations in thickness and spacing are therefore inherent possibilities of this construction, even when the image gradient itself is computed accurately. These distortions arise before any Hölder combination or image-dependent thresholding is applied.

\subsection{Experimental Observations and Loss of Geometric Control}

\begin{figure}[htbp!]
\centering
\begin{subfigure}[b]{0.495\linewidth}
\centering
\includegraphics[width=\linewidth]{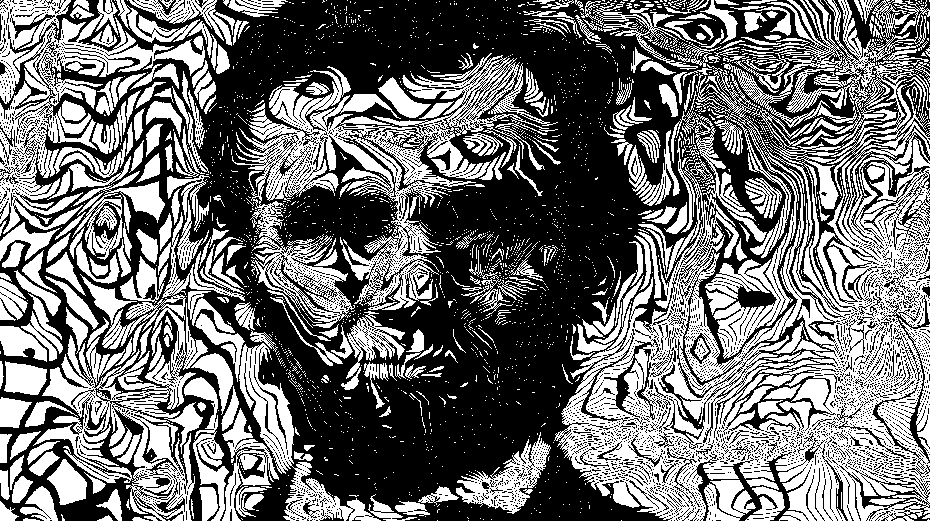}
\caption{Single pencil of lines.}
\label{fig_classical_gradient0}
\end{subfigure}
\hfill
\begin{subfigure}[b]{0.495\linewidth}
\centering
\includegraphics[width=\linewidth]{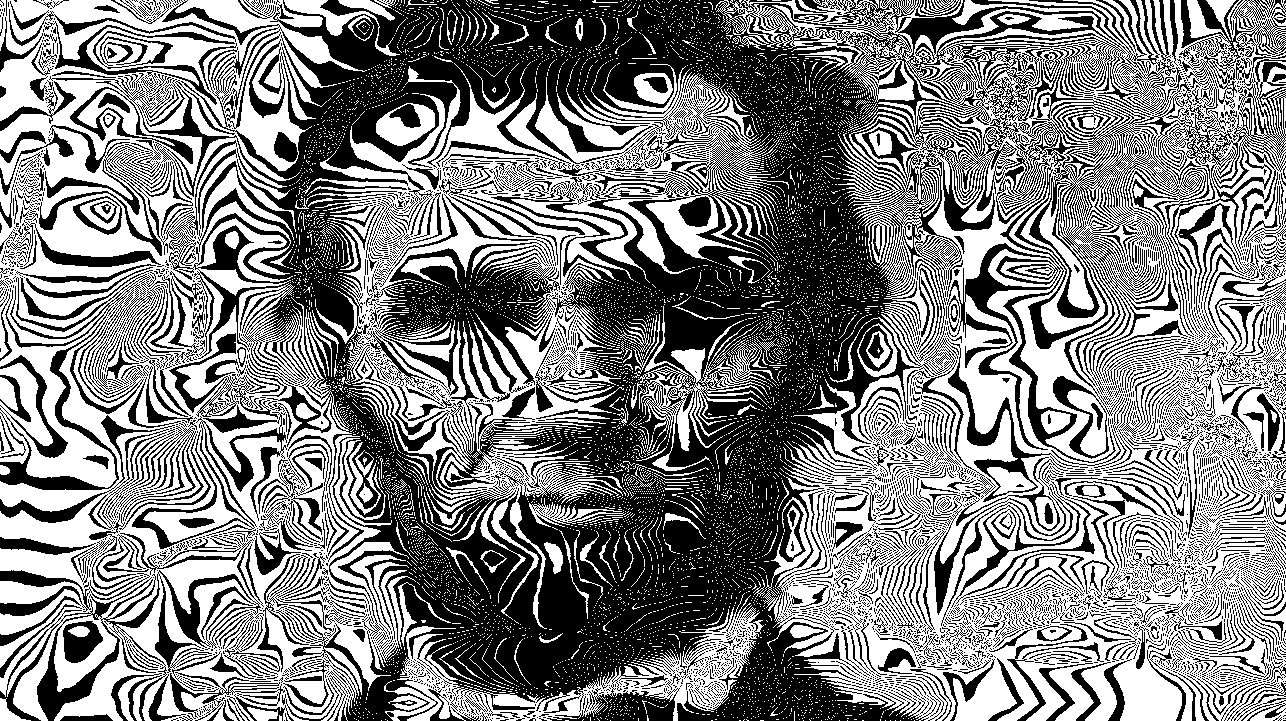}
\caption{Another single pencil of lines.}
\label{fig_classical_gradient3}
\end{subfigure}
\hfill
\begin{subfigure}[b]{0.495\linewidth}
\centering
\includegraphics[width=\linewidth]{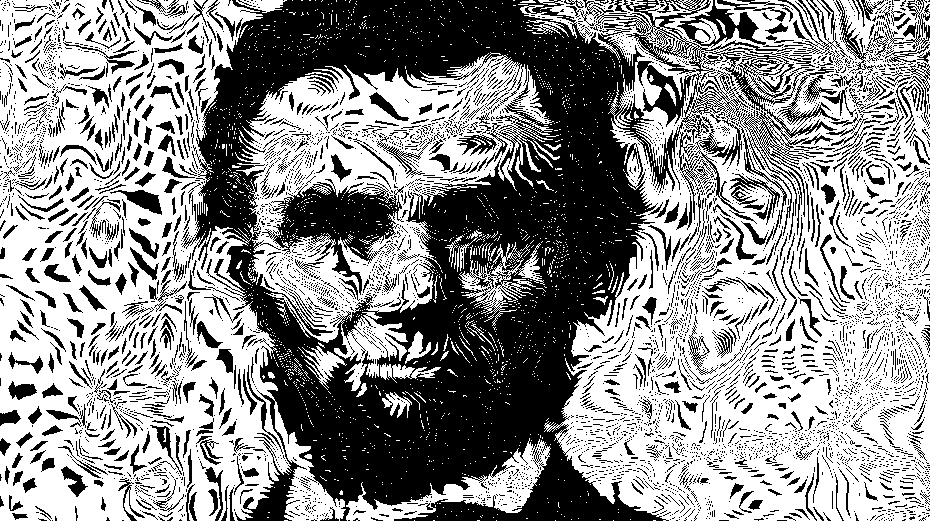}
\caption{Three pencils, $p=2$.}
\label{fig_classical_gradient1}
\end{subfigure}
\hfill
\begin{subfigure}[b]{0.495\linewidth}
\centering
\includegraphics[width=\linewidth]{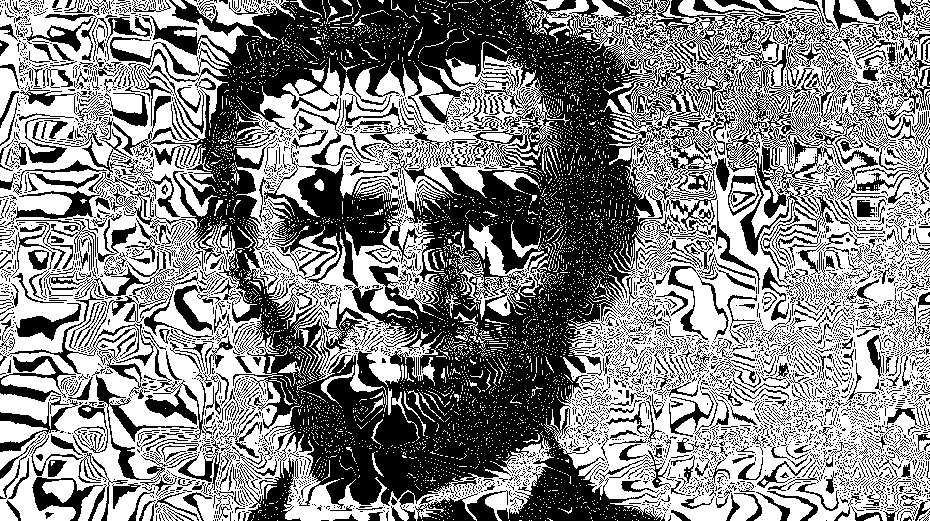}
\caption{Three pencils, $p\rightarrow+\infty$.}
\label{fig_classical_gradient2}
\end{subfigure}
\caption{Examples of artistic dithering using continuously varying classical-gradient orientations. The single-pencil examples exhibit substantial variations in stroke thickness and spacing. The three-pencil examples, obtained with a quadratic Hölder mean and its maximum limit, exhibit similarly distorted and fragmented marks. Although the parameters change the resulting images, their intended geometric effects are obscured by uncontrolled spatial deformation of the underlying pencils. These examples illustrate the difficulty of obtaining predictable hatching and point-like patterns through direct substitution of continuously varying image-gradient directions into stationary pencil equations.}
\label{fig_classical_gradient}
\end{figure}

We experimented with several continuous gradient estimators, including conventional centered finite differences, sums of directional differences over a variable number of uniformly sampled directions, and a biased variant intended to stabilize weak gradients. We also used mipmap-based smoothing of the input image during directional sampling. Increasing the number of samples and smoothing the image did not eliminate the undesirable geometric behavior in our experiments. Figure~\ref{fig_classical_gradient} shows representative results obtained using continuous classical-gradient orientations.

The single-pencil example in Figure~\ref{fig_classical_gradient0} is particularly informative because the observed geometric distortions cannot be attributed to the interaction of multiple pencils. The variations in thickness, densely compressed lines, and regions of broad irregular marks are consistent with the behavior predicted by Equation~\eqref{eq_coherence_stroke_width}. In Figures~\ref{fig_classical_gradient1} and~\ref{fig_classical_gradient2}, combining three pencils using different Hölder exponents changes the appearance but does not recover the predictable geometric structures associated with stationary three-pencil configurations in Figure~\ref{fig_3lines}. The problem is therefore not merely an undesirable visual appearance; it is also a loss of effective artistic control. The number of pencils, their nominal period, and the Hölder exponent remain meaningful parameters of the mathematical construction, but their intended effects can be overwhelmed by the deformation induced by spatially varying orientations. The examples provide qualitative evidence of this difficulty; they do not constitute a controlled quantitative comparison of gradient estimators or measured stroke widths.

\subsection{Piecewise Stationary Dithering Functions}

These observations motivate the finite-direction selection method introduced in Equation~\eqref{eq_directional_contrast}. Rather than continuously rotating the pencil equations according to a numerically estimated gradient, we select a direction from a finite set of candidates. Each candidate orientation determines a stationary pencil or a stationary combination of pencils. For a fixed period, reference point, angular offsets, and Hölder exponent, we can therefore regard the adaptive construction as a selection among a finite family of dithering functions.

Let $\Omega_k=\{\mathbf{x}\in\Omega:k^*(\mathbf{x})=k\}$ be the set of points selecting candidate direction $k$, with ties resolved deterministically. These sets partition the image domain, although individual sets need not be connected. Within the interior of any such region, the pencil normal is constant up to reversal, and reversal does not alter its periodic unsigned distance function. For each pencil with fixed unit normal $\mathbf{n}_{k,i}$ and reference point $\mathbf{p}_i$, we obtain
\begin{equation}
F_{k,i}(\mathbf{x})=\mathbf{n}_{k,i}\cdot(\mathbf{x}-\mathbf{p}_i),\qquad \|\nabla F_{k,i}(\mathbf{x})\|=1.
\label{eq_coherence_piecewise_distance}
\end{equation}
The additional Jacobian term in Equation~\eqref{eq_coherence_variable_gradient} vanishes within these regions. Consequently, individual pencils retain their stationary metric properties locally, including their intended spacing and their stroke thickness at a fixed threshold. With $N$ pencils, the overall construction can be expressed as
\begin{equation}
G_k(\mathbf{x})=H_p\left(g_{k,0}(\mathbf{x}),\ldots,g_{k,N-1}(\mathbf{x})\right),\qquad g(\mathbf{x})=G_{k^*(\mathbf{x})}(\mathbf{x}).
\label{eq_coherence_piecewise_holder}
\end{equation}
We refer to these as \emph{piecewise stationary dithering functions}. The individual pencil functions preserve their local distance interpretation; the Hölder combination itself need not be a Euclidean distance function. Nevertheless, within a region of constant orientation the combination retains the geometry of its corresponding stationary pattern, allowing the number of pencils and the Hölder exponent to exert their intended geometric influence.

Our method does not guarantee globally connected strokes, smooth transitions between orientation regions, or uniform thickness when thresholds vary spatially. Discontinuities and fragmentation may still occur at region boundaries, and weak directional contrasts may produce unstable selections. Its advantage is that orientation-induced deformation is avoided within regions of constant orientation rather than being distributed throughout them. Quantizing a conventional gradient into a finite set of orientations would also provide this local property. Our direct maximum-contrast selection additionally permits finite-scale asymmetric directional comparisons and control over the number of sampled directions. The present observations motivate this choice for artistic purposes without establishing that it is universally superior to other discrete orientation-selection strategies.

\subsection{Implications for Artistic Hatching and Etching}

The distinction between image-gradient accuracy and geometric coherence has important artistic implications. An artist may intentionally vary stroke thickness, spacing, or orientation, but these variations should ideally arise from controllable artistic decisions rather than unintended distortions of the underlying representation. In our framework, selecting a stationary pattern allows changes in threshold, period, number of pencils, and Hölder exponent to remain meaningful mechanisms for shaping marks and controlling tonal appearance. The objective of directional estimation is therefore not necessarily to reproduce the most accurate continuous image-gradient field, but to select orientations that support geometrically coherent and controllable patterns. This also explains why increasing the numerical accuracy of a continuous gradient estimator need not improve the artistic result. A possible alternative for future work is to construct globally coherent curved pencil coordinates or to locally approximate distances by dividing level-set residuals by the magnitude of their gradients. Such corrections may improve stroke thickness near regular level sets, but do not by themselves guarantee uniform spacing, continuity, or predictable geometry of combined pencils.

\ifdefined\FloatBarrier\FloatBarrier\fi

\section{Results and Artistic Exploration}

Having established the mathematical formulation, computational methodology, and local geometric properties of piecewise stationary pencils, we now examine the visual results of the framework. Our objective is to explore the controllability of the underlying patterns rather than to evaluate faithful tonal reproduction. We begin with stationary configurations, for which the geometric effects of the parameters can be observed directly, and then examine locally adaptive patterns and the simultaneous manipulation of pattern geometry and tonal appearance. The examples provide qualitative illustrations of the framework's capabilities rather than quantitative performance measurements.

\subsection{Stationary Hatching and Pointillist Patterns}

\begin{figure}[htbp!]
\centering
\begin{subfigure}[b]{0.31\textwidth}  \fbox{\includegraphics[width=0.99\textwidth]{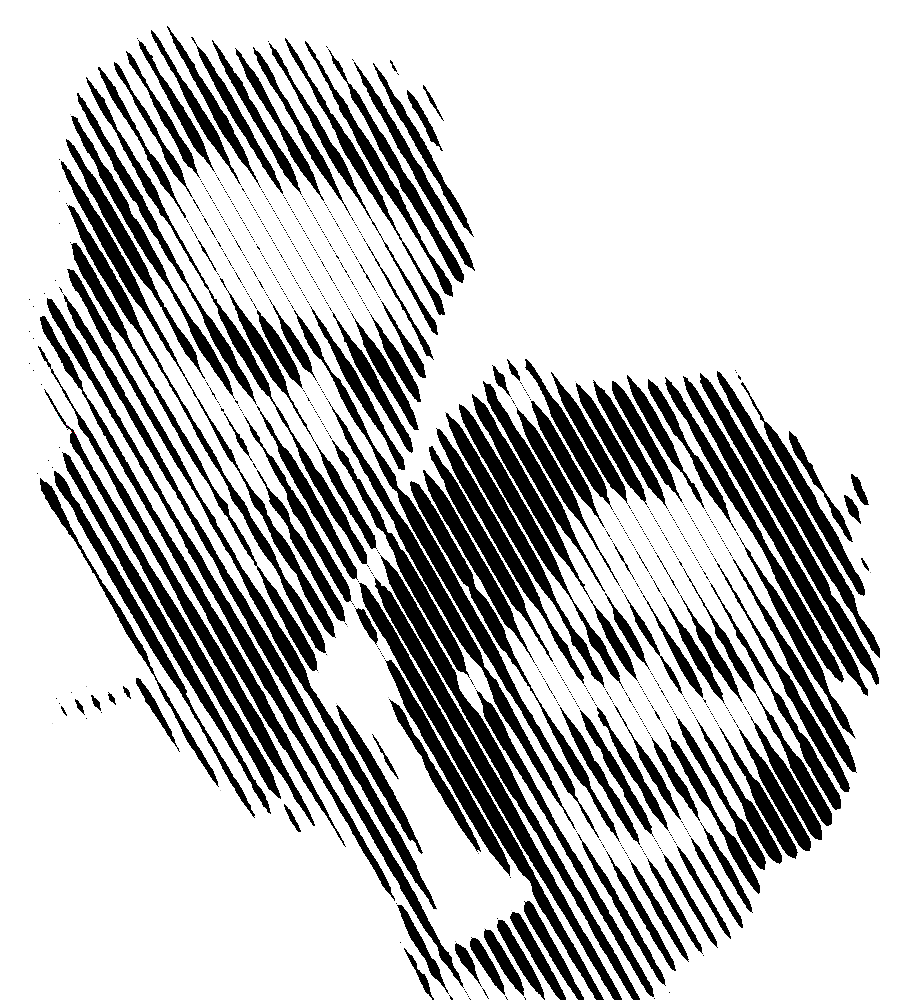}}
  \caption{A stationary hatching image obtained by a single pencil of lines.  }
\label{fig:teaser6}
\end{subfigure}
\hfill
\begin{subfigure}[b]{0.31\textwidth} 
\centering
\fbox{\includegraphics[width=0.99\textwidth]{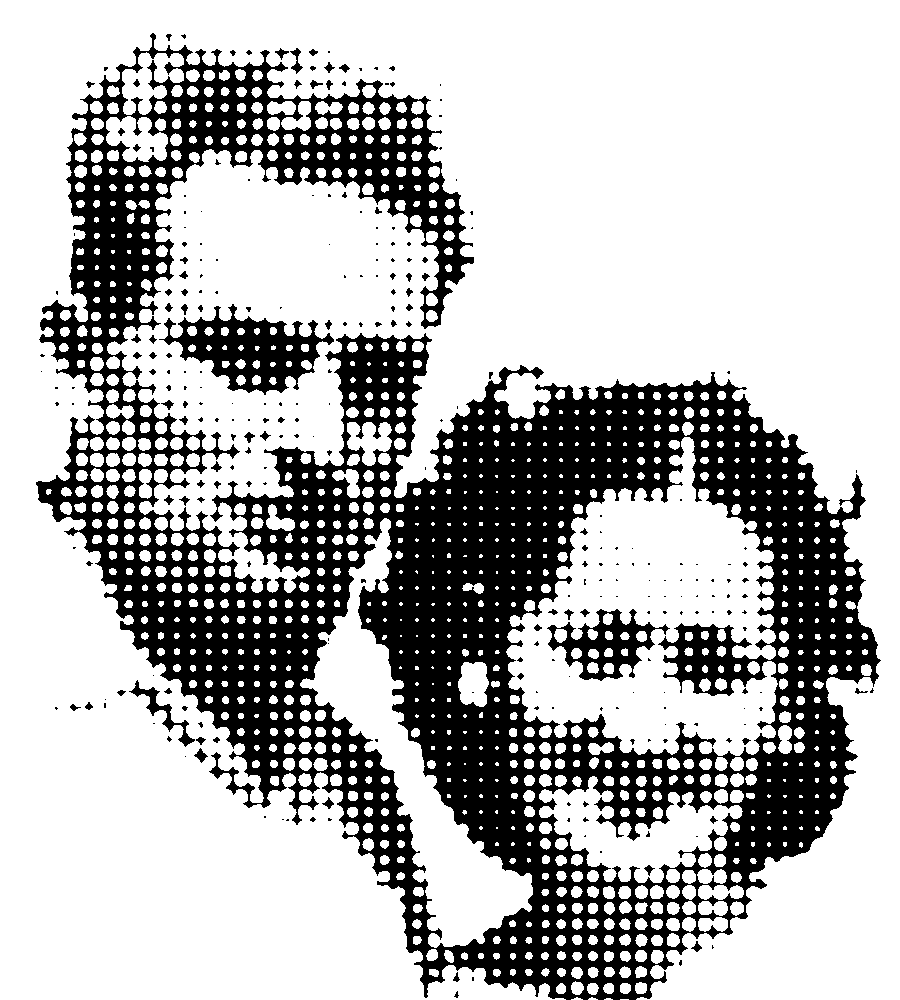}}
  \caption{A stationary pointillist image with a square grid.  }
\label{fig:teaser7}
\end{subfigure}
\hfill
\begin{subfigure}[b]{0.31\textwidth} \fbox{\includegraphics[width=0.99\textwidth]{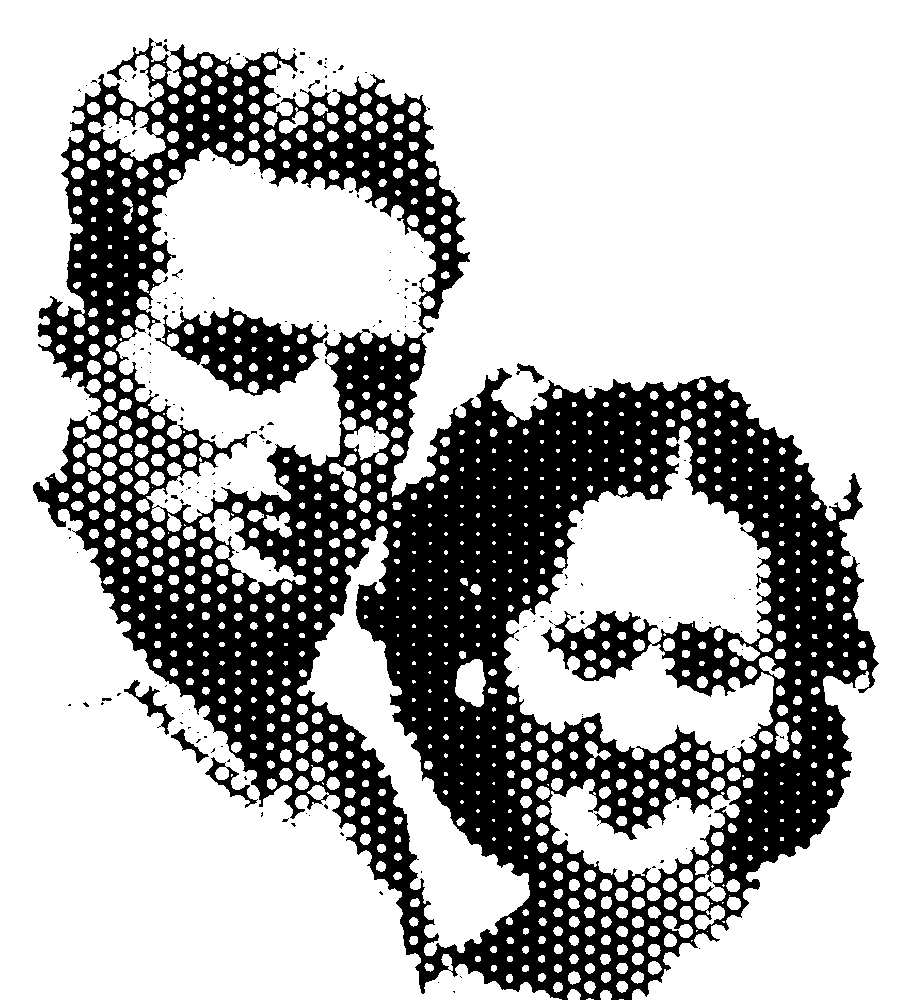}}
  \caption{A stationary pointillist image with a hexagonal grid. }
\label{fig:teaser5}
\end{subfigure}
\caption{Examples of stationary dithering. Figure~\ref{fig:teaser6} uses a single pencil of equidistant parallel lines, producing hatching with a uniform orientation. Figure~\ref{fig:teaser7} combines two perpendicular pencils using the root-mean-square mean ($p=2$), producing a square arrangement of dot-like elements reminiscent of traditional screening and some digital halftoning patterns. Figure~\ref{fig:teaser5} combines three equally spaced pencil orientations using the same mean, producing a hexagonal arrangement. The orientations remain fixed to reveal the underlying geometry. Two perpendicular pencils with unequal periods can instead produce a rectangular lattice.}
\label{fig:babamannem}
\end{figure}

Figure~\ref{fig:babamannem} compares a single stationary pencil with two and three pencils combined using the root-mean-square mean ($p=2$). The single pencil produces parallel hatching, whereas the two- and three-pencil configurations produce dot-like elements arranged on square and triangular lattices, respectively. The latter arrangement is often described visually as hexagonal. These examples illustrate the role of pencil arrangement as an artistic control independent of the exponent: even with the same exponent, different arrangements generate distinct geometric organizations of the marks. In contrast to the continuously varying orientations examined in the preceding section, the stationary configurations retain their underlying lattice geometry while the input image modulates the coverage of the marks.

\subsection{Geometric Transformations through Hölder Averages}

\begin{figure}[htbp!]
\centering
\begin{tabular}{ccc}
\fbox{\includegraphics[width=0.30\textwidth]{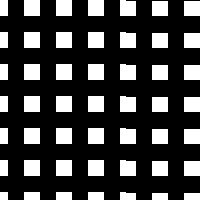}}&
\fbox{\includegraphics[width=0.30\textwidth]{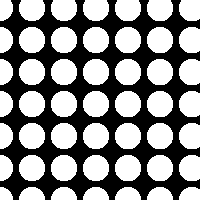}}&
\fbox{\includegraphics[width=0.30\textwidth]{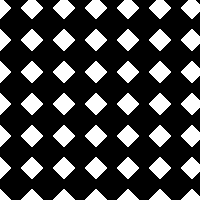}}\\
$p\rightarrow \infty$ & $p=2$ & $p=1$  \\
\fbox{\includegraphics[width=0.30\textwidth]{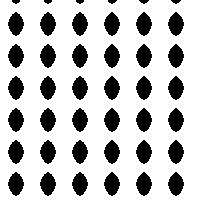}}&
\fbox{\includegraphics[width=0.30\textwidth]{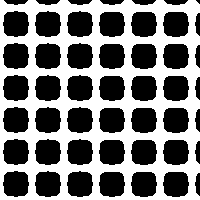}}&
\fbox{\includegraphics[width=0.30\textwidth]{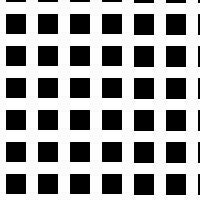}}\\ $p=0$ &$p=-1$ &$p\rightarrow -\infty$
\end{tabular}
  \caption{Hölder dithering patterns constructed from two perpendicular pencils with equal periods and a common reference point $\mathbf{p}_0=\mathbf{p}_1=(0,0)$. Each example is obtained by dithering a constant-valued image. Varying $p$ changes both the geometry and the relative coverage of black and white regions.}
\label{fig_2lines}
  \end{figure}

\begin{figure}[htbp!]
\centering
\begin{tabular}{ccc}
\fbox{\includegraphics[width=0.30\textwidth]{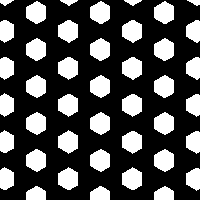}}&
\fbox{\includegraphics[width=0.30\textwidth]{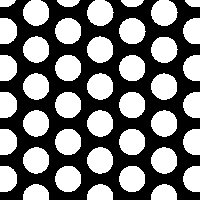}}&
\fbox{\includegraphics[width=0.30\textwidth]{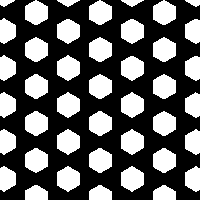}}\\
$p\rightarrow \infty$ & $p=2$ & $p=1$ \\
\fbox{\includegraphics[width=0.30\textwidth]{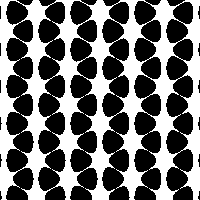}}&
\fbox{\includegraphics[width=0.30\textwidth]{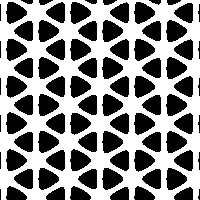}}&
\fbox{\includegraphics[width=0.30\textwidth]{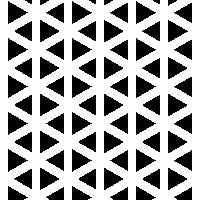}}\\
$p=0$  & $p=-1$ &$p\rightarrow -\infty$
\end{tabular}
  \caption{Hölder dithering patterns constructed from three pencils with equal periods, angular increments of $\pi/3$, and $\mathbf{p}_0=\mathbf{p}_1=\mathbf{p}_2=(0,0)$. Each example is obtained by dithering a constant-valued image. The underlying triangular lattice supports a hexagonal arrangement of visual elements.}
\label{fig_3lines}
  \end{figure}

Figures~\ref{fig_2lines} and~\ref{fig_3lines} isolate the effects of the exponent using constant-valued input images. In the two-pencil configuration, the maximum produces square white openings in a black network, while the quadratic mean produces rounded openings and the arithmetic mean produces diamond-like openings. As the exponent decreases further, the black regions become isolated marks and ultimately the minimum produces a union of white bands. The three-pencil configuration exhibits a corresponding range of transformations, including polygonal and rounded openings and triangular arrangements of black marks. These examples demonstrate that the exponent affects not only the shape of individual elements but also their connectivity and coverage. The exact transitions depend on the threshold and pencil configuration, as discussed in the theoretical framework.

\subsection{Rotational Symmetry and Pattern Complexity}

\begin{figure}[htbp!]
\centering
\begin{tabular}{ccc}
\fbox{\includegraphics[width=0.30\textwidth]{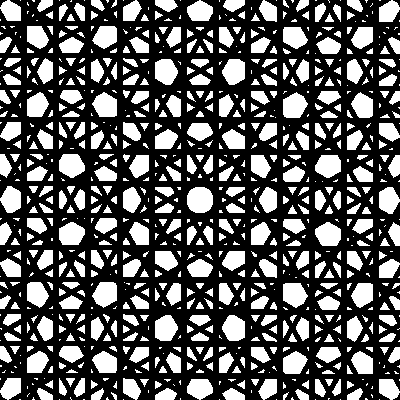}}&
\fbox{\includegraphics[width=0.30\textwidth]{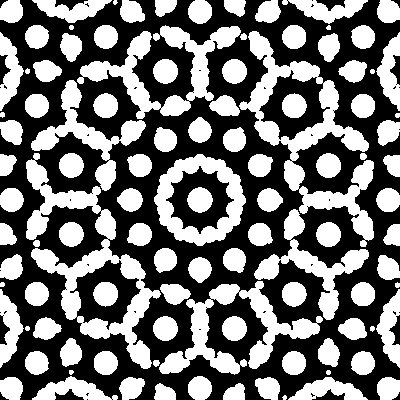}}&
\fbox{\includegraphics[width=0.30\textwidth]{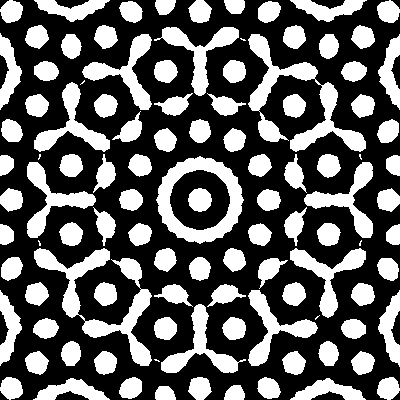}}\\
$p\rightarrow \infty$ & $p=2$ & $p=1$  \\
 \fbox{\includegraphics[width=0.30\textwidth]{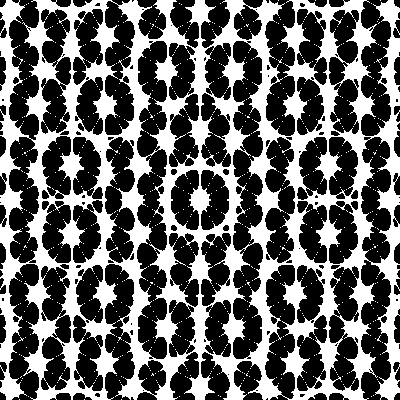}}&
\fbox{\includegraphics[width=0.30\textwidth]{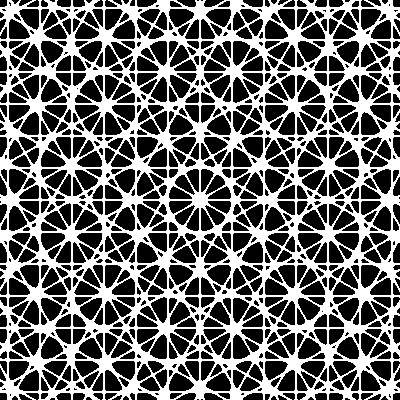}}&
\fbox{\includegraphics[width=0.30\textwidth]{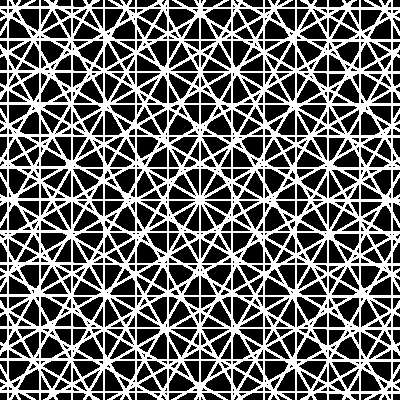}}\\
 $p=0$ & $p=-1$ &$p\rightarrow -\infty$
\end{tabular}
  \caption{Hölder dithering patterns constructed from six pencils with equal periods, angular increments of $\pi/6$, and a common reference point at the origin. These conditions give the underlying functions rotational symmetry about the origin. Rotational symmetry does not imply translational periodicity.}
\label{fig_6lines}
  \end{figure}

Figure~\ref{fig_6lines} extends the exploration to six equally rotated pencils. The resulting patterns exhibit a richer organization of interlocking and isolated elements than the two- and three-pencil examples. Their underlying functions retain the rotational symmetry established in the theoretical framework, without requiring a common two-dimensional translational lattice. Varying the exponent changes the visual organization while leaving the pencil orientations, periods, and reference points fixed. These examples suggest applications beyond conventional image hatching, including decorative patterns and ornamental textures.

\subsection{Locally Adaptive Artistic Dithering}

\begin{figure*}[htbp!]
\centering
        \begin{subfigure}[t]{0.32\textwidth}
      \fbox{\includegraphics[width=0.95\textwidth]{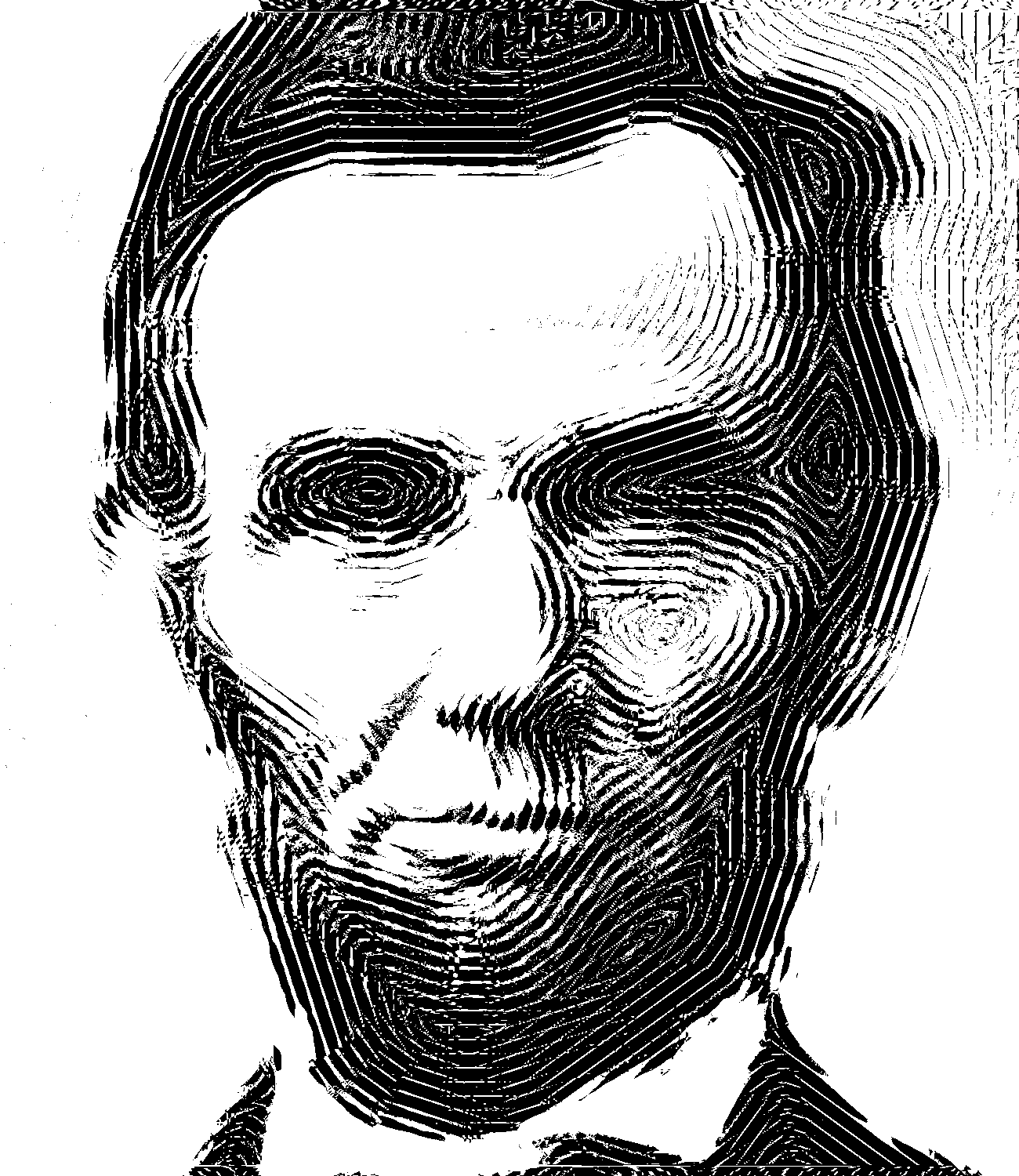}}
        \caption{Single-pencil dithering with lines perpendicular to local gradient orientations.}
        \label{fig_lincoln7}
    \end{subfigure}
    \hfill
    \begin{subfigure}[t]{0.32\textwidth}
      \fbox{\includegraphics[width=0.95\textwidth]{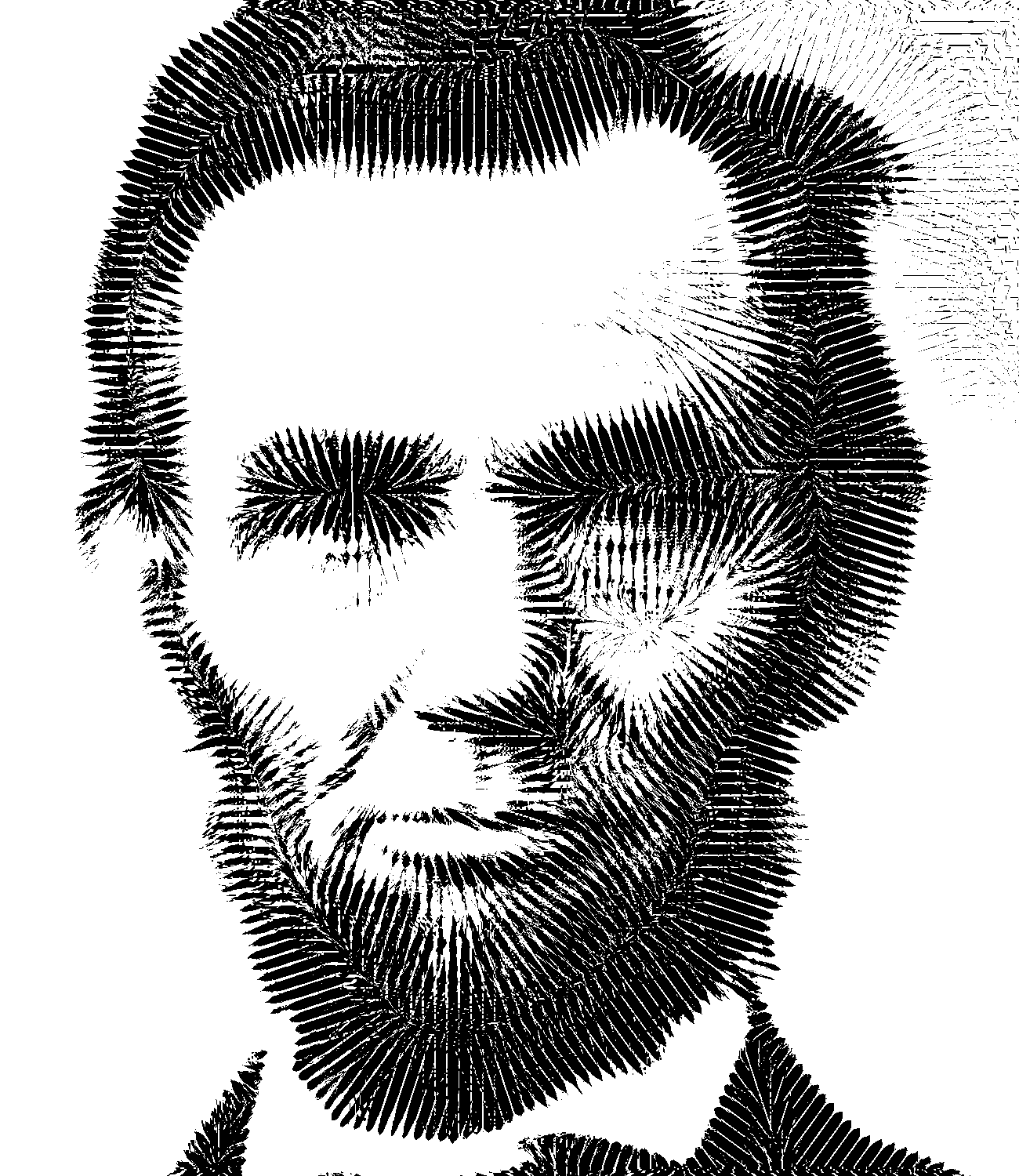}}
        \caption{Single-pencil dithering with lines perpendicular to local tangent orientations.}
        \label{fig_lincoln8}
    \end{subfigure}
    \hfill
    \begin{subfigure}[t]{0.32\textwidth}
      \fbox{\includegraphics[width=0.95\textwidth]{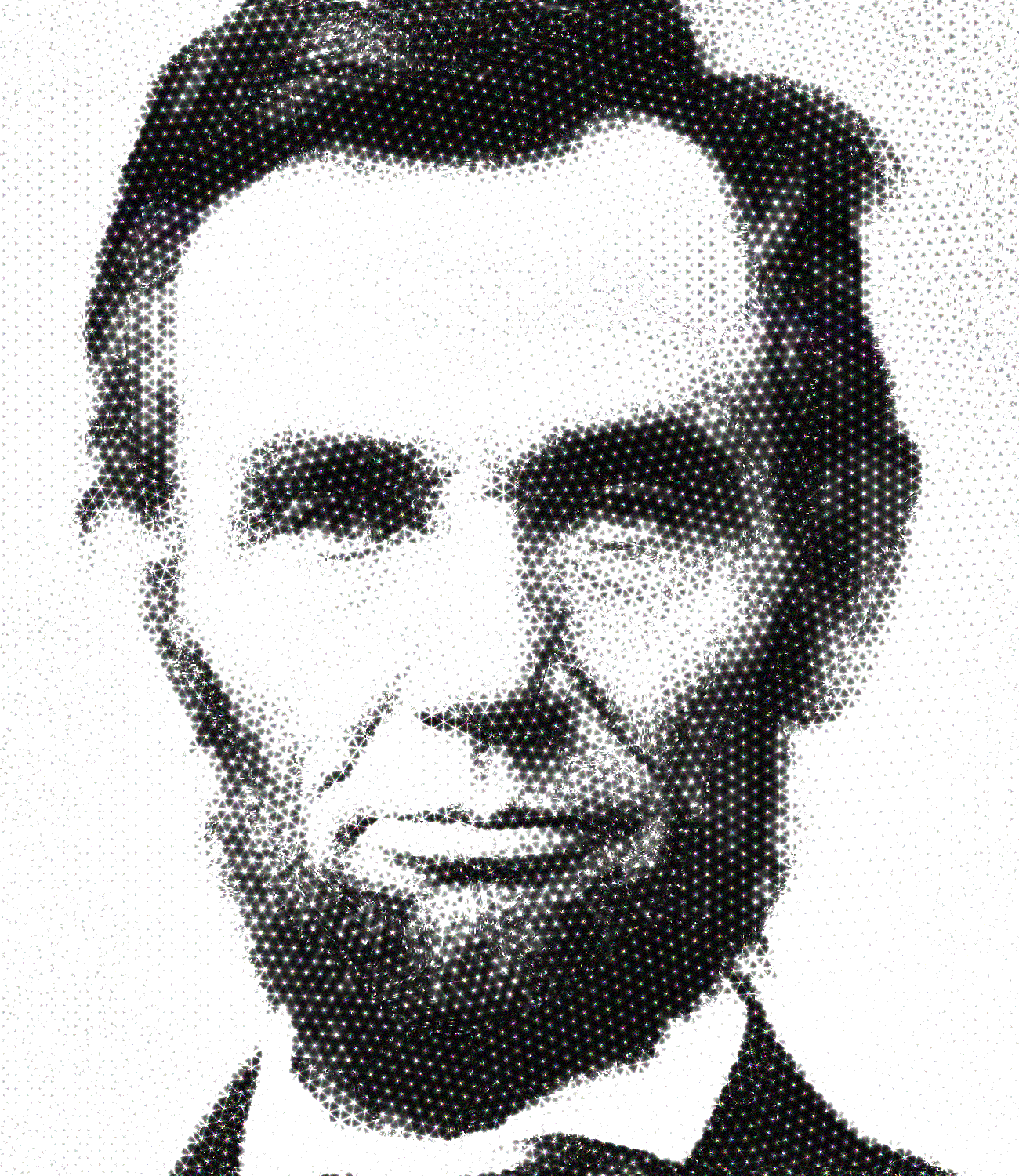}}
        \caption{Composite dithering with three pencils and primary hatching perpendicular to local gradient orientations.}
        \label{fig_lincoln5}
    \end{subfigure}
    \hfill
  \caption{Locally adaptive dithering. Figures~\ref{fig_lincoln7} and~\ref{fig_lincoln8} use a single pencil oriented perpendicular to the estimated gradient and tangent, respectively, with 30 evenly spaced directional sampling steps. Figure~\ref{fig_lincoln5} uses a spatially weighted interpolation between the minimum and maximum of three equally rotated pencil functions, with seven directional sampling steps. The latter example uses neither gray-level layers nor deliberate randomness.}
\label{fig_teaser8}
\end{figure*}

Figures~\ref{fig_lincoln7} and~\ref{fig_lincoln8} compare single-pencil hatching perpendicular to the gradient and tangent, respectively. The examples in Figures~\ref{fig_teaser}, \ref{fig_teaser_jolie}, and~\ref{fig_teaser4} also exploit locally adapted pencil orientations, combining two pencils for crosshatching and pointillist effects. Figure~\ref{fig_lincoln5} illustrates a composite three-pencil pattern constructed using the interpolation described in Equation~\ref{eq_interpolate}. The directional sampling steps reported for the Lincoln examples describe their settings; the supplied two-pencil shader is an implementation of the same general approach and does not reproduce every setting of those earlier examples.

The locally adaptive examples illustrate how a finite family of stationary patterns can be used to respond to image features. In Figures~\ref{fig_lincoln7} and~\ref{fig_lincoln8}, changing the orientation offset produces distinct directional interpretations of the same image. Within regions of constant selected orientation, the individual pencils retain their metric properties, as established in the preceding section. This does not imply that strokes connect seamlessly across region boundaries. Figure~\ref{fig_lincoln5} further illustrates how interpolation between minimum and maximum functions can combine different geometric characters within a single rendering. Since the composite example also uses a different number of directional sampling steps, these images should not be interpreted as a controlled comparison of angular discretization alone.

\subsection{Simultaneous Control of Pattern Geometry and Tonal Appearance}

\begin{figure}[htbp!]
\centering
\begin{subfigure}[b]{0.24\textwidth}
\includegraphics[width=0.99\textwidth]{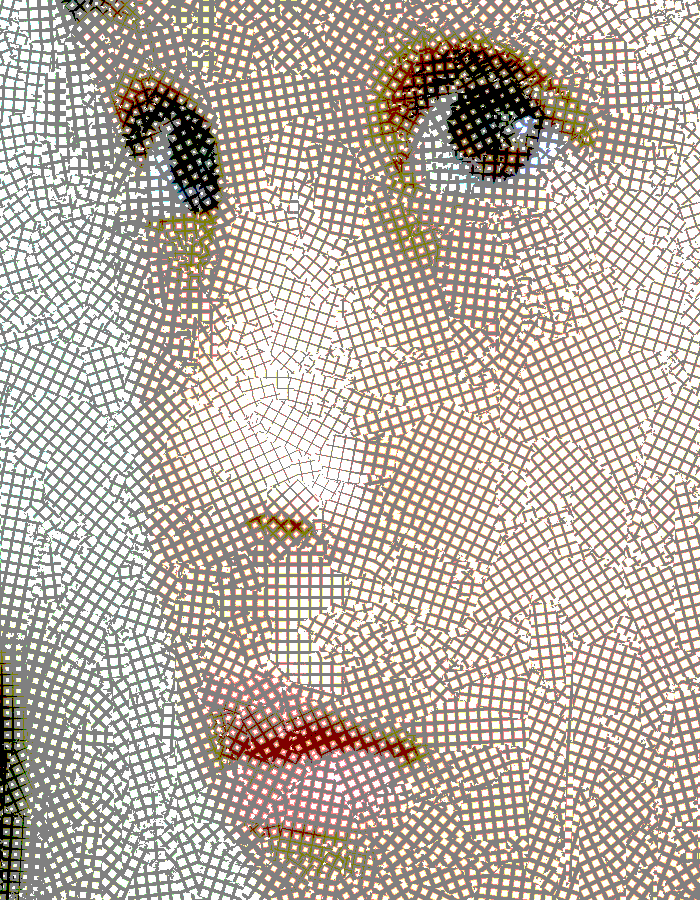}
        \caption{$p \rightarrow \infty$}
        \label{fig_asli0}
    \end{subfigure}   
          \hfill
\begin{subfigure}[b]{0.24\textwidth}
\includegraphics[width=0.99\textwidth]{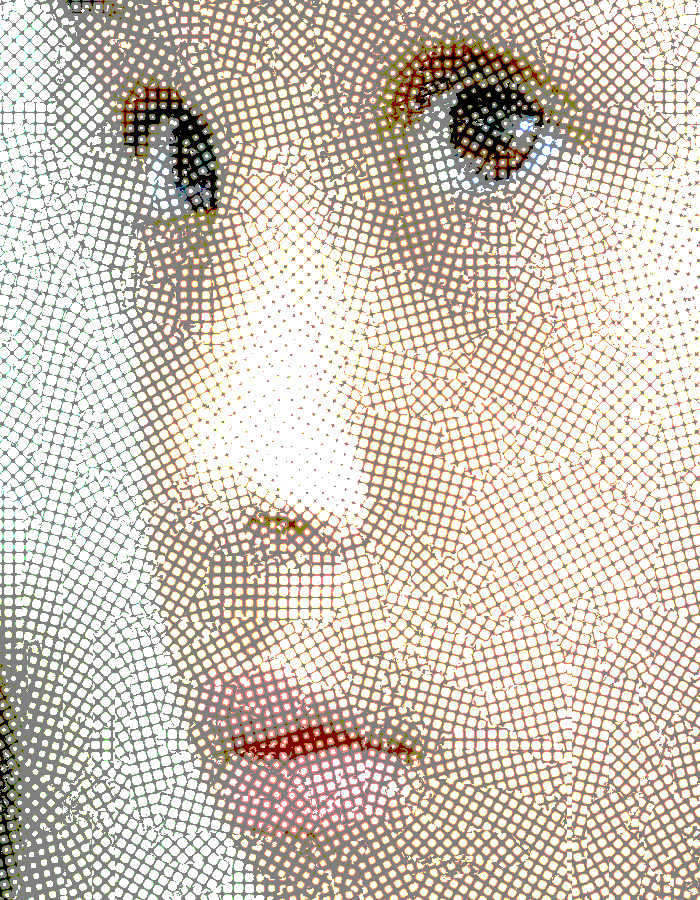}
        \caption{$p =4$}
        \label{fig_asli1}
    \end{subfigure}   
          \hfill
          \begin{subfigure}[b]{0.24\textwidth}
\includegraphics[width=0.99\textwidth]{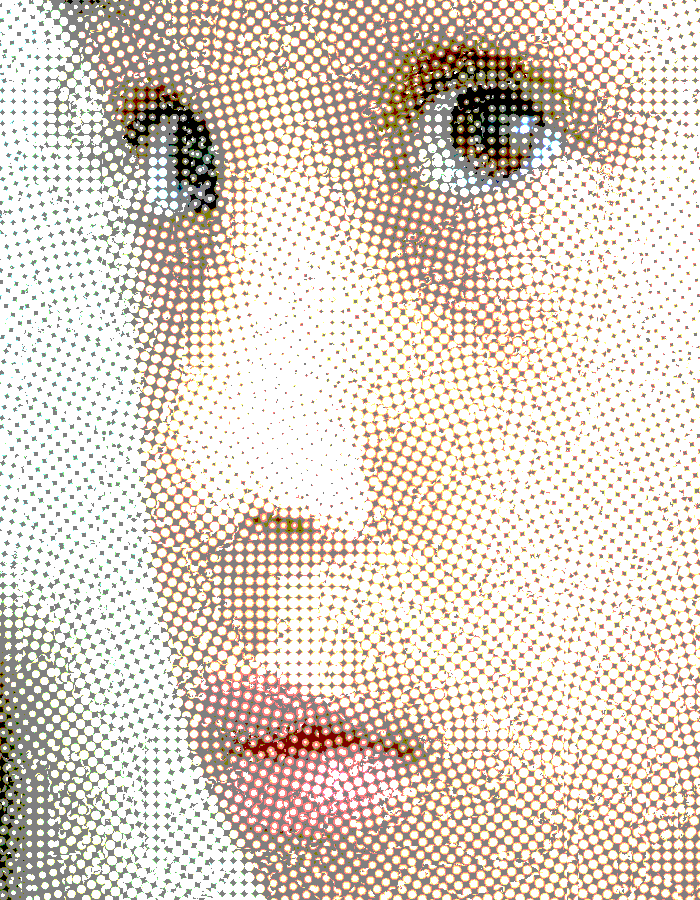}
        \caption{$p =2$}
        \label{fig_asli2}
    \end{subfigure}   
          \hfill
          \begin{subfigure}[b]{0.24\textwidth}
\includegraphics[width=0.99\textwidth]{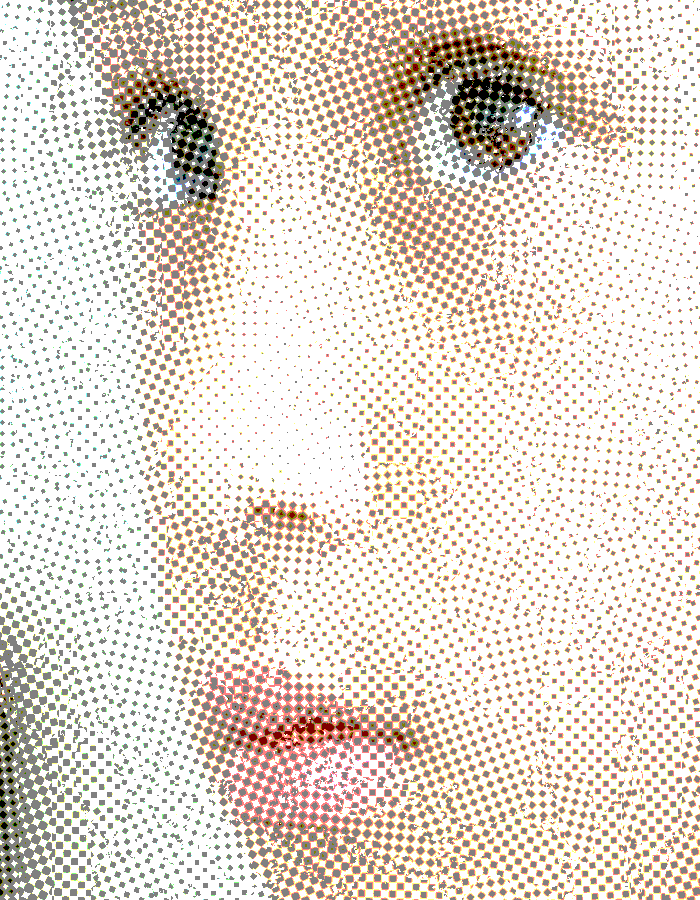}
        \caption{$p =1$}
        \label{fig_asli3}
    \end{subfigure}   
          \par\medskip
          \begin{subfigure}[b]{0.24\textwidth}
\includegraphics[width=0.99\textwidth]{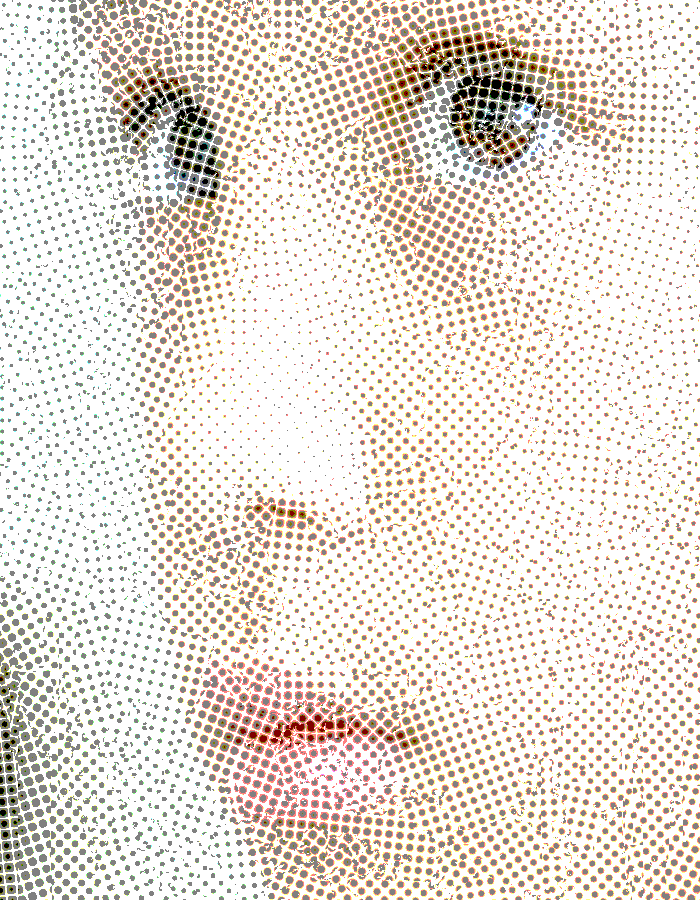}
        \caption{$p =-1$}
        \label{fig_asli4}
    \end{subfigure}   
          \hfill
\begin{subfigure}[b]{0.24\textwidth}
\includegraphics[width=0.99\textwidth]{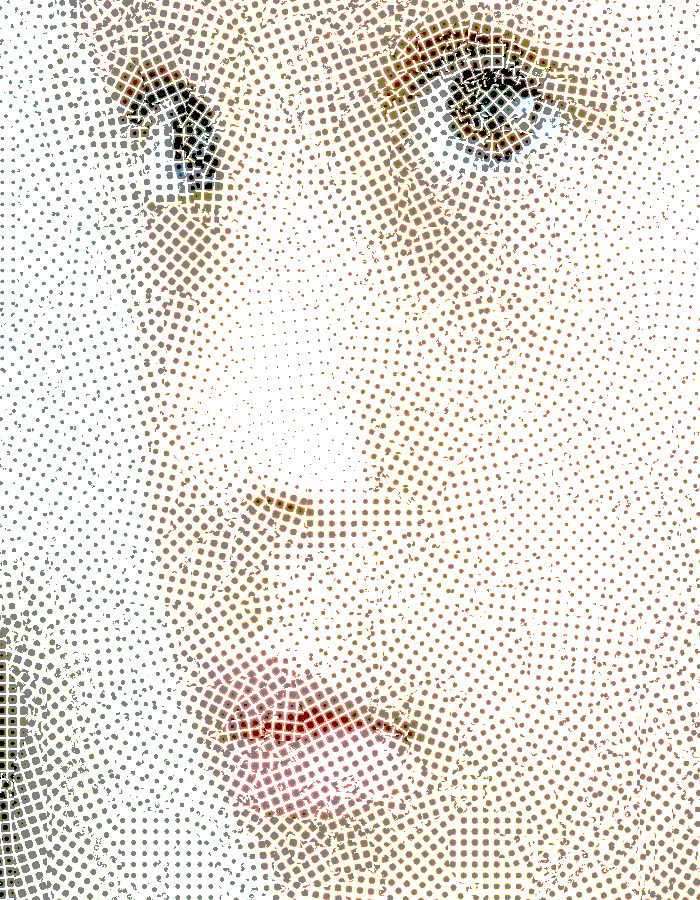}
        \caption{$p =-2$}
        \label{fig_asli5}
    \end{subfigure}   
          \hfill
          \begin{subfigure}[b]{0.24\textwidth}
\includegraphics[width=0.99\textwidth]{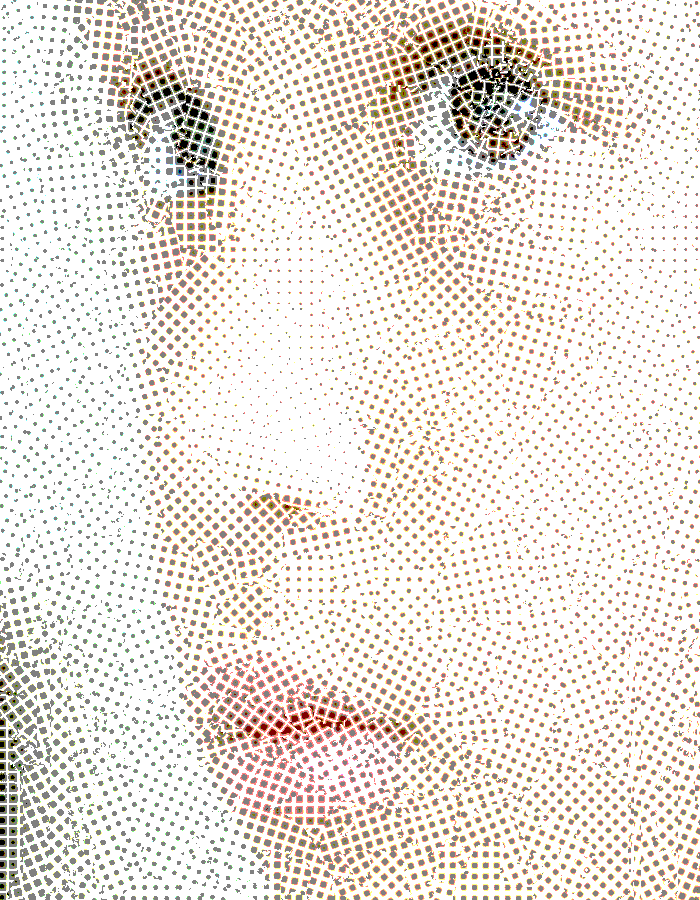}
        \caption{$p =-4$}
        \label{fig_asli6}
    \end{subfigure}   
          \hfill
          \begin{subfigure}[b]{0.24\textwidth}
\includegraphics[width=0.99\textwidth]{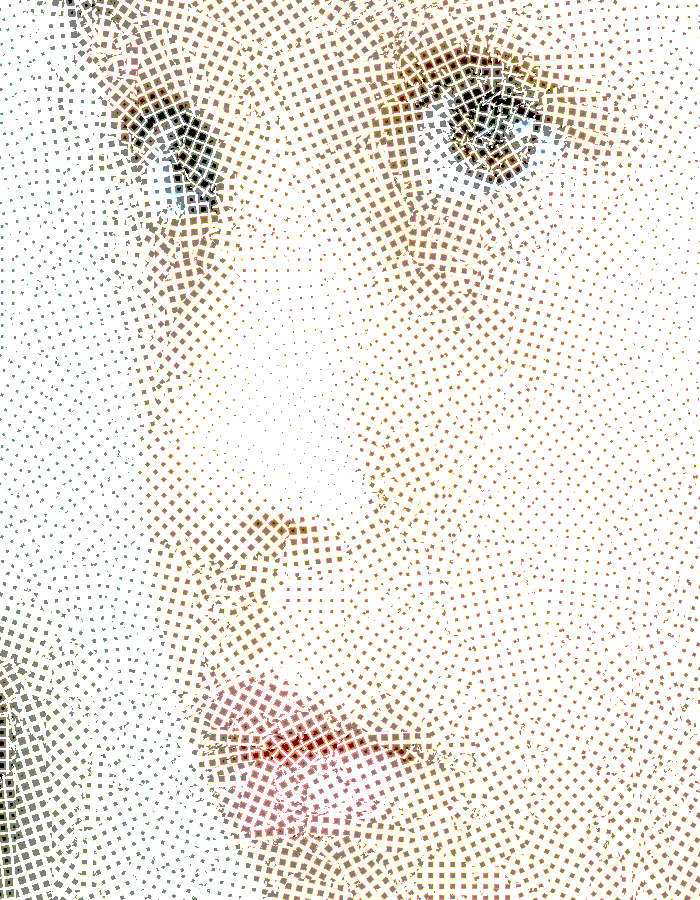}
        \caption{$p \rightarrow -\infty$}
        \label{fig_asli7}
    \end{subfigure}   
          \hfill
  \caption{Effects of varying the Hölder exponent on the geometry and tonal appearance of a portrait rendered using the functional dithering framework. The sequence ranges from the maximum operation ($p\rightarrow+\infty$) to the minimum operation ($p\rightarrow-\infty$), with six intermediate finite exponents. The change in the geometry of the marks is accompanied by a change in overall tonal appearance, providing a useful form of artistic control.}
\label{fig_asli}
\end{figure}

The monotonicity established in Equation~\ref{eq_black_nesting} explains why the exponent is an effective interactive control. Figure~\ref{fig_asli} presents a sequence of eight renderings ranging from the maximum to the minimum operation. In addition to the changes in the arrangement and shapes of the visual elements, the sequence illustrates substantial changes in overall tonal appearance. The ideal binary model establishes the coverage ordering when the input and other parameters remain fixed. The layer averages used for gray-level and color rendering similarly have non-increasing channel values as the exponent increases, provided their scales and perturbations are fixed and the means are evaluated exactly. This provides a mathematical basis for the observed artistic behavior without requiring a separate tonal correction for every exponent.

\begin{figure}[htbp!]
\centering
\begin{subfigure}[b]{0.99\textwidth}
\fbox{\includegraphics[width=0.23\textwidth]{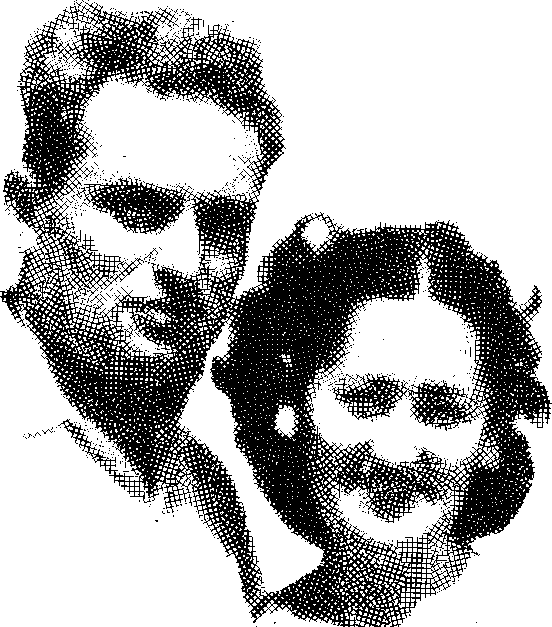}}\hfill
\fbox{\includegraphics[width=0.23\textwidth]{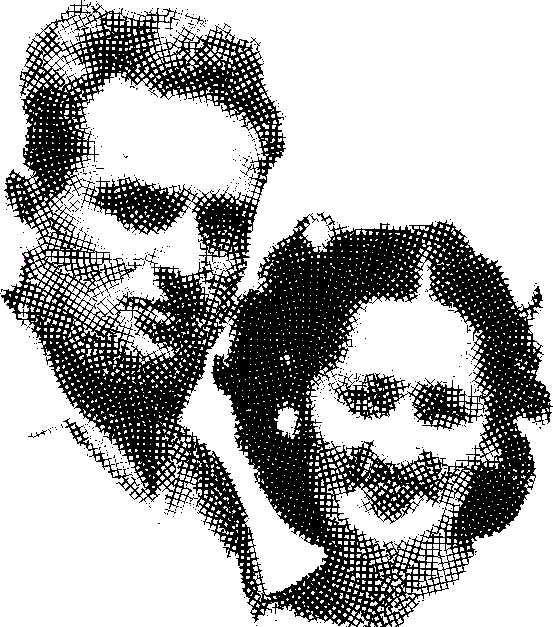}}\hfill
\fbox{\includegraphics[width=0.23\textwidth]{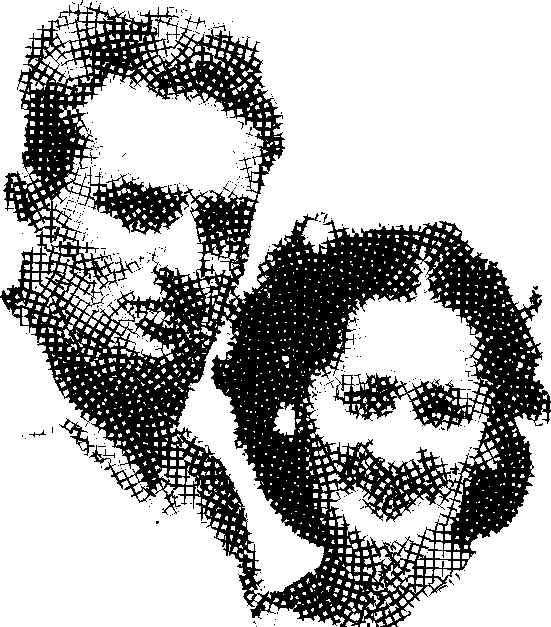}}\hfill
\fbox{\includegraphics[width=0.23\textwidth]{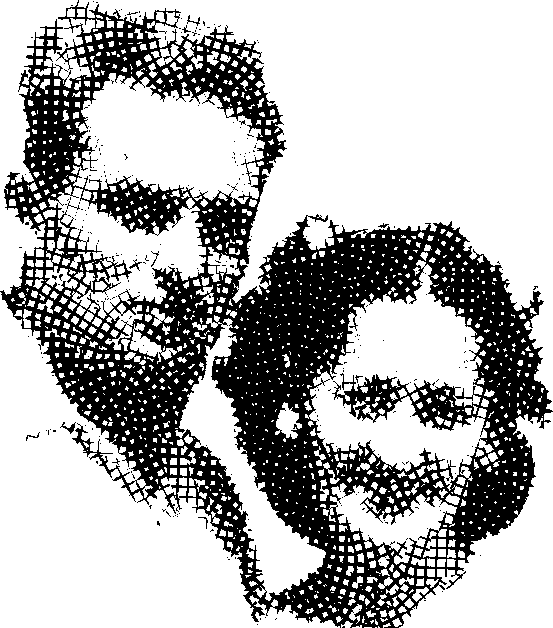}}\hfill
        \caption{Two sets of mutually perpendicular pencil of lines for $p\rightarrow \infty$ with varying periods}
        \label{fig_ab10}
    \end{subfigure}   
          \hfill
\begin{subfigure}[b]{0.99\textwidth}
\fbox{\includegraphics[width=0.23\textwidth]{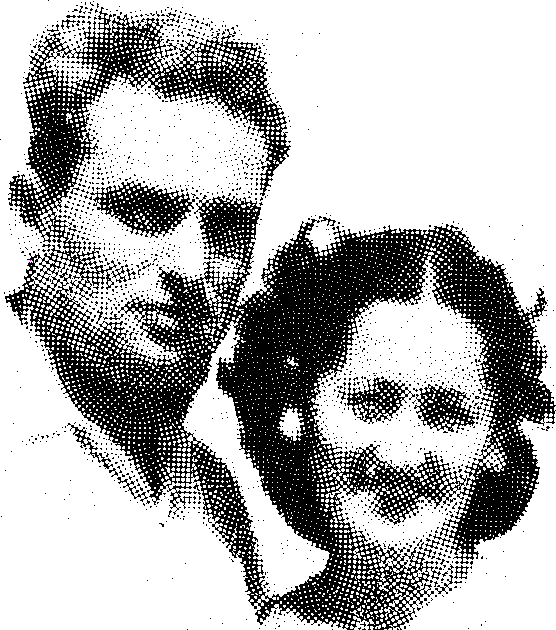}}\hfill
\fbox{\includegraphics[width=0.23\textwidth]{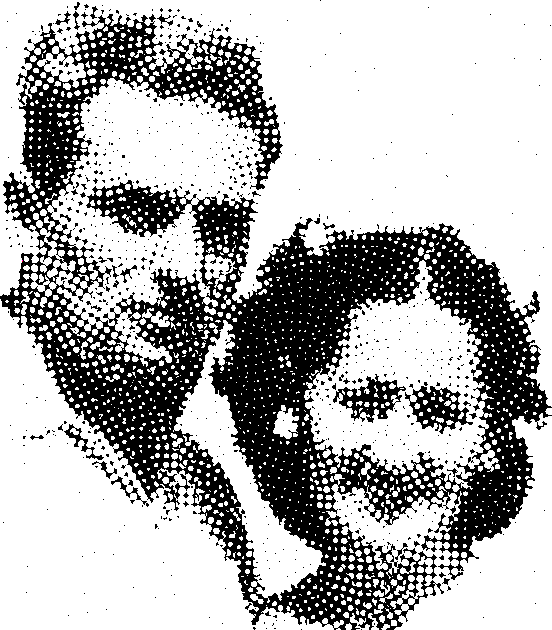}}\hfill
\fbox{\includegraphics[width=0.23\textwidth]{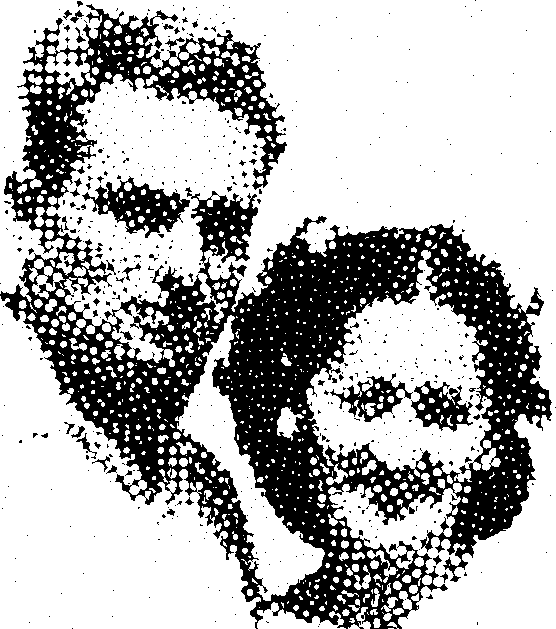}}\hfill
\fbox{\includegraphics[width=0.23\textwidth]{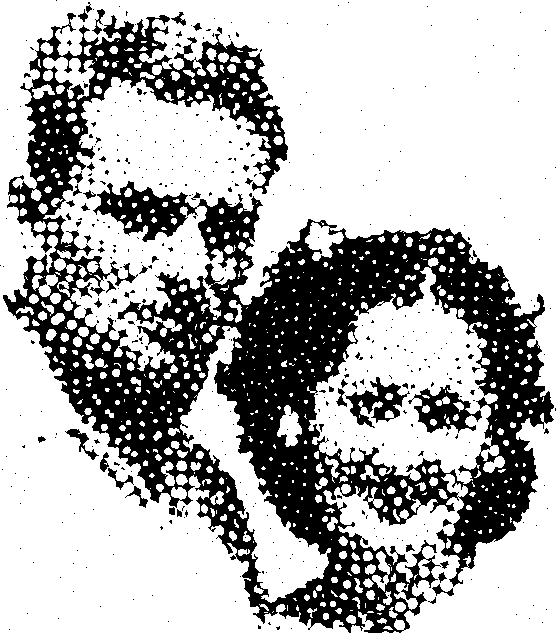}}\hfill
        \caption{Two sets of mutually perpendicular pencil of lines for $p=2.0$ with varying periods}
        \label{fig_ab20}
    \end{subfigure}   
          \hfill
  \caption{Effects of varying the periods.}
\label{fig_babamannem1}
\end{figure}

\begin{figure}[htbp!]
\centering
\begin{subfigure}[b]{0.99\textwidth}
\fbox{\includegraphics[width=0.23\textwidth]{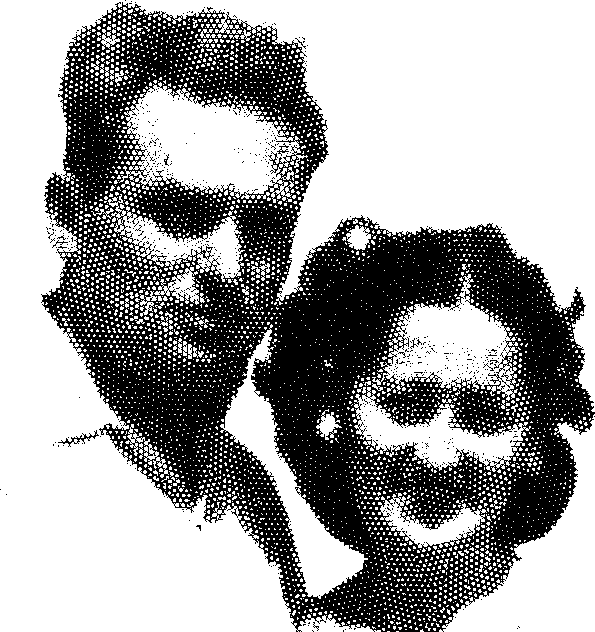}}\hfill
\fbox{\includegraphics[width=0.23\textwidth]{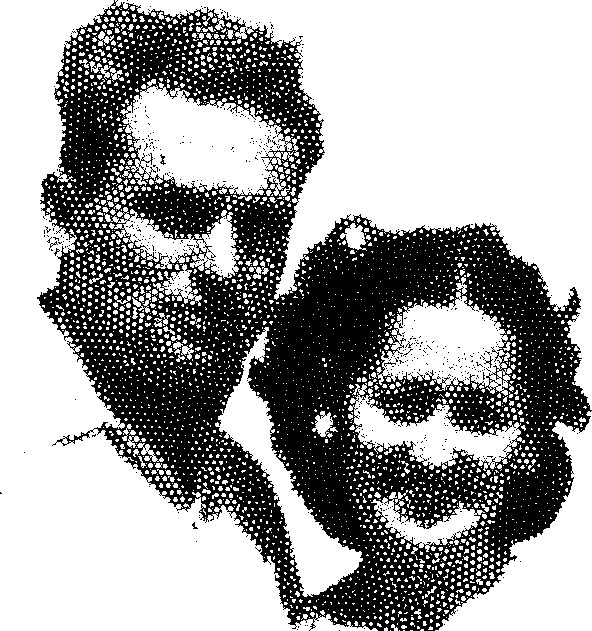}}\hfill
\fbox{\includegraphics[width=0.23\textwidth]{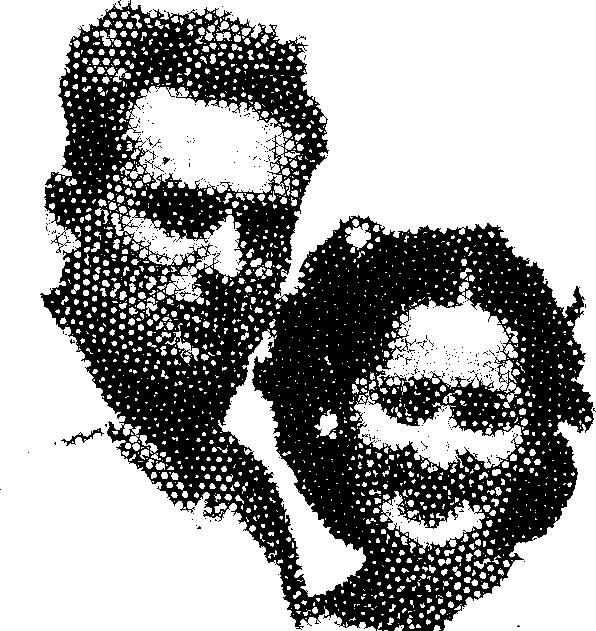}}\hfill
\fbox{\includegraphics[width=0.23\textwidth]{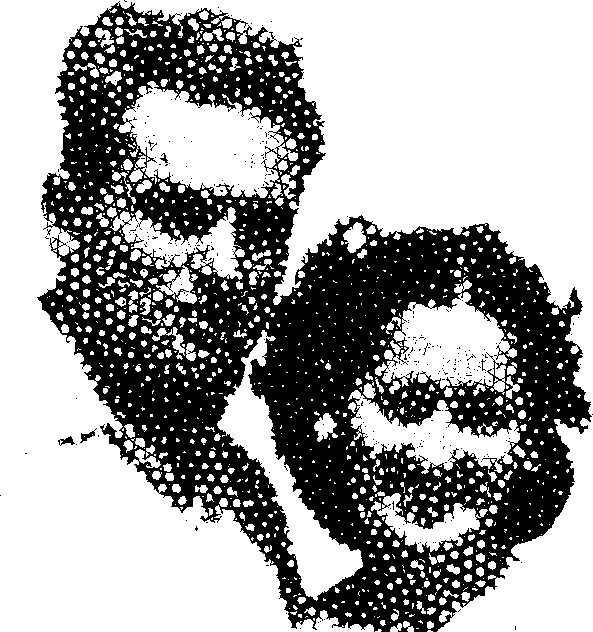}}\hfill
        \caption{Three equally rotated pencils of lines for $p=2.0$ with varying periods}
        \label{fig_ab50}
    \end{subfigure}   
          \hfill
\begin{subfigure}[b]{0.99\textwidth}
\fbox{\includegraphics[width=0.23\textwidth]{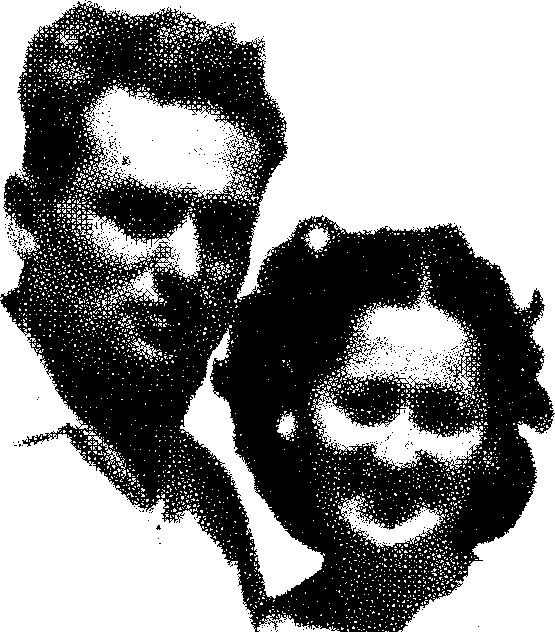}}\hfill
\fbox{\includegraphics[width=0.23\textwidth]{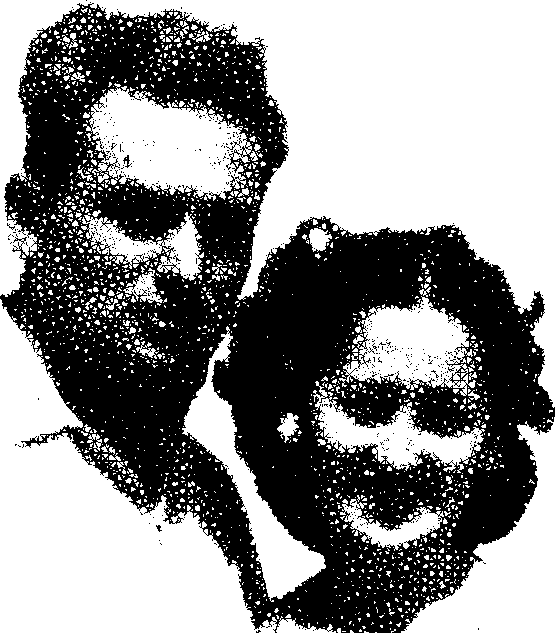}}\hfill
\fbox{\includegraphics[width=0.23\textwidth]{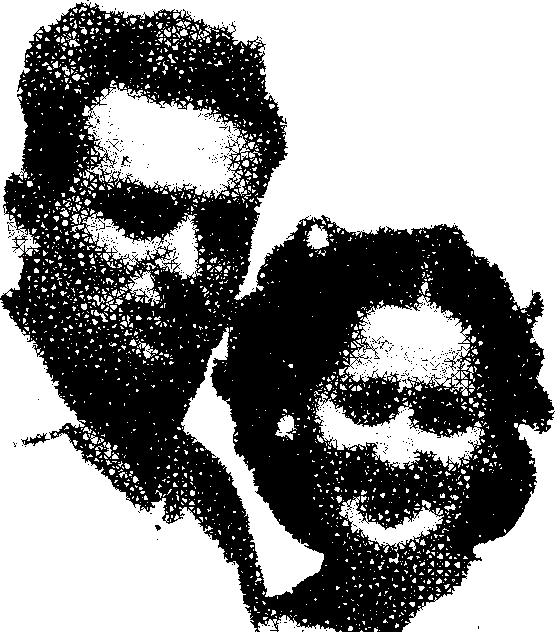}}\hfill
\fbox{\includegraphics[width=0.23\textwidth]{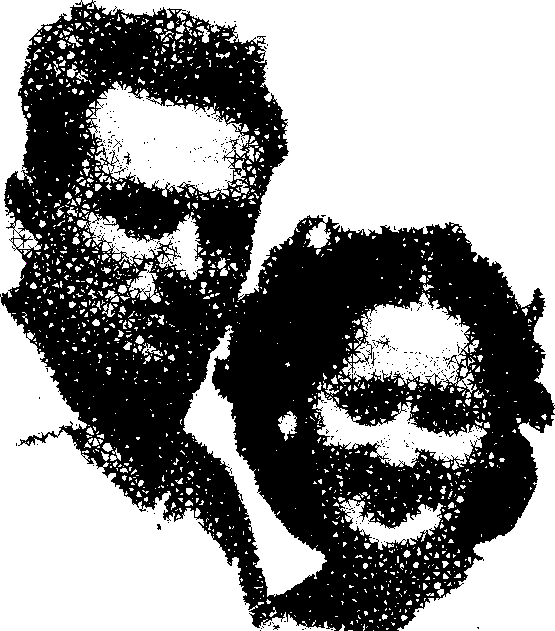}}\hfill
        \caption{Five equally rotated pencils of lines for $p=2.0$ with varying periods}
        \label{fig_ab60}
    \end{subfigure}   
          \hfill
  \caption{Effects of varying the periods.}
\label{fig_babamannem2}
\end{figure}

The examples in Figures~\ref{fig_teaser}, \ref{fig_teaser4}, and~\ref{fig_teaser5} extend this exploration to monochromatic and color renderings. The combination of threshold layers and controlled irregularity provides additional ways of changing the appearance of marks without requiring a separate dithering matrix for each visual style. These examples emphasize artistic flexibility rather than a measured improvement in tonal reproduction.
Figures~\ref{fig_babamannem1} and ~\ref{fig_babamannem2} provide additional examples illustrating the effects of varying the pencil periods.

\ifdefined\FloatBarrier\FloatBarrier\fi

\section{Discussion, Limitations, Future Work, and Conclusion}

Although the examples presented in the preceding sections demonstrate the artistic versatility of our functional dithering framework, they also reveal several fundamental challenges in achieving the expressive freedom of traditional artistic practices. In this section, we examine these challenges from both artistic and computational perspectives. In particular, we discuss the limitations of procedurally constructed geometric patterns, the distinction between geometric and tonal discretization, the relationship between local directional adaptation and spatial coherence, and the artistic role of geometric irregularities. These observations motivate several directions for future research that could further expand the expressive capabilities of free-form artistic dithering.

\subsection{Artistic Freedom and Functional Representations}

\begin{figure}[htbp!]
\centering
\begin{subfigure}[t]{0.31\textwidth}
    \centering    \includegraphics[height=0.99\textwidth,width=\linewidth,keepaspectratio]{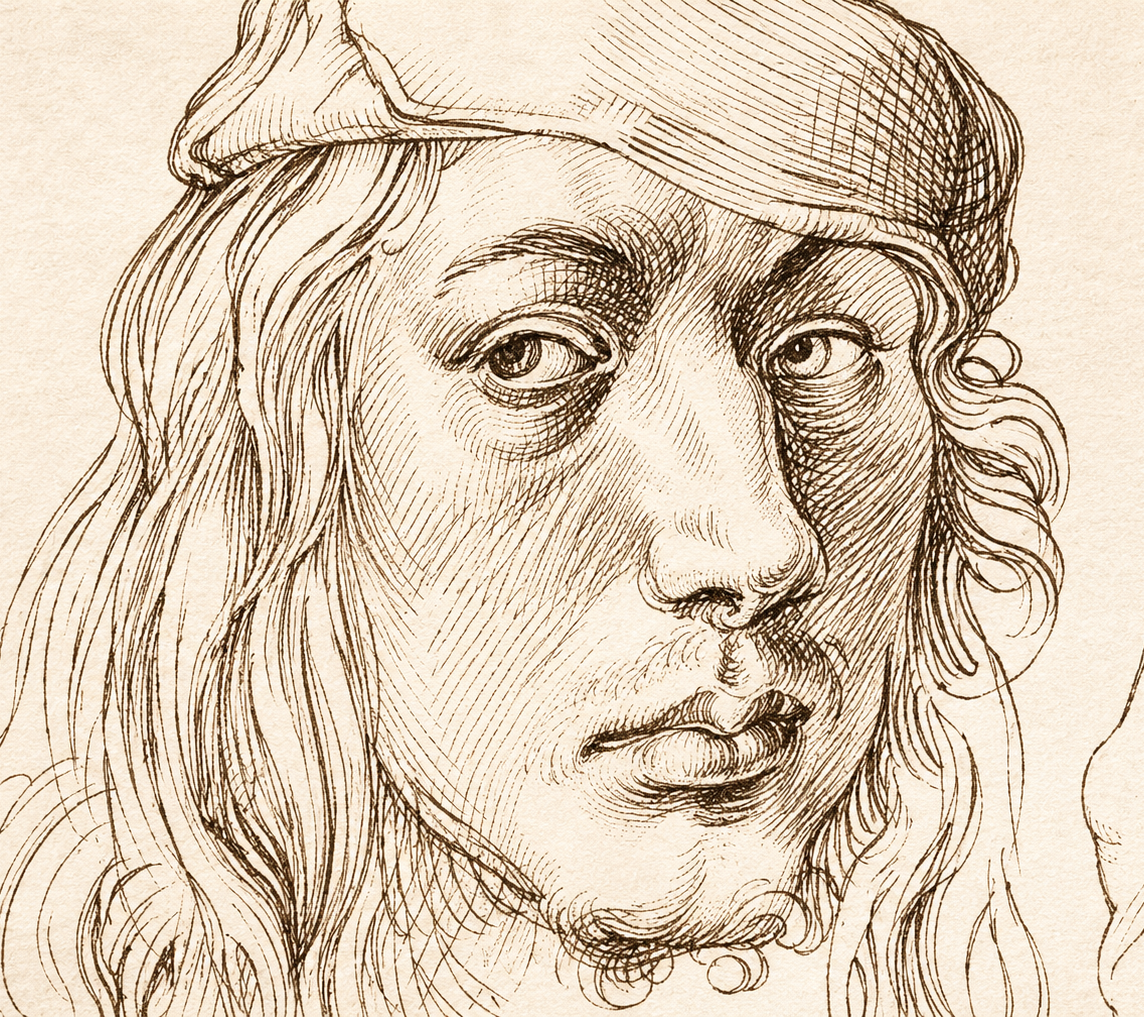}
    \caption{Albrecht Dürer, \emph{Self-Portrait, Study of a Hand and a Pillow} (detail), 1493. Pen and brown ink. Metropolitan Museum of Art, New York.}
    \label{fig_expressive_durer}
\end{subfigure}
\hfill
\begin{subfigure}[t]{0.31\textwidth}
    \centering   \includegraphics[height=0.99\textwidth,width=\linewidth,keepaspectratio]{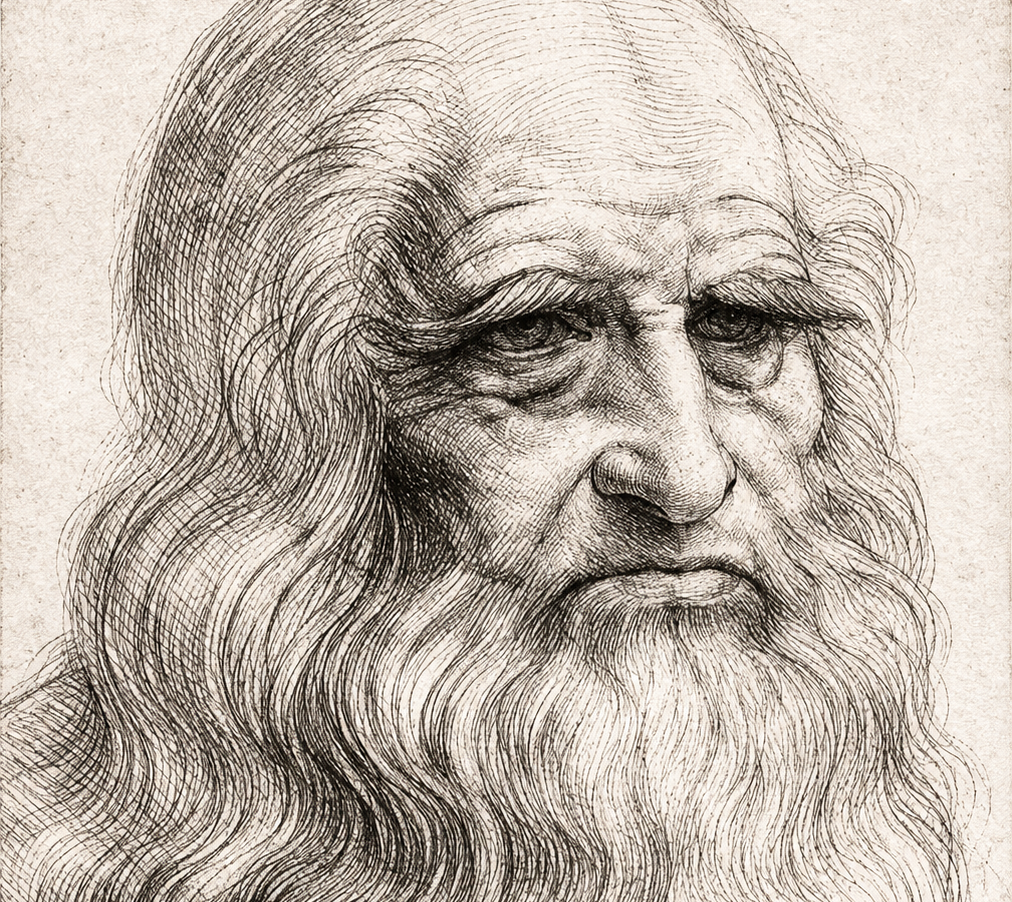}
    \caption{Leonardo da Vinci, presumed self-portrait (detail), ca.~1512. Red chalk. Biblioteca Reale, Turin.}
    \label{fig_expressive_leonardo}
\end{subfigure}
\hfill
\begin{subfigure}[t]{0.31\textwidth}
    \centering   \includegraphics[height=0.99\textwidth,width=\linewidth,keepaspectratio]{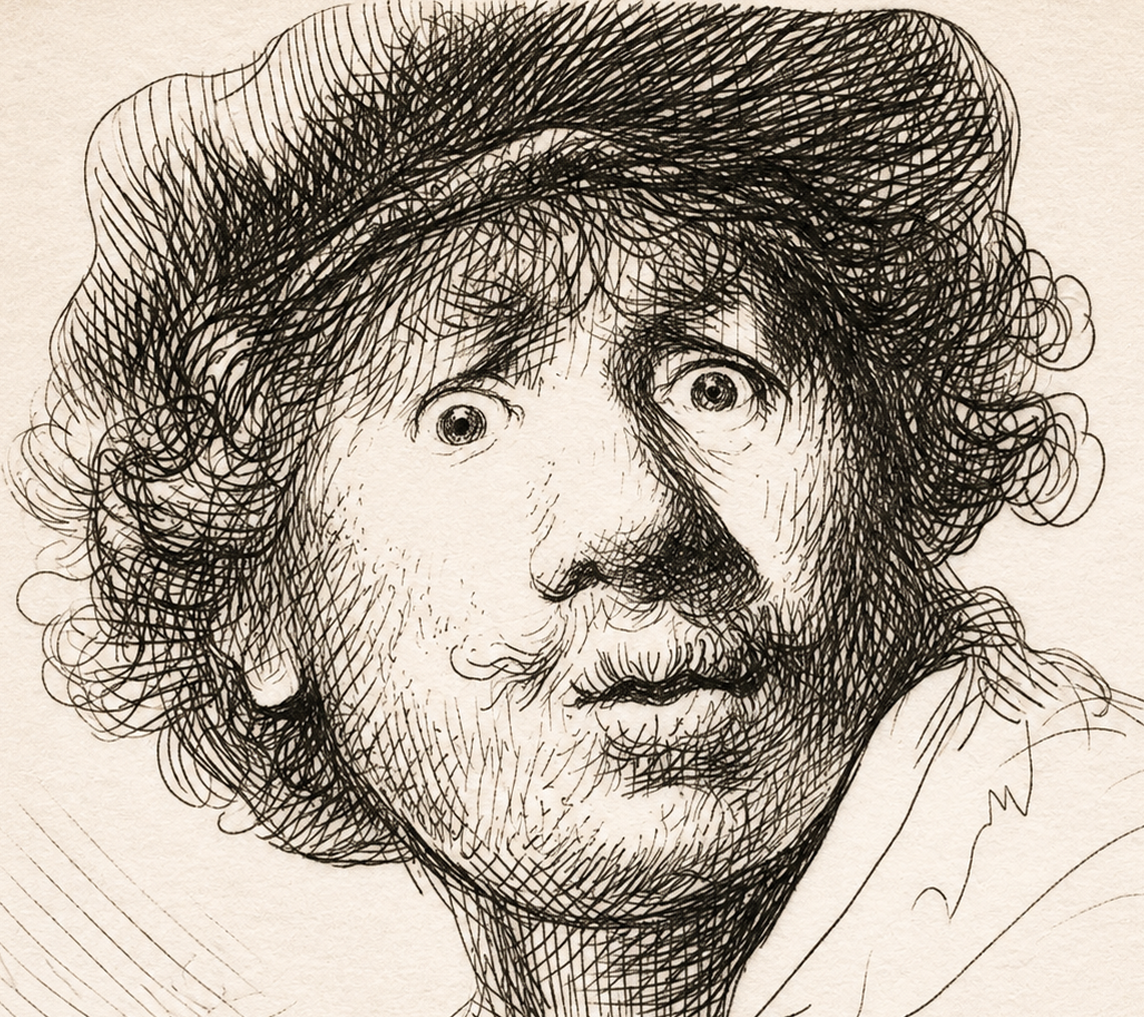}
    \caption{Rembrandt van Rijn, \emph{Self-Portrait in a Cap, Open Mouthed} (detail), 1630. Etching. Norton Simon Museum, Pasadena.}
    \label{fig_expressive_rembrandt}
\end{subfigure}
\caption{Three historical examples of expressive free-form hatching and crosshatching, illustrating the remarkable geometric and tonal freedom available to traditional artists. Dürer's form-following strokes emphasize the apparent three-dimensional structure of the face; Leonardo's flowing lines and delicate tonal variations convey both form and material characteristics; and Rembrandt's spontaneous, irregular, and overlapping strokes create a particularly expressive visual character. Such artistic representations involve considerable variation in stroke orientation, curvature, length, thickness, spacing, intensity, and overlap, often without an underlying regular geometric structure. Reproducing this degree of artistic freedom computationally remains a significant challenge. The images are AI-assisted visual reconstructions based on reproductions of the original artworks and are intended for illustrative purposes rather than precise analysis of the original strokes.}
\label{fig_expressive_hatching}
\end{figure}

Although our approach is described as free-form artistic dithering, the freedom provided by a procedural representation should be distinguished from the freedom available to a traditional artist. Figure~\ref{fig_expressive_hatching} presents illustrative reconstructions of three historical portraits that suggest the remarkable range of expressive possibilities of hatching and crosshatching. The form-following marks associated with Dürer, the flowing curves and delicate tonal variations associated with Leonardo, and the irregular, overlapping marks associated with Rembrandt illustrate different artistic approaches to representing form. Since the images have been enhanced using generative AI, we use them to motivate the discussion rather than to analyze the precise geometry of the original marks.

An artist can freely determine where an individual stroke begins and ends, how it curves, how its thickness and intensity change, and how it interacts with other strokes. The orientation and spacing of marks need not be uniform, even within a small region. Moreover, these choices need not be determined solely by the local intensity or gradient of the depicted image. Artists may deliberately emphasize or suppress particular features, change the character of strokes between adjacent regions, and introduce marks for compositional or expressive reasons. Obtaining this degree of freedom through a compact procedural representation remains a difficult problem.

Our approach provides a different kind of freedom. Instead of explicitly placing each mark, we construct entire families of patterns using generalized Hölder averages of implicitly represented pencils of lines. The stationary examples in Figures~\ref{fig_2lines}, \ref{fig_3lines}, and~\ref{fig_6lines} illustrate how the exponent and the number of pencils affect the geometry of repeating patterns. The locally adaptive examples in Figures~\ref{fig_teaser4}, \ref{fig_teaser5}, and~\ref{fig_teaser8} illustrate how this geometric control can be applied to artistic image representation. The functional formulation extends the range of pattern geometries that can be obtained from a fixed arrangement of pencils without requiring a different lattice or a separately designed dithering matrix for each pattern.

More complex and customized patterns can be constructed by using unions and intersections as guiding principles. However, designing a particular motif in this way requires some understanding of implicit geometric modeling. Explicit artistic screening techniques \citep{ostromoukhov1995artistic} offer more direct control over the placement of features and can produce patterns that are difficult to obtain with our current primitives. Conversely, the implicit approach allows patterns to be modified dynamically by changing parameters such as the orientations $\theta_i$, the number of pencils $N$, and the Hölder exponent $p$. Neither representation is universally preferable: explicit representations offer direct control over individual features, whereas functional representations facilitate continuous transformations and real-time adaptation of entire families of patterns. Combining these complementary forms of control is an interesting direction for future work.

\subsection{Beyond Binary Discretization}

A second important observation is that geometric discretization does not necessarily imply tonal discretization. Although binary thresholding provides a convenient mathematical model for hatching, crosshatching, and halftoning, the marks produced by physical image-making processes need not have only two intensity levels. Even in printing, where a binary ink-coverage model is useful, the actual appearance of a mark can vary with ink transfer, paper texture, and other physical properties of the medium. In pencil drawing, pressure and the interaction between the pencil and the paper can produce substantial variations in intensity within individual strokes. Overlapping pencil marks can also produce darker regions without a corresponding increase in their geometric coverage. Similar considerations apply to colored pencil drawings, in which overlapping marks can vary in color as well as intensity.

The current framework already supports some of this flexibility. The gray-level and color rendering examples in Figures~\ref{fig_teaser}, \ref{fig_teaser4}, and~\ref{fig_teaser5} demonstrate that procedurally constructed patterns need not be restricted to uniformly black marks on white backgrounds. The variation of the Hölder exponent illustrated in Figure~\ref{fig_asli} also provides control over the overall tonal appearance in conjunction with changes in pattern geometry. These capabilities are particularly useful in artistic applications, where tonal variation is not necessarily an error to be compensated for.

Nevertheless, a more explicit mathematical treatment of the appearance of individual marks is needed. An implicit function could define the geometric support of a stroke or dot, while an associated appearance function could independently specify its intensity, color, opacity, or internal texture. Additional compositing operations could describe the interaction of overlapping marks. Such a formulation would distinguish changes in the geometric coverage of marks from changes in their appearance and would provide a more systematic foundation for simulating pencil, ink, and other drawing media. The present results motivate this extension, but do not constitute a complete physical model of these media.

\subsection{Directional Adaptation and Spatial Coherence}

One significant limitation of the current approach arises when adjusting dithering functions to local image features. The direction-estimation process provides locally defined orientations, and direction quantization can group neighboring image locations into regions with a common orientation. In regions where the estimated directions change slowly, interpolation between nearby orientations or their associated dithering functions can produce visually smooth transitions. When directions change abruptly, however, neighboring locations can evaluate substantially different patterns, producing discontinuities, fragmented marks, and noise. Such effects are visible in some regions of Figures~\ref{fig_teaser}, \ref{fig_teaser4}, and~\ref{fig_teaser8}.

A small gradient magnitude is one important source of directional instability. In nearly uniform regions, even small intensity variations can cause large changes in the estimated gradient direction. Rapid changes in the gradient direction can also occur in regions with strong gradients, particularly near complex image structures. Thus, gradient magnitude alone is not sufficient to determine whether the local direction estimates are compatible. Applying a low-pass filter before estimating gradients can reduce abrupt variations, but it does not resolve the underlying directional ambiguity in flat regions.

Discretizing the estimated directions into a limited number of orientations provides a useful alternative. The examples in Figures~\ref{fig_lincoln7} and~\ref{fig_lincoln8} use thirty evenly spaced rotational steps, whereas the composite pattern in Figure~\ref{fig_lincoln5} uses seven. Reducing the number of orientations can encourage larger regions of uniform directionality, although it may also produce noticeable boundaries between groups of orientations and does not by itself eliminate instability in flat regions. Interpolation can help where adjacent directions are sufficiently similar; abrupt transitions may instead require a different treatment that respects the local image structure.

More fundamentally, local directional adaptation does not necessarily imply global geometric coherence. A smoothly varying direction field alone does not guarantee that independently evaluated pencils will form continuous strokes with compatible positions and spacing. The relative phases of the local patterns also matter. Developing mechanisms that coordinate orientation, spacing, and phase across neighboring regions, or that explicitly construct continuous curves following a direction field, could help address this problem. Such mechanisms should ideally preserve the interactive nature of the current formulation.

\subsection{Controlled versus Uncontrolled Irregularity}

Traditional artistic marks are rarely perfectly regular, and deviations in their shape, orientation, spacing, and intensity often contribute to the expressive character of an illustration. Our framework can introduce controlled irregularity directly through variations in the periods $T_i$ and positions $\mathbf{p}_i$ of the underlying pencils, as illustrated in Figures~\ref{fig_teaser0} and~\ref{fig_teaser4}. This provides a mechanism for exploring appearances reminiscent of handcrafted illustrations without explicitly specifying each individual mark.

However, deliberate irregularity should be distinguished from unwanted irregularity caused by unstable direction estimates or incompatible neighboring dithering functions. The former can serve an artistic objective, while the latter can obscure local image structure or disrupt the continuity of strokes. Consequently, the objective is not necessarily to eliminate all noise, but to provide more effective control over its magnitude, character, and spatial distribution. Such control would allow artists to introduce irregularity where it is visually desirable while maintaining coherent patterns elsewhere.

\subsection{Future Work}

An immediate research direction is adaptive control of the number of gradient directions based on gradient magnitude and local directional consistency. Regions with strong, coherent gradients could employ a finer discretization of orientations, while nearly uniform regions could use fewer orientations or directions propagated from neighboring regions. In regions with abrupt directional changes, feature-aware interpolation or separate treatment of adjacent regions may be more appropriate. We have not systematically tested adaptive direction quantization, and its effectiveness remains to be evaluated.

Another direction is the extension of the framework to non-uniformly spaced and curved pencils, as well as other implicit geometric primitives. These extensions could support asymmetric, non-uniform, and quasi-periodic arrangements and could provide more coherent form-following strokes. Hybrid implicit and explicit representations could also offer direct artistic control over individual strokes while retaining the advantages of functional pattern construction. A more explicit separation of geometric support and mark appearance, as discussed above, would further expand the range of drawing and printing effects that can be represented.

Spatial control of the Hölder exponent and other pencil parameters could provide gradual or deliberately abrupt transitions between different artistic styles within a single image. Extending these capabilities to animation and video would also be interesting, although temporal coherence would require additional mechanisms to prevent undesirable fluctuations in pattern appearance between frames.


The evaluation presented in this paper is primarily qualitative. Comparative studies with existing artistic screening, non-photorealistic rendering, and tiling methods, including Voronoi-based and Fourier-based approaches, would help identify the strengths and limitations of the functional representation. Mathematical evaluations of pattern geometry and structural coherence, together with performance measurements and user studies involving artists and designers, could provide a more systematic assessment. Since faithful tonal reproduction is not our primary objective, such evaluations should also consider artistic expressiveness, geometric controllability, and interactive usability. Further investigation of asymmetric patterns and unintended artifacts associated with particular parameter choices would be valuable.

Although our primary focus is artistic image representation, the construction and manipulation of geometric patterns using generalized Hölder averages may also be useful in procedural texture generation, real-time graphics, architectural ornamentation, computational design, digital fabrication, and mathematical tiling theory. These potential applications would benefit from further mathematical and experimental investigation.

\subsection{Conclusion}

We have introduced Hölder Dither, a functional framework for free-form artistic dithering that uses generalized Hölder averages of normalized distance functions associated with pencils of equidistant parallel lines. By replacing predefined dithering matrices with procedurally evaluated functions, the framework provides continuous control over pattern geometry and tonal appearance, while spatially varying parameters enable adaptation to local image features. Its pointwise formulation is particularly suitable for parallel GPU-based implementations and interactive exploration of monochromatic and color rendering styles.

Our results demonstrate that implicit geometric modeling can provide useful artistic flexibility without requiring the explicit construction of individual marks. At the same time, the expressive freedom of traditional artists suggests important opportunities for further research, particularly in the representation of non-binary marks, the construction of spatially coherent strokes, and the control of geometric irregularity. Rather than attempting to reproduce traditional artistic practices exactly, we seek to develop mathematical and computational tools that expand the possibilities for artistic exploration. In this sense, Hölder Dither continues the historical interplay between artistic and engineering techniques discussed throughout this paper.

\section{AI Usage and Disclosure}

Generative AI tools, particularly ChatGPT, were used to assist in the preparation of this manuscript, including language editing, organization of the narrative, translation of the author's previously published Turkish cartoons into English, and discussions concerning the clarity and presentation of mathematical concepts. The original research ideas, mathematical framework, artistic objectives, computational methods, and experiments were developed by the author. All AI-assisted content was reviewed and revised by the author, who takes full responsibility for the final content of the manuscript.
Generative AI was also used to assist in converting the author's original GLSL shaders into standalone browser-based applications and developing their user interfaces.

\paragraph{Historical Image Reconstructions.} The historical photographic examples (five images shown in Figures~\ref{fig_talbot_pencil_of_nature}, \ref{fig_historical_Halftone0}, and \ref{fig_historical_Halftone1}) presented in this paper are AI-assisted visual reconstructions based on surviving historical images. Generative AI was used to improve visual clarity, reconstruct typography where applicable, reduce visible deterioration, and introduce or emphasize illustrative screening patterns. Although efforts were made to preserve the original compositions and historical appearances, some visual details, textual elements, and screening patterns may differ from those of the original publications. 

In particular, the screening patterns added to the examples from Talbot's \emph{The Pencil of Nature} were not present in the original mounted photographic prints. These reconstructions shown in Figures~\ref{fig_talbot_open_door} and~\ref{fig_talbot_ladder} are intended solely for educational and illustrative purposes and should not be regarded as authentic archival reproductions or evidence of the precise technical characteristics of the original printing processes.

\paragraph{Historical Artwork Reconstructions.} Generative AI was also used to improve the visual clarity of the three historical artwork reproductions presented in Figure~\ref{fig_expressive_hatching}. The objective was to reduce visible paper deterioration, discoloration, and reproduction artifacts while enhancing the visibility of fine hatching and crosshatching marks. Although efforts were made to preserve the original compositions and artistic characteristics, the reconstruction process may have modified, removed, or introduced individual strokes and altered their tonal properties. The original media identified in the subcaptions describe the source artworks rather than the reconstructed images. These reconstructions are intended to illustrate the general expressive possibilities of traditional hatching and crosshatching rather than to provide faithful archival reproductions or evidence of the precise geometry of the original marks.

\bibliographystyle{unsrtnat}
\bibliography{references}

\end{document}